\documentclass[onecolumn]{aastex63}

\newcommand\lya{Ly$\alpha$}
\newcommand\kms{km\,s$^{-1}$}
\newcommand\oii{[\ion{O}{2}]}

\submitjournal{AJ}

\shorttitle{DESI Instrument Overview}
\shortauthors{DESI Collaboration}
\graphicspath{{./}}

\begin{document}

\title{Overview of the Instrumentation for the Dark Energy Spectroscopic Instrument}

\correspondingauthor{DESI Spokespersons}
\email{spokespersons@desi.lbl.gov}


\author{B.~Abareshi}
\affiliation{NSF's National Optical-Infrared Astronomy Research Laboratory, 950 N. Cherry Avenue, Tucson, AZ 85719, USA}

\author{J.~Aguilar}
\affiliation{Lawrence Berkeley National Laboratory, 1 Cyclotron Road, Berkeley, CA 94720, USA}

\author{S.~Ahlen}
\affiliation{Physics Dept., Boston University, 590 Commonwealth Avenue, Boston, MA 02215, USA}

\author[0000-0002-3757-6359]{Shadab Alam}
\affiliation{Institute for Astronomy, University of Edinburgh, Royal Observatory, Blackford Hill, Edinburgh EH9 3HJ, UK}

\author[0000-0002-5896-6313]{David~M.~Alexander}
\affiliation{Centre for Extragalactic Astronomy, Department of Physics, Durham University, South Road, Durham, DH1 3LE, UK}
\affiliation{Institute for Computational Cosmology, Department of Physics, Durham University, South Road, Durham DH1 3LE, UK}

\author{R.~Alfarsy}
\affiliation{Institute of Cosmology \& Gravitation, University of Portsmouth, Dennis Sciama Building, Portsmouth, PO1 3FX, UK}

\author{L.~Allen}
\affiliation{NSF's National Optical-Infrared Astronomy Research Laboratory, 950 N. Cherry Avenue, Tucson, AZ 85719, USA}

\author{C.~Allende~Prieto}
\affiliation{Instituto de Astrof\'{i}sica de Canarias, C/ Vía L\'{a}ctea, s/n, 38205 San Crist\'{o}bal de La Laguna, Santa Cruz de Tenerife, Spain}

\author{O.~Alves}
\affiliation{Department of Physics, University of Michigan, Ann Arbor, MI 48109, USA}

\author[0000-0003-3638-2584]{J.Ameel}
\affiliation{Department of Physics, University of Michigan, Ann Arbor, MI 48109, USA}
\affiliation{Physics Department, University of Michigan Ann Arbor, MI 48109, USA}

% ...

\collaboration{500}{(DESI Collaboration)}

\begin{abstract}

The Dark Energy Spectroscopic Instrument (DESI) has embarked on an ambitious five-year survey to explore the nature of dark energy with spectroscopic measurements of 40 million galaxies and quasars. DESI will determine precise redshifts and employ the Baryon Acoustic Oscillation method to measure distances from the nearby universe to beyond redshift $z > 3.5$, as well as employ Redshift Space Distortions to measure the growth of structure and probe potential modifications to general relativity. In this paper we describe the significant instrumentation we developed to conduct the DESI survey. The new instrumentation includes a wide-field, $3.2^\circ$ diameter prime-focus corrector that focuses the light onto 5020 robotic fiber positioners on the 0.812\,m diameter, aspheric focal surface. This high density is only possible because of the very compact positioner design, which allows a minimum separation of only 10.4\,mm. The positioners and their fibers are evenly divided among ten wedge-shaped `petals.' Each petal is connected to one of ten spectrographs via a contiguous, high-efficiency, nearly 50\,m fiber cable bundle. Two fibers per petal direct light into a separate system to monitor the continuum sky brightness. The ten identical spectrographs each use a pair of dichroics to split the light into three wavelength channels, and each is optimized for a distinct wavelength and spectral resolution that together record the light from $360 - 980$\,nm with a spectral resolution that ranges from 2000 to 5000. We describe the science requirements, their connection to the technical requirements on the instrumentation, the management of the project, and interfaces between subsystems. DESI was installed at the 4\,m Mayall telescope at Kitt Peak National Observatory, and we also describe the facility upgrades to prepare for DESI and the installation and functional verification process. DESI has achieved all of its performance goals, and the DESI survey began in May 2021. Some performance highlights include root-mean-squared positioner accuracy of better than $0.1''$, signal-to-noise ratio (SNR) per $\sqrt{\text{\AA}} > 0.5$ for a $z > 2$ quasar with flux $0.28 \times 10^{-17}\,\mathrm{erg\,s^{-1}\,cm^{-2}\,\text{\AA}^{-1}}$ at 380\,nm in 4000\,s, and median $\mathrm{SNR} = 7$ of the \oii\ doublet at $8 \times 10^{-17}\,\mathrm{erg\,s^{-1}\,cm^{-2}}$ in a 1000\,s exposure for emission line galaxies at $z = 1.4 - 1.6$. We conclude with additional highlights from the on-sky validation and commissioning of the instrument, key successes, and lessons learned. 

\end{abstract}


\section{Introduction} \label{sec:intro}

The goal of the Dark Energy Spectroscopic Instrument (DESI) is to determine the nature of dark energy through the most precise measurement of the expansion history of the universe ever obtained \citep{levi13}. DESI was designed to meet the definition of a Stage IV dark energy survey with only a 5-year observing campaign. The Stage IV definition was developed by the Dark Energy Task Force \citep[DETF;][]{albrecht06} to quantify the uncertainty on the dark energy equation of state parameter $w_0$ and its evolution $w_a$. The DETF Figure of Merit is the reciprocal of the area of the error ellipse in the $w_0 - w_a$ plane. DESI is a project of the U.S. Department of Energy (DOE) Office of Science, and the project used DOE funds combined with contributions from private foundations and partners to build substantial new instrumentation that can meet this definition with a survey of at least 9000 deg$^2$. The more ambitious baseline survey is to obtain spectroscopic measurements of 40 million galaxies and quasars in a 14,000 deg$^2$ footprint in five years.

DESI will measure the expansion history or distance--redshift relationship from the local universe to redshift 3.5 through precise measurements of the Baryon Acoustic Oscillation (BAO) scale. The BAO scale is a standard ruler that corresponds to a fixed comoving physical size at all redshifts. The BAO scale originates from perturbations in the early universe that excited sound waves in the primordial photon--baryon fluid prior to recombination \citep[e.g.,][]{peebles70,sunyaev70,bond84}. After recombination occurred at $z \sim 1100$, the sound speed decreased abruptly and the waves stalled. The result was a small excess of baryonic matter at a fixed physical scale of approximately 150 Mpc. This scale is detectable in the late-time clustering of the Universe, and it forms a distinctive pattern in the temperature anisotropies and polarization of the CMB that have been exquisitely mapped in a variety of experiments \citep[e.g.,][]{hinshaw13,ade16}. 

The BAO scale was first measured by \citet{eisenstein05} with data from the Sloan Digital Sky Survey (SDSS) and by \citet{cole05} with data from the 2dF Galaxy Redshift Survey. Numerous, subsequent studies have measured the BAO scale at a range of redshifts \citep[e.g.,][]{jones09,blake11,kazin14,hinton17,alam17,bautista17,dumasdesbourboux17}. Many of these studies were based on progressively larger and larger samples that culminated in the SDSS Sixteenth Data Release \citep[DR16,][]{ahumada20}, which contained more than 2.6 million galaxy and quasar redshifts, and more than four million spectra total, obtained over nearly two decades. The aggregate precision of the corresponding expansion history measurements that culminated with the SDSS Extended Baryon Oscillation Spectroscopic Survey \citep[eBOSS;][]{dawson16} is $\sim 1\%$ \citep{alam21}. Forecasts for DESI \citep{desi16a} predict a factor of approximately five to ten improvement on the size of the error ellipse of the dark energy equation of state parameters $w_0$ and $w_a$ relative to Stage-III as defined by the final eBOSS cosmology results. The range in the potential improvement largely depends on the smallest scales that are ultimately included in the measurements of the broadband power spectra of galaxies.

In addition to the expansion history and dark energy, DESI will also measure the growth of cosmic structure, provide new information on the sum of the neutrino masses, study the scale dependence of primordial density fluctuations from inflation, and test potential modifications to the general theory of relativity. DESI's measurements of large scale structure will include anisotropies in galaxy clustering, commonly referred to as Redshift Space Distortions (RSD), which probe the growth of structure \citep{kaiser87}. Measurements of RSD from anisotropies in the correlation function \citep[e.g.,][]{howlett15} are often combined with BAO measurements to obtain joint constraints on both the growth of structure and cosmological parameters \citep{tamone20, demattia20}. The aggregate precision on RSD from SDSS, BOSS, and eBOSS is $4.78\%$ \citep{alam21}, and the same work constrains the sum of neutrino masses to $\Sigma m_{\nu} < 0.115$ eV (95\% confidence). 

To achieve a gain in precision of a factor of five to ten over existing data sets, DESI will target 40 million galaxies and quasars. The targets for DESI are split into four target classes. In order of increasing average redshift these are: the Bright Galaxy Survey (BGS) targets, Luminous Red Galaxies (LRGs), Emission Line Galaxies (ELGs), and Quasars (QSOs). The BGS is a magnitude-limited sample of approximately 14 million galaxies with a median redshift of $z \sim 0.2$. The LRG sample utilizes the distinctive absorption lines of the most massive galaxies to extend to $z=1$. DESI plans to measure redshifts for approximately 8 million LRGs. The ELG sample of luminous star-forming galaxies extends to $z = 1.6$ and relies on the identification of the \oii\ emission-line doublet at rest-frame 3726, 3729\,\AA\ for a secure redshift measurement, and this sets the spectral resolution requirement for the longest-wavelength channel. The ELG sample also requires detection of a $10^{-16}\,\mathrm{ erg\,s^{-1}\,cm}^{-2}$ \oii\ doublet with signal-to-noise ratio (SNR) of $\sim 7$. DESI plans to measure redshifts for 17 million ELGs. The combination of the ELG sample size, surface density, and flux limit together set most of the high-level requirements of the survey. QSOs will be observed as both direct tracers of the matter distribution and, at $z > 2.1$, used to probe the intervening matter distribution via the intergalactic neutral Hydrogen absorption that forms the Lyman\,$\alpha$ (\lya) forest. DESI plans to measure 2.8 million QSOs, including 0.8 million at $z > 2.1$. Measurement of the \lya\ forest above $z > 2.1$ set the performance requirements at the blue end of the wavelength range.

The number of targets in each class is set by multiple factors. These include their relative value for cosmological constraints, the availability of targets, and observation considerations. The LRG and QSO samples are the highest priority for dark time, and $z>2.1$ QSOs are observed multiple times to improve measurements of the Ly$\alpha$ forest. The ELGs are the next priority for dark time. The BGS sample is observed when the sky is too bright for observations of the three fainter target classes. Given the modest density of BGS targets, DESI is also observing approximately 10 million stars (the Milky Way Survey or MWS) in conjunction with the BGS observations. All DESI target selection is based on the public Legacy Surveys \citep{dey19}. Preliminary target selection details have been published for the BGS sample \citep{ruizmacias20}, LRGs \citep{zhou20}, ELGs \citep{raichoor20}, QSOs \citep{yeche20}, and the MWS \citep{allende20}. 

The plan to observe 40 million galaxies and quasars in just five years requires an enormous increase in the number of measurements per unit time (survey speed) relative to previous experiments. The closest comparable project to DESI is SDSS, which measured about 4 million spectra over the course of nearly 20 years \citep{ahumada20}. The three key instrumentation parameters that set the survey speed are the number of spectra per observation, the exposure time of each observation, and the time between successive exposures (or the inter-exposure time). SDSS obtained their data with a 2.5\,m telescope with a $3^\circ$ diameter field of view \citep{gunn06} combined with a pair of multi-object, fiber fed spectrographs \citep{smee13} at Apache Point Observatory in New Mexico. Through the original SDSS and SDSS-II surveys, this instrumentation observed 640 spectra at a time (320 per spectrograph).  This system was expanded to 1000 spectra per exposure with an upgrade for BOSS, with the addition of key new technologies such as high-efficiency volume phase holographic gratings and CCD detectors with excellent red sensitivity. SDSS observations with these spectrographs typically included three 15\,min science exposures per field and required an additional 15\,min for field acquisition and calibration and approximately $3-5$\,min to change the fiber plug plates and slit head between fields. Observations for eBOSS consequently required somewhat over an hour to obtain approximately 1000 spectra in good conditions. In contrast, DESI can obtain 5000 spectra per observation, has a 1000\,s science exposure time in nominal conditions, and only requires approximately 2\,min to change between fields.

This substantial increase in survey speed is due to several critical characteristics of the DESI instrumentation, most notably high throughput combined with a larger telescope, a wide field of view, and robotic fiber positioners. The new DESI corrector has six lenses, each approximately a meter-diameter in diameter, that focus an 8\,deg$^2$ field of view onto an aspheric focal surface that is 0.812\,m in diameter. This focal surface is densely populated with 5020 robotic fiber positioners that have a minimum center-to-center separation of only 10.4 mm. The focal plate assembly is divided into ten, wedge-shaped petals. Each petal contains 502 fiber positioners, and a bundle of 500 fibers from each petal connects to one of ten high-efficiency, bench-mounted spectrographs that are maintained in a climate-controlled environment off the telescope. The remaining two fiber positioners per petal are connected to a dedicated sky continuum monitor system. Each of the ten spectrographs has a pair of dichroics that split the light into three wavelength channels, and each channel has a distinct spectral resolution that ranges from $2000 - 3000$ in the shortest-wavelength channel to $4000 - 5000$ in the longest-wavelength channel. This instrumentation is installed at the 4\,m Nicholas U. Mayall telescope at Kitt Peak National Observatory, which is operated by NSF's National Optical-Infrared Astronomy Research Laboratory (NOIRLab). The superb throughput of the instrumentation, combined with the larger telescope aperture,  allows for shorter science exposures relative to SDSS, even though the DESI targets are typically fainter. The stability of the system, combined with the robotic fiber positioning system, is the reason DESI can change between fields in only $\sim2$\,min. 

\begin{figure*}[ht!]
\includegraphics[width=7in]{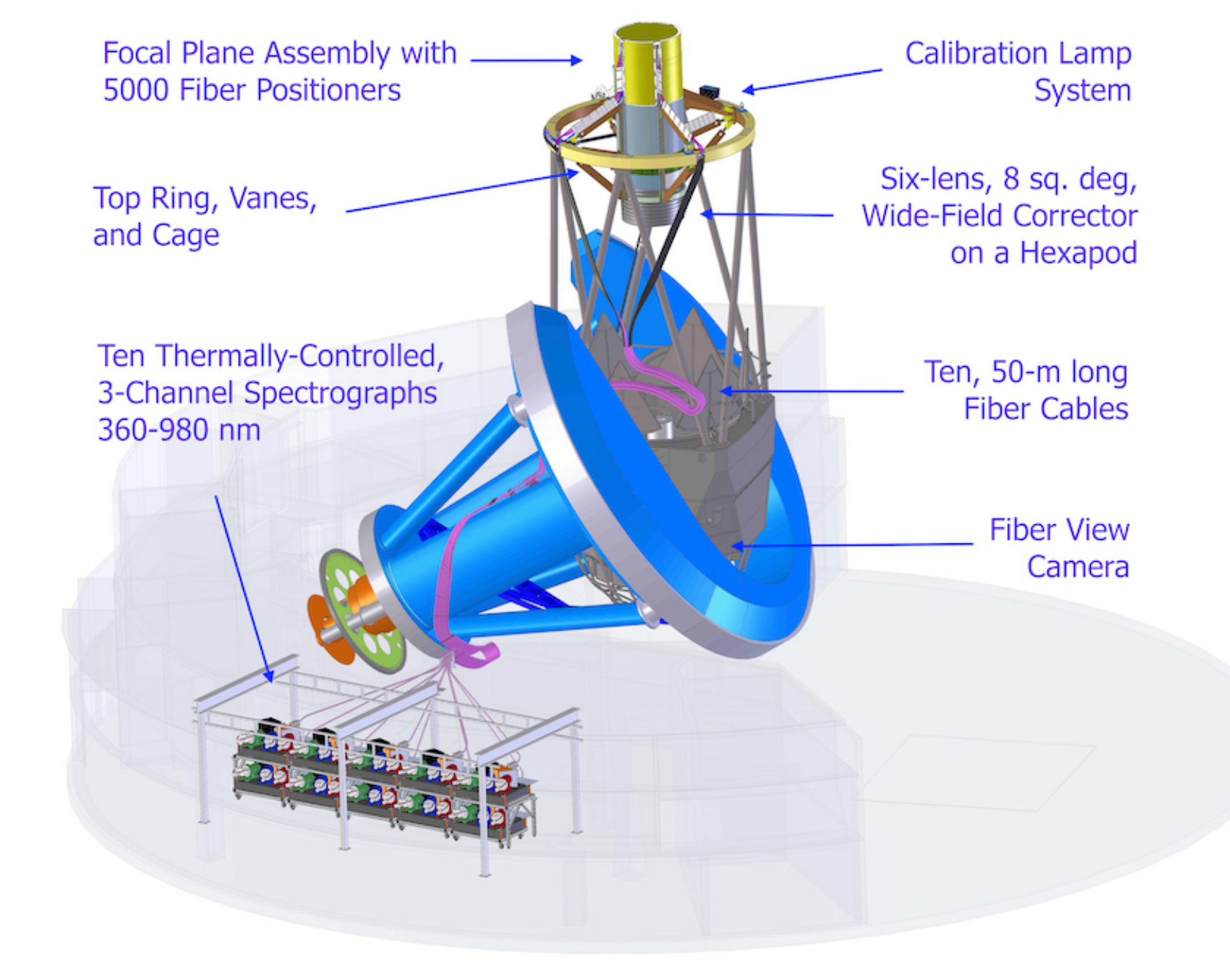}
\caption{Model of the 4\,m Mayall telescope with the new instrumentation built for the DESI project. All of the main components are labeled and described in this paper. The major subsystems are the new, 8\,deg$^2$ corrector (see \S\ref{sec:corr}), the new top ring, vanes and cage that support the corrector (see \S\ref{sec:corrsupp}), the focal plate assembly (see \S\ref{sec:fps}), the fiber system (see \S\ref{sec:fibers}), and the spectrograph system (see \S\ref{sec:spec}). Also labeled are the Fiber View Camera (see \S\ref{sec:fvc}) and the Calibration Lamp System (see \S\ref{sec:cals}). The ten spectrographs are located in a thermally-controlled environment called the ``Shack'' (see \S\ref{sec:shack}) that was custom built in the Large Coud\'e Room of the Mayall building. \label{fig:overview}}
\end{figure*}

This paper presents an overview of the instrumentation for DESI. There are also four companion papers that provide further details on specific hardware components: the corrector and the corrector support system (Miller et al. 2022), the focal plane system (Silber et al. 2022), the fiber system (Poppett et al.\ 2022), and the spectrograph system (Jelinsky et al.\ 2022). Section~\ref{sec:reqs} describes the inception of the survey, the key science requirements, the technical requirements developed from these science requirements, and the management structure we used to develop the instrument across many institutions distributed across five continents. Sections \ref{sec:corr} through \ref{sec:ics} describe the instrument hardware subsystems and Figure~\ref{fig:overview} shows an overview of these components. Each of these sections begins with a subsection that describes the key technical requirements for that subsystem. In the case of the corrector system \S\ref{sec:corr} and support structure \S\ref{sec:corrsupp}, focal plane system \S\ref{sec:fps}, fiber system \S\ref{sec:fibers}, and the spectrograph system \S\ref{sec:spec}, these sections are brief summaries of the more detailed papers. In Section \ref{sec:ics} we describe the Instrument Control System (ICS) software, which serves as the central nervous system of the instrumentation. Section \ref{sec:datasystems} provides a brief description of the data systems, which includes target selection, survey design, and the spectroscopic pipeline, with an emphasis on the components most relevant to the demonstration of instrument performance. We made numerous updates and other changes to the Mayall infrastructure to prepare for DESI, and we describe that work in \S\ref{sec:mayall}. Section \ref{sec:precmx} describes two technology demonstration and risk reduction efforts that used the Mayall telescope prior to the start of commissioning with the complete instrument. In \S\ref{sec:ait} we describe the subsystem acceptance process for each major subsystem, the installation phase, and functional verification after integration. We describe some performance results from commissioning in \S\ref{sec:results}. These include the superb total throughput of 30 -- 40\%, fiber positioning accuracy of on order $0.1''$ rms, excellent guider performance, the exceptional point spread function (PSF) stability, and our success with minimization of the inter-exposure time. We also show examples of the excellent sky subtraction, and spectra of ELGs near our flux and redshift limits. These results are followed by a discussion of some successes and lessons learned in \S\ref{sec:lessons} and a brief summary in  \S\ref{sec:sum}.

\begin{deluxetable*}{llr}
\tablenum{1}
\tablecaption{Select Major Milestones of the DESI Project\label{tab:dates}}
\tablewidth{700pt}
\tablehead{
\colhead{CD/WBS} & \colhead{Milestone} &  \colhead{Date}
}
\startdata
CD-0 & Approve Mission Need & 9/18/12 \\
CD-1 & Approve Alternative Selection and Cost Range & 3/19/15 \\
CD-2 & Approve Performance Baseline & 9/17/15 \\
CD-3 & Approve Start of Construction & 6/22/16 \\
CD-4 & Approve Project Completion & 5/11/20 \\
\hline
1.6 & EM Spectrograph Start & 9/26/13 \\
1.4 & Positioner Downselect & 6/16/14 \\
1.2 & C2 and C3 lens boule start & 6/30/14 \\
1.2 & C1 and C4 lens boule start & 8/28/14 \\
1.2 & ADC1 and ADC2 lens boule start & 1/26/15 \\
1.1 & Technical Design Report & 6/1/15 \\
1.8 & Preliminary Spectro Pipeline running end-to-end on DESI Sims & 7/20/16 \\
1.6 & EM Spectrograph Fully Verified at Vendor & 1/18/17 \\
1.2 & ADC Lenses Ground and Polished & 2/1/17 \\
1.6 & First Spectrograph Delivered & 5/15/18 \\
1.5 & First Fiber Cable with Spool Boxes Fabricated & 5/25/17 \\
1.3 & Corrector Barrel Fabrication Complete & 7/11/17 \\
1.2 & Fused Silica Lenses Ground and Polished & 10/27/17 \\
1.2 & Fused Silica Lenses Coated & 12/18/17 \\
1.2 & Borosilicate Lenses (ADC) Coated & 11/8/17 \\
1.9 & Start of Mayall Shutdown for Installation & 2/12/18 \\
1.3 & New Top Ring Delivered & 4/19/18 \\ 
1.2 & Lenses Aligned in Corrector Barrel & 6/26/18 \\
1.2 & Corrector Re-Assembled and Ready for Installation & 7/25/18 \\
1.7 & ICS Complete & 8/31/18 \\
1.6 & Spectrograph Thermal Enclosure Verified & 10/12/18 \\
1.6 & Rack System for Spectrographs Installed & 10/24/18 \\
1.8 & Target Selection Pipeline Operational & 11/8/18 \\
1.9 & Corrector Installed & 11/30/18 \\
1.9 & Commissioning Instrument Installed & 3/27/19 \\
1.4 & First Focal Plate Petal Delivered & 5/8/19 \\
1.4 & First Focal Plate Petal Installed & 6/25/19 \\
1.4 & Complete Focal Plate Assembly Delivered and Verified & 7/19/19 \\
1.9 & Focal Plate Assembly Installed & 7/24/19 \\
1.9 & Start of Commissioning & 10/22/19 \\
1.9 & All Equipment Delivered and Verified & 12/20/19 \\
1.9 & Commissioning Tasks Complete & 3/16/20 \\
    & Pause nighttime operations due to COVID-19 & 3/16/20 \\
    & Restart of nighttime operations & 11/20/20 \\
    & Start of Survey Validation & 12/14/20 \\
    & Start of the Main Survey & 5/14/21
\enddata
\tablecomments{Select major milestones of the DESI Project, including the five Critical Decision (CD) phases for a DOE project. Column 1 lists the Work Breakdown Structure (WBS) element at Level 2 that tracked the milestone, column 2 provides a brief description of the milestone, and column 3 provides the date on which we recorded successful completion of the milestone. Descriptions of the WBS elements are provided in \S\ref{sec:reqs}. Other acronyms are: EM: Engineering Model; ADC: Atmospheric Dispersion Corrector; ICS: Instrument Control System.}
\end{deluxetable*}

\section{Science Requirements and Management} \label{sec:reqs}

The key goal that drove the development of DESI was to conduct a spectroscopic survey that would meet the definition of a Stage IV dark energy survey in only five years. Early estimates showed that a survey of on order 30 million objects would meet the Stage IV definition and identified that one of the most challenging requirements would be the measurement of emission line galaxies at least as faint as $10^{-16}\,\mathrm{ erg\,s^{-1}\,cm}^{-2}$ to $z \sim 1.6$. The five year duration was motivated by considerations of hardware reliability, the engagement of scientists to conduct and analyze the data, and the timing relative to other surveys. We used these considerations to produce a conceptual design for the instrument that included a large prime-focus corrector, a focal plane with robotic fiber positioners, fiber optics cables, and bench spectrographs. We also developed survey simulations in parallel with the instrument design to evaluate potential sites. 

The first subsection describes the early inception and development timeline of the DESI project. This is followed by subsections that describe the top level (Level 1 and 2) science and survey requirements and then the corresponding Level 3 technical requirements that flow from or were imposed to meet these science requirements. This includes the rationale for the requirements. The Level 3 requirements were then used to develop Level 4 requirements on the various subsystems, which are described at the start of the section that describes each subsystem. One measure of the thoroughness of this early development work is that the Level 1, 2, and 3 requirements were written in 2014 and were not changed for the duration of the project development and construction. The last subsection describes the management of the instrument development. 

\subsection{Inception and Development Timeline} \label{sec:start}

The formal start of the project that became DESI occurred with DOE approval of Critical Decision Zero (CD-0) or Mission Need on September 18, 2012. DOE subsequently selected Lawrence Berkeley National Laboratory (LBNL) as the lead laboratory for DESI and appointed the LBNL Project Director. Extensive design studies began at this time and showed that the instrument would need a field of view of $2 - 3^\circ$ diameter, many thousands of fiber positioners, high throughput, and excellent PSF stability for sky subtraction. This period also coincided with significant research and development (R\&D) and systems engineering activity. The R\&D included the development of fiber positioners, study of alternative positioner technologies, work on spectrograph PSF stability that would be critical for sky subtraction, and further development of low-noise amplifiers for the CCDs. During this time period we had many conversations with vendors about the corrector lenses, spectrograph optics, and volume-phase-holographic (VPH) gratings. We also received generous support from the Gordon and Betty Moore Foundation and the Heising-Simons Foundation. This early financial support enabled us to begin procurements of critical, long-lead components, namely the glass for the corrector and the first spectrograph.

The systems engineering approach included an extensive model for all contributions to throughput and to system noise. Our throughput model included all of the relevant hardware components. These included the primary mirror and spectrograph collimator reflectance, transmission of optical elements, and anti-reflection (AR) coatings, as well as vignetting, blur due to optical misalignments, defocus between the fiber tips and the detectors used for guiding and wavefront sensing, fiber system focal ratio degradation, lateral alignment errors, and the gratings and detectors. The noise included sky noise, scattered light in the telescope, corrector, and spectrograph, fiber cross talk, and the detector noise. The project maintained a throughput and noise budget for each subsystem and a margin on all quantities. The allocations were initially based on estimates either derived from communications with vendors or based on calculations, and these estimates were replaced over time with measurements of the as-built components. Examples of measurements include the coatings and material absorption in the elements of the corrector, fiber system, and spectrographs, and the spectrograph detectors.

The project immediately began trade studies to support site selection. The two critical aspects of the instrument that especially influenced site selection were the total mass at prime focus and the focal ratio of the primary mirror. An instrument capable of observing tens of millions of galaxies in just five years would be very massive, and only older-generation telescopes were expected to be able to support the expected mass of the instrumentation at prime focus. Numerous design studies showed that correctors with a $2 - 3^\circ$ diameter field of view were only feasible for telescopes with a primary mirror of $f/2.5$ or slower. These considerations narrowed the list of potential telescopes down to a short list in the 4\,m class. Given this aperture size, the flux limit requirement, and survey duration, the instrument would clearly require extremely high throughput and at least a very substantial fraction of all of the dark time at the site. These technical factors and discussions about the programmatic availability of various candidates led to the selection of the 4\,m Mayall telescope in 2013. The Mayall telescope is a Ritchey-Chr\'etien (RC) design that was built to support a prime focus and a coud\'e focus, in addition to a hyperbolic secondary for the wide-field RC focus, and therefore could support substantial instrumentation at prime focus. The telescope is part of the Kitt Peak National Observatory and is located approximately 80\,km WSW of Tucson, Arizona at an altitude of approximately 2021\,m on the land of the Tohono O'odham Nation. 

The next two milestones were CD-1 (Alternative Selection and Cost Range) on March 19, 2015, and CD-2 (Performance Baseline) on September 17, 2015. During these phases we produced both a detailed design for the survey and the final instrument design, which were published as \citet{desi16a} and \citet{desi16b}, respectively. We also developed detailed performance parameters including science and system requirements, a project execution plan, risk registry, risk management plan, and other documents that are a standard part of DOE program and project management. 

At CD-2 the total project cost from DOE was set at \$56.328M as spent dollars and the date for project completion (CD-4) was set to the end of FY21. This total does not include \$19M in foreign and private (including university group) contributions, nor does it include the costs to operate the survey. U.S. Congressional approval for the start of DESI was included in the FY15 Energy and Water appropriations legislation. CD-2 approval also included authorization to begin procurements of long-lead items in advance of CD-3 (Start of Construction). Construction formally began with CD-3 approval on June 22, 2016. DESI achieved CD-4 (Project Completion) approval on May 8, 2020, one year ahead of schedule and \$2M under budget. This date marked the formal completion of the instrument development and construction phase and readiness to begin the survey. Table~\ref{tab:dates} summarizes key dates and milestones for the DESI Project. Table~\ref{tab:contribs} lists the institutions that contributed to the instrumentation, along with their principle responsibilities. 

\begin{deluxetable*}{ll}
\tablenum{2}
\tablecaption{Institutional Instrumentation Responsibilities\label{tab:contribs}}
\tablewidth{700pt}
\tablehead{
\colhead{Institution} & \colhead{Responsibilities} 
}
\startdata
LBNL        & Project Management; Project Office; Design; Lead for Focal Plane, Fiber System,  Spectrographs, \\ 
            & and Data Systems; NERSC; NIR Detectors and Electronics; Positioner R\&D \\ 
University of Arizona &	Blue Detectors \\
UC Berkeley & Optical design; Spectrograph Acquisition; Dichroics; VPH gratings; Corrector Lens Acquisition \\
Boston University   & Petal Fabrication and Metrology \\
CEA Saclay  & Cryostats \\
Durham University   & 	Fiber System \\
EPFL        & Positioner Components \\
Fermilab    & Corrector Support System and Hexapod; Telemetry Database; PlateMaker, Detectors \\
IFAE, ICE, CIEMAT, IFT        & Guide/Focus/Alignment Arrays \\
LPNHE       & Calibration System, Spectrograph Testing \\
University of Michigan	& Positioners Assembly and Testing; Petal Electronics \\
NOIRLab	    & Installation; Mayall Upgrades; Facility Operations; Data Transfer and Backup\\
Ohio State University & 	Instrument Control System; Commissioning Instrument; Spectrograph Mechanisms; \\ 
            & Rack and Shack; Sky Monitor \\
CPPM, LAM, OHP	        & Spectrograph Testing \\
UC Irvine	& Dynamic Exposure Time Calculator; Sky Monitor \\
UC Santa Cruz & Lead for Commissioning \\
University College London & 	Corrector \\
Yale University	& Fiber View Camera; Fiducials; Focal Plane Imaging \\
\enddata 
\tablecomments{Contributions to instrumentation by institution through CD-4. For each institution listed in Column (1), their contribution(s) are listed in Column (2). Acronyms are: LBNL: Lawrence Berkeley National Laboratory; CEA Saclay: Commissariat \`a l'\'energie atomique at Saclay; EPFL: \'Ecole polytechnique f\'ed\'erale de Lausanne; Fermilab: Fermi National Accelerator Laboratory; IFAE: Institut de F\'isica d'Altes Energies; ICE: Institut de Ciències de l’Espai; CIEMAT: Centro de Investigaciones Energéticas, Medioambientales y Tecnológicas; IFT: Instituto de Física Teórica; LPNHE: Laboratoire de Physique Nucleaire et de Hautes Energies; NOIRLab: NSF's National Optical-Infrared Astronomy Research Laboratory; CPPM: Centre de Physique des Particules de Marseille; LAM: Laboratoire d'Astrophysique de Marseille; OHP: Observatoire de Haute-Provence (now the Observatoire des Sciences de l'Univers Institut Pyth\'eas).}
\end{deluxetable*}

\subsection{Top Level Science and Survey Requirements} \label{sec:topscireq}

There are four Level 1 science requirements that motivate the survey dataset. These requirements set the survey area, the aggregate precision for how well DESI will measure the isotropic cosmic distance scale $R(z)$ from BAO at $0 < z < 1.1$, the precision for how well DESI will measure the Hubble parameter at $1.9 < z < 3.9$, and the allowable size of systematic errors from instrumental and observational methods for the angular diameter distance and the Hubble parameter. The Level 1 requirements are listed in Table~\ref{tab:level12}. 

The four Level 1 requirements were used to develop Level 2 requirements on the survey dataset. These are a series of requirements on the average density, on random, systematic, and catastrophic redshift errors, and on redshift completeness for the three dark time target classes. The \lya\ QSO sample includes specific requirements on the SNR and on the contribution of calibration errors on the flux power spectrum. There are specific fiber completeness requirements on each target class that impact the density of positioners on the focal plane as well as the relative priorities of targets. The Level 2 requirements are also listed in Table~\ref{tab:level12}. These were the requirements during the development of the instrumentation. Some modifications to the Level 2 requirements were made after the instrumentation was complete, most notably the addition of requirements for BGS.

\subsection{Top Level Technical Requirements} \label{sec:toptechreq}

\begin{deluxetable*}{ll}
\tablenum{3}
\tablecaption{Level 1 and 2 Imposed and Derived Technical Requirements \label{tab:level12}}
\tablewidth{700pt}
\tablehead{
\colhead{Number} & \colhead{Text} 
}
\startdata
L1   & {\bf Scientific Requirements} \\
L1.1 & The DESI survey shall cover at least 9000 deg$^2$. The baseline survey with margin covers 14,000 deg$^2$. \\
L1.2 & The DESI galaxy and low-$z$ quasar survey will measure the isotropic cosmic distance scale $R(z)$ from the BAO \\ 
     & method to 0.28\% precision aggregated over the redshift bin $0.0 < z < 1.1$ and to 0.39\% precision in the redshift \\
     & bin $1.1 < z < 1.9$. For the baseline survey with margin, the distance scales will be measured to 0.22\% and \\ 
     & 0.31\% precision.\\ 
L1.3 & DESI will measure the Hubble parameter at $1.9 < z < 3.7$ from the BAO method to 1.05\%; 0.84\% for the \\ 
     & larger baseline survey. \\
L1.4 & The galaxy survey at $z < 1.5$ shall be capable of separately determining $D_A(z)$ and $H(z)$ from the BAO without  \\ 
     & instrumental and survey uncertainties degrading the performance available from the sky. In particular, the  \\ 
     & systematic errors from the instrument and observational methods must not exceed 0.16\% for $D_A$ and 0.26\% for $H$. \\
L2   & {\bf Survey Data Set Requirements} \\
L2.1 & DESI will conduct a spectroscopic survey of luminous red galaxies (LRGs), emission line galaxies (ELGs), and \\ 
     & quasars (QSOs) that provide continuous coverage in redshift out to $z \sim 3.7$. \\
L2.2 & {\bf Luminous Red Galaxies} \\
L2.2.1 & The average density with redshift $0.4 < z < 1.0$ shall be at least 300 deg$^{-2}$.\\
L2.2.2 & The random redshift error in a $\sim$Gaussian core shall be less than $\sigma_z = 0.0005 (1 + z)$ (150\,\kms\,rms). \\
L2.2.3 & Systematic inaccuracy in the mean redshift shall be less than $\Delta z = 0.0002 (1+z)$ (60\,\kms). \\
L2.2.4 & Catastrophic redshift failures exceeding 1000\,\kms\,shall be $<5$\%. \\
L2.2.5 & The redshift completeness shall be $>95$\% for each pointing averaged over all targets that receive fibers.\\
L2.3 & {\bf Emission Line Galaxies} \\
L2.3.1 & The average density of successful observations shall be at least 1280 deg$^{-2}$ for $0.6 < z < 1.6$.\\
L2.3.2 & The random redshift error in a $\sim$Gaussian core shall be less than $\sigma_z = 0.0005 (1 + z)$ (150\,\kms\,rms). \\
L2.3.3 & Systematic inaccuracy in the mean redshift shall be less than $\Delta z = 0.0002 (1+z)$ (60\,\kms). \\
L2.3.4 & Catastrophic redshift failures exceeding 1000\,\kms\,shall be $<5$\%. \\
L2.3.5 & The redshift completeness shall be $>90$\% for each pointing averaged over all targets above the \oii\ flux limit.\\
L2.4 & {\bf Tracer Quasars} \\
L2.4.1 & The average density of successful observations shall be at least 120 deg$^{-2}$ for $z < 2.1$.\\
L2.4.2 & The random redshift error in a $\sim$Gaussian core shall be less than $\sigma_z = 0.0025 (1 + z)$ (750\,\kms\,rms). \\
L2.4.3 & Systematic inaccuracy in the mean redshift shall be less than $\Delta z = 0.0004 (1+z)$ (120\,\kms). \\
L2.4.4 & Catastrophic redshift failures exceeding $1000\, \mathrm{km\,s}^{-1}$ shall be $<5$\%. \\
L2.4.5 & The redshift completeness shall be $>90$\% for each pointing averaged over all targets.\\
L2.5 & {\bf Ly-$\alpha$ Quasars} \\
L2.5.1 & The average density of successful observations at $z > 2.1$ and $r < 23.5$ mag shall be at least 50 deg$^{-2}$.\\
L2.5.2 & The redshift accuracy shall be $\sigma_z = 0.0025 (1 + z)$ (equivalent to $750\, \mathrm{km\,s}^{-1}$ rms). \\
L2.5.3 & The catastrophic redshift failures shall be $<2$\%. \\
L2.5.4 & The $S/N$ per Angstrom (observer frame) shall be greater than 1 in the Ly$\alpha$ forest for $g = 23$ mag and scale with \\ 
       & flux for brighter quasars. \\
L2.6 & {\bf Spectrophotometric Calibration} \\
L2.6.1 & The Ly$\alpha$ QSO fractional flux calibration errors shall have power less than $1.2\,\mathrm{km\,s}^{-1}$ at $k \sim 0.001\,\mathrm{s\,km}^{-1}$. \\
L2.7 & {\bf Fiber Completeness} \\
L2.7.1 & The fraction of targets that receive a fiber shall be at least 80\%. \\
L2.8 & {\bf Target Selection} \\
L2.8.1 & The LRG target density shall be 350 per deg$^2$, with at least 300 per deg$^2$ successfully measured. \\
L2.8.2 & The ELG target density shall be 2400 per deg$^2$, with at least 1280 per deg$^2$ successfully measured. \\
L2.8.3 & The low-$z$ tracer QSO target density shall be 170 per deg$^2$, with at least 120 per deg$^2$ successfully measured. \\
L2.8.4 & The Ly$\alpha$ QSO target density shall be 90 per deg$^2$, with at least 50 per deg$^2$ successfully measured. \\
\enddata
\tablecomments{Level 1 Scientific Requirements and Level 2 Survey Data Set Requirements. The top-level requirements (Level 1) motivated the survey dataset (Level 2) and the experimental implementation (Level 3, see Table~\ref{tab:level3}). These are the requirements during the development of the instrumentation. We made some modifications to the Level 2 requirements after the instrumentation was complete, most notably to add requirements for BGS.}
\end{deluxetable*}

\begin{deluxetable*}{ll}
\tablenum{4}
\tablecaption{Level 3 Imposed and Derived Technical Requirements \label{tab:level3}}
\tablewidth{700pt}
\tablehead{
\colhead{Number} & \colhead{Text} 
}
\startdata
L3.1 & {\bf Implementation} \\
L3.1.1 & The spectral range shall be $360 - 980$\,nm. \\
L3.1.2 & Spectral Resolution shall be (a) $> 1500$ at wavelengths $360 < \lambda < 555$\,nm, (b)$> 3000$ at wavelengths $555 < \lambda < 656$\,nm, \\ 
       & and (c) $> 4000$ at wavelengths $656 < \lambda < 980$\,nm. \\
L3.1.3 & The median $S/N = 7$ flux limit will be $10, 9, 9, 8$, and $9 \times 10^{-17}\,\mathrm{erg\,s^{-1}\,cm}^{-2}$ in redshift bins of $0.6-0.8, 0.8-1.0$,  \\
       & $1.0-1.2, 1.2-1.4$ and $1.4-1.6$ for an \oii\ doublet emission line in an ELG with an exponential half-light radius of \\ 
       & $0.45''$ observed in $1.1''$ seeing with the sky spectrum under median, dark-sky, photometric conditions. \\ 
L3.1.4 & The fiber density shall be less than 700 per square degree. \\
L3.1.5 & The field of view shall be no less than 7.65 square degrees. \\
L3.1.6 & The spectroscopic PSF shall be characterized for all fibers in each science exposure over the full wavelength range such   \\
       & that the PSF bias shall not exceed 1\%. \\
L3.2 & {\bf Programmatic} \\
L3.2.1 & The 9,000 deg$^2$ survey shall complete in 4 years including 6 months commissioning and validation. A goal is \\ 
       & 14,000 deg$^2$ survey in the same period but not more than 5 years plus commissioning and validation. \\ 
L3.2.2 & A target galaxy and QSO catalog shall be assembled to a depth of $r = 23.4$\,mag with astrometric errors not exceeding \\ 
       & 100\,mas rms for each target class. \\
L3.3 & {\bf Environmental} \\ 
L3.3.1 & Median seeing shall be assumed to be $1.1''$ FWHM, characterized by a galaxy Moffat profile with $\beta=3.5$. \\
L3.3.2 & DESI shall meet all of its requirements while observing with zenith angles between 0 and $60^\circ$. \\
L3.3.3 & Telescope guiding accuracy shall be assumed to be 100\,mas rms. \\
\enddata
\tablecomments{Level 3 Imposed and Derived Technical Requirements, which are motivated by the top-level science requirements (Level 1) and the survey dataset requirements (Level 2) that are listed in Table~\ref{tab:level12}. The L3.1 Implementation requirements are the top-level requirements on the instrumentation, including the overall throughput, fiber density, and spectral range. The L3.2 Programmatic requirements constrain the survey design. The L3.3 Environmental requirements describe the expected performance of the instrument and telescope. We made some modifications to the Level 3 requirements after the instrumentation was complete}
\end{deluxetable*}

The L2 requirements flowed down to eleven L3 requirements that are listed in Table~\ref{tab:level3}. These are divided into three categories: implementation requirements, programmatic requirements, and environmental requirements. The six implementation requirements directly flow to the Level 4 technical requirements on the subsystems. The two programmatic requirements relate to the survey area and the input catalog. The three environmental requirements include the image quality, range of zenith angle, and telescope guiding. 

DESI utilizes a spectral range of $360 - 980$\,nm (requirement L3.1.1) because of the Level 1 redshift range requirement. Specifically, the short wavelength cutoff of 360\,nm aids in the identification of the Balmer break at $364.6$\,nm in very low redshift interlopers that could contaminate the target classes. This limit also enables the detection of the \lya\ emission line at the observed wavelength of $377$\,nm at $z = 2.1$, the entire \lya\ forest between \lya\ and Ly$\beta$ at $z > 2.5$, and the use of the Cd $361.0$\,nm arc lamp line for wavelength calibration. The long-wavelength cutoff enables detection of the \oii\ doublet in ELGs up to $z =1.6$, which corresponds to the observed wavelength of $970$\,nm. This long-wavelength cutoff also enables more secure redshift measurements for other targets, most notably the rest-frame $400$\,nm break feature in LRGs can be measured up to $z \sim 1.4$.

The spectral resolution $\lambda/\Delta\lambda$ requirement L3.1.2 flows from the requirements on redshift accuracy, precision, and on the fraction of catastrophic redshifts, especially for the ELG sample. The resolution is sufficient to identify the \oii\ doublet as a pair of lines for galaxies at $z > 0.49$. This is important as the \oii\ doublet will usually be the only detected spectral feature and resolution of the doublet is critical for an unambiguous redshift measurement. The spectral resolution may be less at lower redshifts ($z< 0.49$) because at those redshifts the $H\alpha$ line is also visible to enable a secure redshift.

The $SNR = 7$ requirement on the \oii\ doublet (L3.1.3) at various fluxes and redshifts sets joint requirements on total instrument throughput, the resolution, read noise, and the environmental conditions. These limits are set in redshift bins because the \oii\ doublet is observed through a thicket of night sky emission features that produce a modulation of the flux limit as a function of redshift within each $\Delta z = 0.2$ redshift bin. The fiber density requirement (L3.1.4) and the field of view requirement (L3.1.5) are motivated by the expected density of targets and the number of spectra required to meet the science requirements. They flow from three Level 2 requirements: the complete survey area $A_{\rm survey}$, the minimum time required to observe any one field or tile $T_{\rm tile}$, and the total survey duration $T_{\rm survey}$. Since a circular field of view does not perfectly tile the sky, DESI initially adopted a correction factor of $F_{\rm over} = 1.21$ for the amount of overlap with neighboring tiles, although subsequently changed to a \citet{hardin01} icosahedral tiling with multiple layers before the start of the survey. Accurate knowledge of the 2-D spectroscopic PSF shape is required to extract the spectra and obtain unbiased measurements of the flux, noise, and resolution. We characterize uncertainty in the PSF measurement as the PSF bias, which can produce artifacts in spectra, such as distortions of absorption or emission features, or lead to mis-modeling of the sky subtraction derived from other spectra in the same exposure. Requirement L3.1.6 sets a limit on the PSF bias. 

The four programmatic requirements set the area and duration of the bright and dark time surveys, the conditions under which the dark time program will be executed, the astrometric errors of the input catalog, and the uniformity of the data. Requirement L3.2.1 on area and duration is driven by total operation costs. This in turn sets requirements on throughput and the number of targets per exposure, and it was also used to set a requirement on the inter-exposure time of two minutes. The inter-exposure time is the time between closing the shutter on one spectroscopic exposure and opening the shutter to start the next. The inter-exposure time requirement impacts nearly every subsystem, as between exposures we read out the previous spectroscopic exposure, move the telescope and dome, adjust the Atmospheric Dispersion Corrector (ADC), adjust the hexapod, adjust the fiber positioners for the next asterism, acquire the new field, and start guiding.  Requirement L3.2.2 quantifies the allowable astrometric errors, which flows down from the target densities for each target class and the total fiber positioning error budget. This requirement is met by the Legacy Surveys dataset \citep{dey19}. 

The three environmental requirements are on the median seeing, zenith distance range, and telescope guiding. The median seeing (requirement L3.3.1) is motivated by the historical performance of the Mayall with the MOSAIC instrument \citep{dey14} and effectively is a requirement that upgrades to the telescope do not negatively impact the seeing. This requirement impacts the optical design, throughput calculation, and exposure time per tile. Requirement L3.3.2 states that DESI shall meet all requirements while observing between zenith angles from $0 - 60^\circ$. This is motivated by the survey footprint and the flexibility needed to target fields at a range of airmass. This requirement impacts the optical design (most notably the ADC), throughput, and exposure time. Finally, L3.3.3 requires the telescope guiding accuracy to be 100\,mas rms. This is also motivated by the historical performance of the Mayall telescope and is a component of the fiber positioning error budget. 

\subsection{Management and Personnel} \label{sec:manage}

The management of the DESI project followed standard Department of Energy guidelines for similar projects. The project was led by the LBNL Project Director, who was responsible and accountable for the successful execution of the construction. The Project Director also represented the project to the DOE, LBNL, and other participating institutions. The LBNL Project Manager was responsible for the management and safe execution of the DESI Project. This includes adherence to the technical, cost, and schedule baselines, and risk and integrated safety management. 

DESI includes two co-Spokespersons. The Spokespersons represent the science collaboration and are responsible for encouraging the productivity of the collaboration, the performance of the science working groups, and advising the Project Director. The Project Director is also advised by an Executive Board, who are appointed by the Project Director from the senior membership of the project and the science collaboration. They provide advice on the scientific scope, mission objectives, operational priorities, and membership issues. 

In addition to the Director and Project Manager, the DESI Project Office included two Project Scientists, an Instrument Scientist, Commissioning Scientist, Systems Engineer, and Safety Officer during the construction phase\footnote{After commissioning was complete, the Commissioning Scientist role changed to a co-Instrument Scientist role.}. The Project Scientists were responsible for maintaining the science requirements, evaluating the scientific impact of any changes to hardware during construction, preparation of software, development of the survey strategy, and the overall plan for commissioning and operations. The responsibilities of the Instrument Scientist were to ensure that the instrument met the technical requirements, evaluate the technical impact of changes or trade-offs during construction, and plan for commissioning and operations. The Commissioning Scientist was responsible for commissioning planning, and the organization of commissioning observations and analysis. The Systems Engineer was responsible for high-level systems engineering and value management oversight of all aspects of the project, such as interface control documents, maintenance of budgets for system-level requirements such as throughput, and technical alternative studies to determine the best design solutions. The DESI Safety Officer was responsible for ensuring that Environment, Health, and Safety concerns are addressed, and that integrated safety management was implemented in all phases of DESI construction. 

The DESI Project Manager leads the Technical Board, which during construction provided information on progress toward scientific and technical goals, along with budget and schedule information, and presently focuses on performance during operations, maintenance, and potential improvements. During construction the Technical Board included the Project Director, Project Manager, Project Scientists, Instrument Scientist, Systems Engineer, WBS Level 2 Managers and Cognizant Scientists, and the Project Safety Officer. Each Level 2 Manager was paired with a Cognizant Scientist to efficiently address the scientific impact of technical issues. The Systems Engineer, along with the Project Director, Project Manager, and Project Scientists formed the Change Control Board, which evaluated changes to technical scope. The Level 2 Managers were responsible for the completion of the tasks in their subsystem. This included the identification of the resources necessary to complete the subsystem within the projected schedule and budget, meet the subsystem design requirements, and conform to interfaces with other subsystems. 

\section{Corrector} \label{sec:corr}

\begin{figure*}[ht!]
 \includegraphics[width=7in]{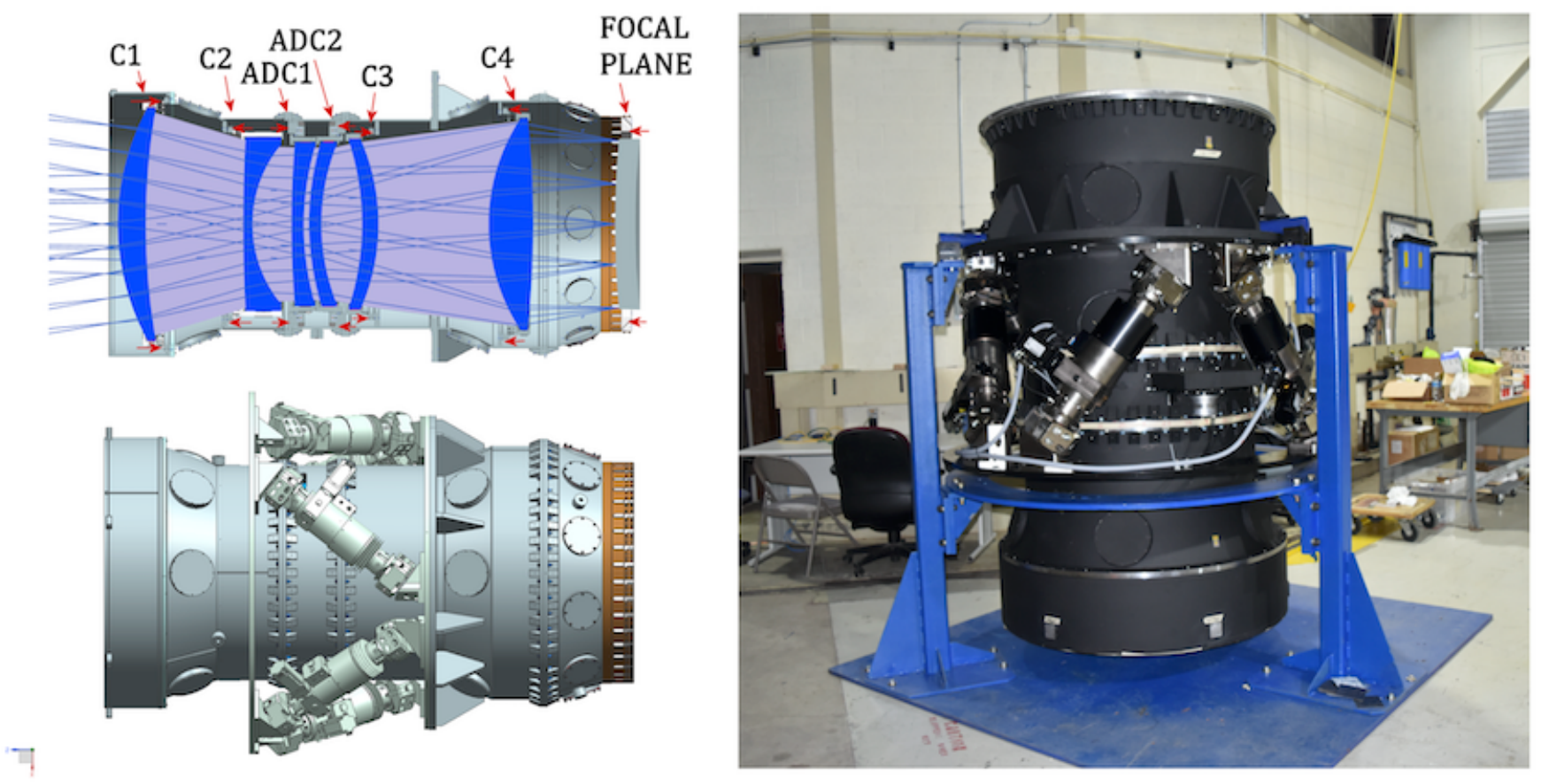}
 \caption{DESI corrector barrel. ({\it Left}) Model of the corrector that emphasize the lenses ({\it top left}) and the barrel design with the hexapod ({\it bottom left}). ({\it Right}) Photo of the reassembled corrector barrel on the ground floor of the Mayall telescope in August 2018. \label{fig:corrbarrel}}
 \end{figure*}

\begin{figure*}[ht!]
 \includegraphics[width=7in]{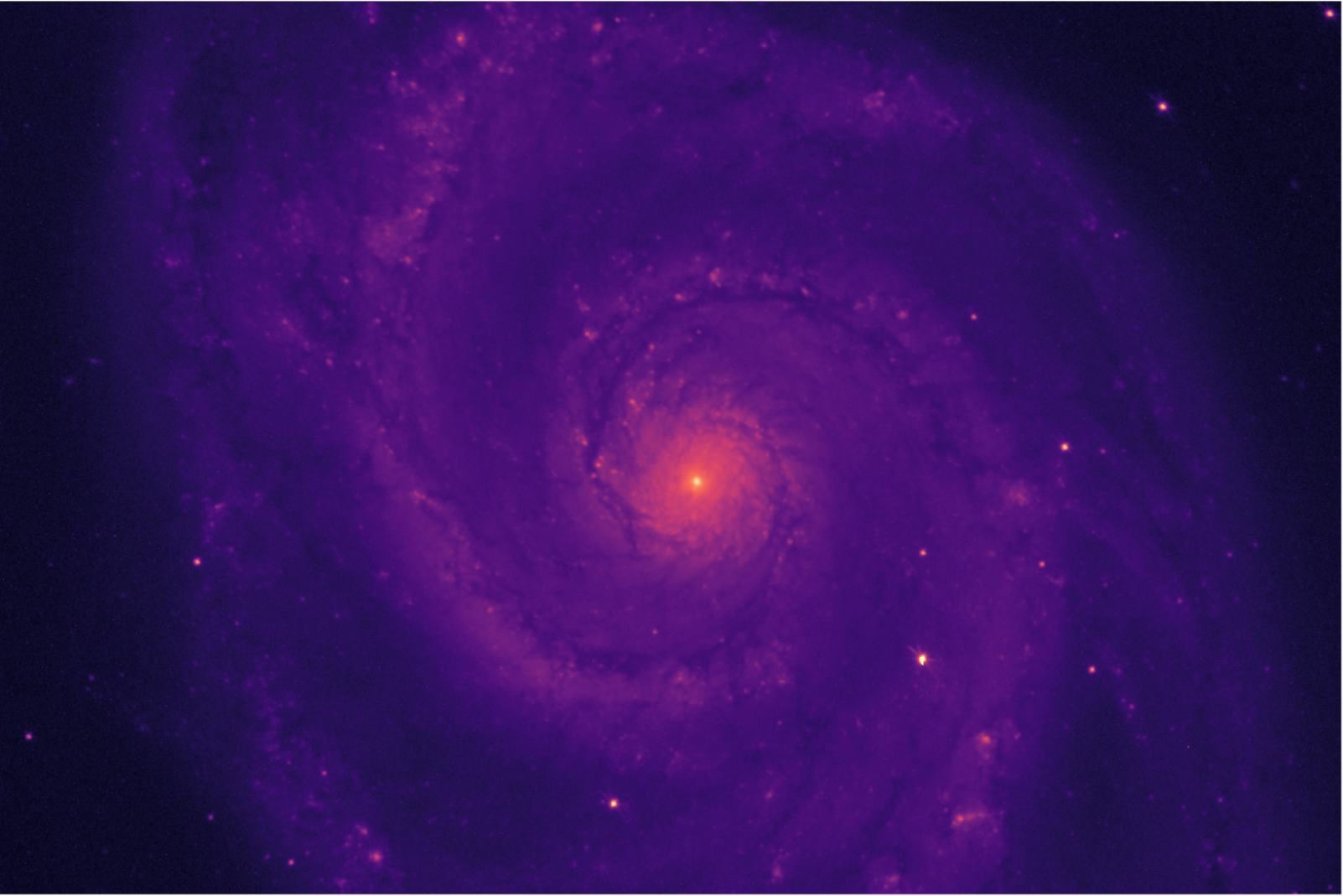}
 \caption{False color image of the nearby galaxy M51 obtained with the corrector on the first night of observations (1 April 2019). This $r-$band image was obtained with the Commissioning Instrument (see \S~\ref{sec:ci}) and shows M51 at the center of the corrector field of view. The image quality is approximately $0.65''$ FWHM. The size of the image is approximately $6.5' \times 4.5'$.  North is up and East is to the left. \label{fig:firstlight}}
 \end{figure*}

\begin{figure*}[ht!]
 \includegraphics[width=7in]{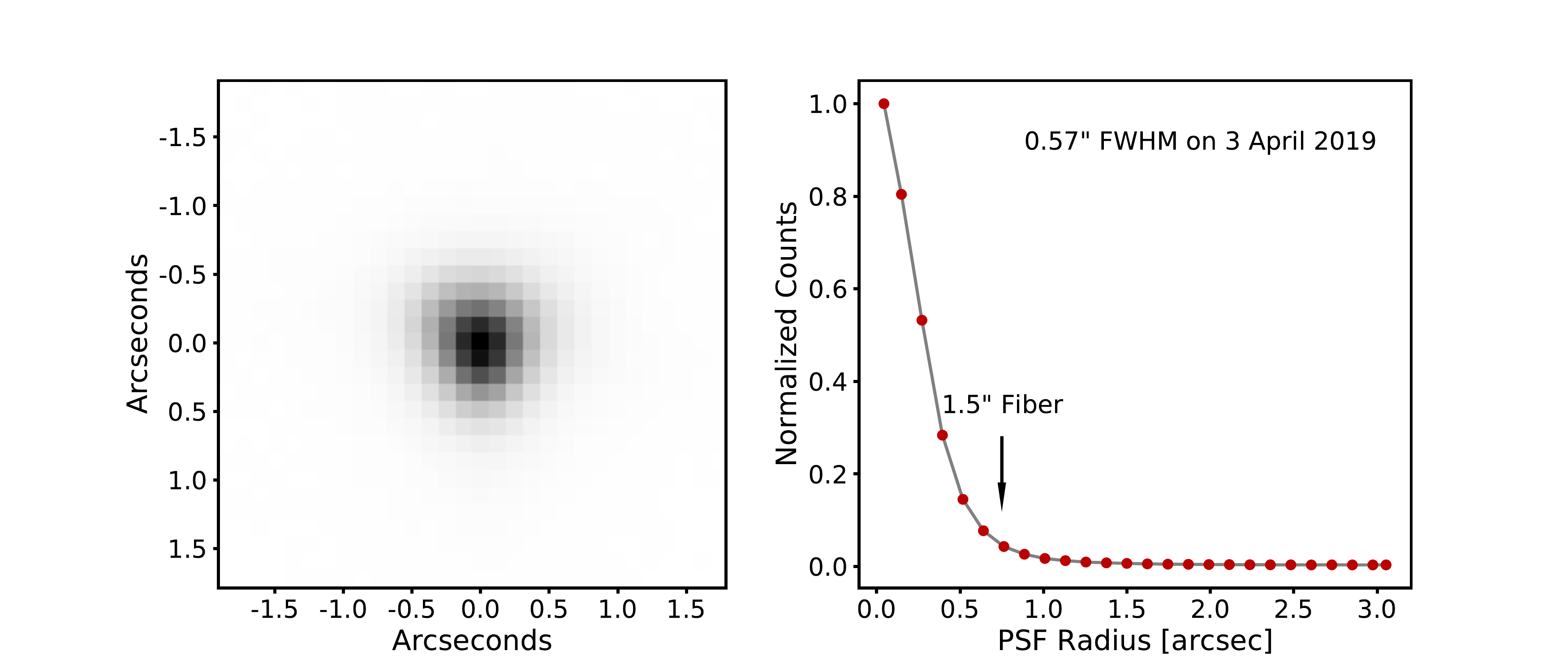}
 \caption{Best delivered image quality obtained with the corrector during commissioning. On 3 April 2019 we obtained somewhat better than $0.6''$ FWHM images on axis. ({\it Left Panel}) This image is the average of six bright stars observed with the Commissioning Instrument, which has a plate scale of approximately $0.13''$ per pixel. ({\it Right Panel}) Average radial profile of the six stars. The vertical arrow marks the size of a nominal $1.5''$ diameter fiber.  \label{fig:corrpsf}}
 \end{figure*}

The new DESI prime focus corrector has six lenses, each approximately a meter in diameter, that change the focal ratio of the telescope from $f/2.8$ to approximately $f/3.9$ and produces a $3.2^\circ$ diameter field of view. Four of the lenses are fused silica (lens elements C1 through C4), and two of them have one aspheric surface. The remaining two lenses form an Atmospheric Dispersion Corrector (ADC; lens elements ADC1, ADC2) that are located between C2 and C3 and were figured from borosilicate glass. These six lenses are held in cells mounted in a steel barrel assembly, and the cells for the two ADC elements rotate to correct for dispersion by the atmosphere. Figure~\ref{fig:corrbarrel} shows the optical and mechanical design of the corrector, as well as a photo of the complete corrector barrel prior to installation. Figure~\ref{fig:firstlight} shows an image of M51 obtained on the first night of observations with the corrector, and Figure~\ref{fig:corrpsf} shows the image quality from several nights later when the seeing reached $0.57''$ full width at half maximum (FWHM). 

The first element of the DESI corrector is approximately 8\,m above the primary mirror of the Mayall telescope. The telescope's primary mirror was polished from a 61\,cm thick disk of fused quartz. The telescope was originally built as a Ritchey-Chr\'etien (RC) design, and thus the mirror has a hyperbolic shape. The unmasked aperture is 3797 mm and the central hole diameter is 1324 mm. The telescope was initially designed to support a prime focus and a coud\'e focus, in addition to a hyperbolic secondary for the wide-field RC focus. The top ring of the telescope was consequently constructed as a pair of concentric rings (a `split ring') such that the inner ring could be flipped around to switch between a secondary mirror and prime focus. The flip ring was eliminated for DESI to save significant weight.

The subsections briefly summarize the technical requirements, optical design, procurement, coating of the corrector lenses, and the system performance. The lens barrel, ADC mechanism, and assembly are described in the following section on the corrector support system. Previous descriptions of the corrector include \citet{sholl12}, \citet{doel16} and \citet{miller18}. Miller, Doel, Gutierrez et al.\ (2022) {\it in preparation} will describe both the corrector and corrector support system in greater detail. 

 \subsection{Technical Requirements}

The corrector optical design was guided by a series of Level 4 requirements, as well as a set of Level 5 specifications with implementation details. The key Level 4 requirements fall into several broad classes: wavelength range, image quality, throughput, and coupling with the fiber system.  The Level 3 requirement on the bandpass of $360-980$\,nm applies to both the image quality and throughput. The image quality requirement on the FWHM at zenith is a field-weighted mean of $<0.4''$ over the full bandpass, and a maximum of $<0.6''$ from $360 - 450$nm and $<0.5''$ from $450 - 980$nm. At $60^\circ$ from zenith, the maximum FWHM is $<0.75''$ and $<0.6''$ over these two wavelength ranges. The image quality requirement was one of the most significant design challenges, as it required the minimization of lateral chromatic aberrations over this entire bandpass. In addition, atmospheric dispersion at $60^\circ$ from zenith corresponds to several arcseconds between the wavelength extremes, which constrained the design of the ADC system. The throughput specification on the optical design only applied to the glass transmission, as there was a separate requirement on the performance of the optical coatings. The glass throughput requirement varied with wavelength, and the coating requirement was an average transmission of $98.5$\% per surface over the entire bandpass. There were also specifications on the homogeneity of the glasses, as inhomogeneity can degrade image quality.

Another significant design challenge was the requirement that the focal surface must be sufficiently large to contain the 5020 fiber positioners, as well as the Guide/Focus/Alignment (GFA) cameras and fiducials, and have an angular field of view of at least $3^\circ$ (driven by requirement L3.1.5, as described above in \S\ref{sec:reqs}). The chief ray deviation at the focal surface was also constrained to be $<0.5^\circ$  on average and $<1^\circ$ maximum in order to obtain good coupling into the fiber cables and avoid substantial tilts of the fiber positioners that would both complicate the focal surface mechanical design and impact the filling factor of the positioners on the focal surface. One compensation to the complexity of the optical design requirements was that the focal surface did not need to be flat. Instead the requirement was that the radius of curvature must be greater than 3000 mm (convex). To limit the complexity of lens fabrication and testing, we imposed requirements to have no more than one aspheric surface per lens and to limit the maximum aspheric departure to no more than 30 mrad. The constraint on the maximum aspheric departure was recommended by potential vendors in order to allow interferometric testing with better than Nyquist fringe sampling. There were some early studies that attempted an all-spherical design, but those could not produce more than half the field of view of the final design. 

 \subsection{Optical Design} \label{sec:optdesign}
 
 The corrector has four large fused silica elements. C1 and C4 are the largest lenses, both more than 1\,m in diameter with two spherical surfaces. C1 has a diameter of 1.14\,m and mass of 201 \,kg, and C4 has a diameter of 1.03\,m and mass of 237\,kg. C2 and C3 are both approximately 0.8\,m in diameter, and the first surface of both lenses is an even asphere. The ADC lenses are $\sim0.8$\,m diameter prisms figured from borosilicate, and each has two wedged, spherical surfaces ($\sim0.25^\circ$) for first-order correction of atmospheric dispersion away from zenith.  The total mass of glass in all six lenses in 864\,kg.\footnote{For comparison, the largest DECam lens was 0.98\,m in diameter and 176\,kg.} There are no bonded joints in the ADC in order to reduce the system complexity and risk. The dispersion magnitude is set by the rotation of the ADC lenses relative to one another, and the direction of the correction is set by their net orientation relative to the parallactic angle. The optical system is asymmetric due to the presence of the wedged, spherical ADC elements. We consequently characterized the design image quality across the field of view to verify performance, and not simply as a function of radius. 
 
 \begin{figure*}[ht!]
 \includegraphics[width=7in]{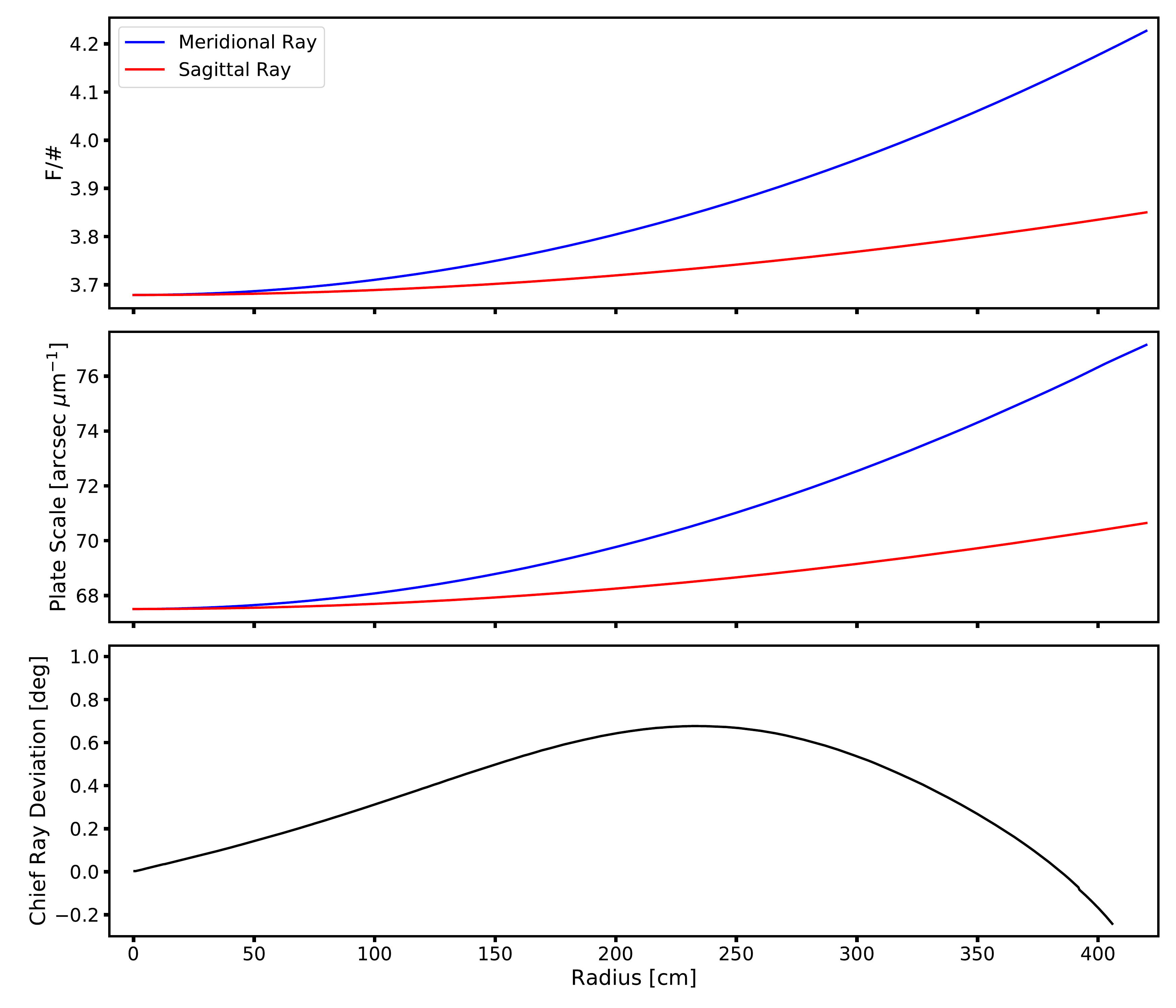}
 \caption{Properties of the corrector optical design as a function of radius. ({\it Top Panel}) Focal ratio of the meridional and sagittal rays. The anamorphic distortion causes the focal ratio to vary from approximately $f/3.68$ at the center of the field to between $f/3.85$ (sagittal) and $f/4.23 $ (meridional) at the edge. ({\it Middle Panel}) Plate scale of the meridional and sagittal rays. While the $107\,\mu$m diameter fibers are $1.585''$ diameter circles on axis, they project to $1.39'' \times 1.51''$ (meridional $\times$ sagittal) ellipses at the edge of the field. ({\it Bottom Panel}) Chief Ray deviation at 500\,nm as a function of radial position. This shows the difference between the absolute angle of the chief ray and the local surface normal, and minimizing this difference is important to achieve excellent coupling into the fibers (see \S\ref{sec:fibers}). \label{fig:corrprops}}
 \end{figure*}
 
 The corrector causes a central obscuration that impacts the throughput, and the finite length of the corrector produces a field-dependent shadow. The design includes some allowance for vignettting of the larger C1 and C4 elements in order to reduce the mass and size of those elements. This vignetting is present starting at $1.45^\circ$ off axis and is 5\% by the edge of the field of view at $1.6^\circ$. 
 
 The efficiency of coupling to the fiber cables is greatest when there is minimal deviation of the chief ray\footnote{The chief ray is notional for an obscured, prime focus configuration. Although the chief ray may be traced, it is blocked in reality by the corrector.} from the local fiber normal. The mounting holes for the fiber positioners, described in \S\ref{sec:fps}, were machined to orient the fiber tip at the optimal angle for their location on the aspheric focal surface, rather than simply normal to the surface. The deviations between the surface normal and the chief ray meet the requirement that they are less than $0.5^\circ$ on average and less than $1^\circ$ at all field positions. The chief ray deviation at 500\,nm is shown in Figure~\ref{fig:corrprops}. 
 
 The design does not include a constraint on the distortion of the pupil, although the shape of the pupil is both useful to quantify the projected size of the fibers on the sky and as input to the optical design of the spectrograph collimator. The focal ratio is $f/3.68$ at the center of the field and increases toward the edge. There is some anamorphic distortion, such that there is a somewhat larger increase in the meridional plane direction than in the sagittal plane. The net effect of this distortion is that while the $107\,\mu$m diameter fibers are $1.585''$ diameter circles on axis, they are $1.39'' \times 1.51''$ (meridional $\times$ sagittal) ellipses at $1.65^\circ$ (420mm) from the center of the field. The variation of the $f/\#$ and plate scale with radius is also shown in Figure~\ref{fig:corrprops}. 
 
 The final optical design, called the ``Echo 22'' design by the project, meets all of the requirements. The two aspheric surfaces are well within the maximum aspheric departure requirement of 30\,mrad, with peak values of $<15$\,mrad and $<11$\,mrad for C2 and C3, respectively, with the maxima at the edge of the lenses. The corresponding magnitudes of the aspheric departure were about 1.1\,mm for C2, 0.3\,mm for C3, and about 0.2\,mm for the focal surface. The optical design of the corrector is presented in Miller, Doel, Gutierrez et al. (2022), {\it in preparation}. 
 
 \begin{figure*}[ht!]
 \includegraphics[width=7in]{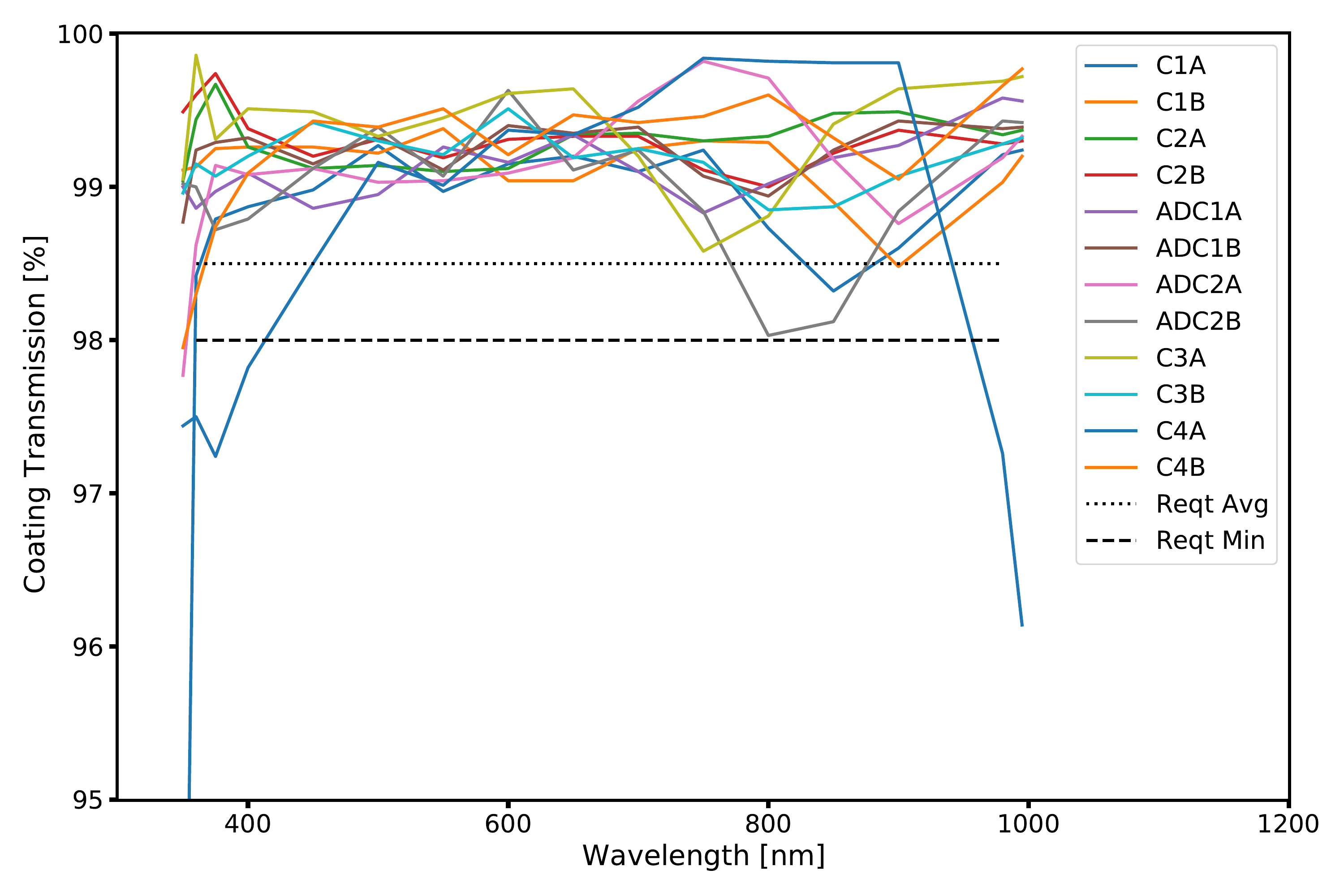}
 \caption{Corrector coating performance. Coating measurements for the six elements of the corrector. Each data point is the average of multiple measurements across each surface. The average requirement per surface was 98.5\% ({\it dotted line}) and the required minimum value was 98\% ({\it dashed line}). C4A did not meet the minimum requirement at the extremes of the passbands, although the other coatings sufficiently exceed the requirement that the net performance exceeds the performance assumed for survey simulations. Surface ``A'' faces the primary mirror for all lenses.  \label{fig:corrperf}}
 \end{figure*}

 \subsection{Procurement, Fabrication, and Coating}
 
  The lens blanks for C1 and C2 were produced by Corning,  the lens blanks for C3, C4, and ADC1 were produced by Ohara Corporation, and the blank for ADC2 was produced by Schott North America, Inc. The lens blank orders were placed between July 2014 and January 2015. All were delivered by the end of 2014 except for ADC1, which was delivered in August 2015, and ADC2, which was delivered in July 2015. Because poor glass homogeneity (variation of glass index) directly degrades image quality, glass homogeneity was a key performance driver for the lens blanks and we gave vendors a maximum inhomogeneity requirement. Furthermore we obtained measurements of the glass homogeneity along multiple sightlines through each blank from the vendors and modeled the impact relative to the beam footprint to determine that the delivered inhomogeneity produced an acceptable amount of blur within the corrector budget.

  The two all-spherical lenses C1 and C4 were polished by L3 Brashear, and the two aspheres C2 and C3 were polished by Arizona Optical Systems. Rayleigh Optical Corporation polished both ADC lenses \citep{miller18}. The polishing vendors supplied all test procedures and data (e.g. interferometric surface maps, wedge measurements) that were used to verify requirements and we performed an independent evaluation as part of the acceptance process.  The two spherical lenses were delivered in January 2016, and the two aspheric lenses were delivered in June and October 2017. The spherical lenses consequently took approximately a year from the order of the blanks through polishing, and the aspheres took about three years. The purchase of the lens blanks and the polishing of C2 and C3 were initiated prior to CD-3 approval for the start of construction with support from the Gordon and Betty Moore Foundation and the Heising-Simons Foundation. Given the long procurement times, this early support was crucial to the timely completion of DESI. 

  The six lenses were coated by Viavi Solutions, Inc.\ with ion-assisted deposition techniques in a 3\,m coating chamber \citep{kennemore18} through funding by the National Council of Science and Technology, Mexico (CONACYT). The size of the lenses and the shapes of the surfaces presented numerous challenges for coating. One was that the radius of curvature of the 12 surfaces ranges from nearly flat to a radius of curvature of $\sim 600$ mm, with a sag of 140\,mm. Another is that the angle of incidence for one lens surface varies from nearly normal to $40^\circ$. The vendor created a coating design with a wider bandwidth than required in order to compensate for the non-uniformity and angle variation and thus was able to employ a single design for all surfaces with the same substrate. The coatings were measured as averages over 110\,mm diameter circular areas. Figure~\ref{fig:corrperf} shows average measurements of the coatings on both surfaces of all six lenses. All of the coatings exceeded the requirement of $>98.5\%$ averaged over the bandpass. At the extremes of the bandpass, one coating is just barely non-compliant with the $>98\%$ minimum. We evaluated this non-compliance and determined that there would be a minimal impact on performance, as the other coatings exceeded the requirements. 
  
  \subsection{Performance} \label{sec:corrperf} 

  The corrector meets all requirements. Requirements such as wavelength range, glass transmission, image quality, field of view, size of the focal surface, chief ray deviation, and performance up to $60^\circ$ from zenith were met by the optical design. The total throughput set requirements on the performance of the coatings, which were met as shown in Figure~\ref{fig:corrperf}. The image quality also set requirements on the alignment of the lenses in their cells, the alignment of the cells within the corrector barrel, and the alignment of the barrel with the primary mirror. These requirements are discussed further in the next section on the corrector support system. The final verification of the corrector performance was the demonstration on sky that the image quality requirements were met. Early data were obtained with an $r-$band filter, as shown in Figures~\ref{fig:firstlight} and \ref{fig:corrpsf}. The performance over the full wavelength range across the entire field of view was ultimately verified with measurements of the total throughput, which is described in \S\ref{sec:throughput}. 

\section{Corrector Support System} \label{sec:corrsupp}

The corrector support system includes the lens cells and barrel, the ADC mechanism, the hexapod,  prime focus cage, vanes, and top ring. This system is responsible for maintaining the optical alignment of the six corrector lenses with the primary mirror for angles up to $60^\circ$ from zenith. This alignment is achieved with a combination of a stiff overall design and active adjustment of the alignment with the hexapod. Figure~\ref{fig:corrbarrel} shows the complete corrector barrel including the lens cells, ADC mechanism, and the hexapod. The subsections briefly summarize the technical requirements, the design of the top ring, vanes, and cage, the hexapod, the corrector barrel and ADC mechanism, the assembly process, and lastly some performance results. As in the previous section on the corrector, the present section is a brief summary of a more detailed description that will be published in Miller, Doel, Gutierrez et al. (2022) {\it in preparation}. 

 \subsection{Technical Requirements}

 There are three main classes of requirements on the corrector support system. The first is that all performance requirements need to be met over zenith distance angles from $0 - 60^\circ$, which flows from the Level 3 technical requirement L3.3.2 described in \S\ref{sec:toptechreq}. This requirement impacts the design and analysis of all of the components through a joint constraint on the mass and stiffness of this system. The project separated this constraint by establishing an overall mass budget for the upper assembly structure. This mass budget applied to all items installed for DESI that rotate about the declination axis. This includes the components in this section (barrel, cage, and ring), the lenses described in \S\ref{sec:optdesign}, the focal plane system in \S\ref{sec:fps}, and a fraction of the mass of the fiber cables (see \S\ref{sec:fibers}). The total budget corresponds to 10,700\,kg at the center of the upper ring, which is 8.5\,m from the declination axis, and corresponds to a minimal change in the mass at the top end of the telescope. That is possible because the larger mass of the corrector and the substantial mass of the focal plane system are approximately compensated by the mass removed with the replacement of the original split top ring with a single ring and the removal of the secondary mirror support structure. As a point of comparison, DECam is more than 2500\,kg heavier than the previous camera and cage assembly on the Blanco telescope \citep{flaugher15}. 
 
 The mass budget and zenith distance range requirements flowed down to requirements on the displacement of the optics due to gravity. There are both static and dynamic lateral and tilt tolerances for the six lenses, and there are additional tolerances between the lenses and cells and between the cells and the barrel. There are also tolerances on the interface flange to the focal plane system. The static tolerances are alignment tolerances and were measured with either a coordinate measuring machine or optical metrology with a rotary table in a climate-controlled environment at a specified temperature. The dynamic tolerances limit flexure due to gravity and were calculated with finite element analysis. The total tolerances include both static and dynamic tolerances and both lens-to-cell and cell-to-barrel tolerances. The total lateral tolerances ranged from $75-200\,\mu$m, and the total tilt tolerances ranged from $105 - 250\,\mu$rad. There are also requirements on gravity sag and tilt of the C3 mounting flange that connects to the hexapod flange.
 
 The ADC requirements included continuous rotation, rotation rate and accuracy, and the stability and lifetime of the mechanisms. The requirements on the continuous rotation and the rotation speed were motivated by the requirement that the inter-exposure sequence not exceed two minutes. Continuous rotation avoids the need for rotations greater than $180^\circ$ between exposures, and a rotation speed of up to $10^\circ$ per minute was adopted to ensure that the ADC motion would not contribute to the inter-exposure time. The lifetime requirement was primarily motivated by the bearings, as the motors can be readily accessed for maintenance. 

 \subsection{Top Ring, Vanes, and Cage} \label{sec:ring}
 
 The new prime focus cage, vanes, and ring were fabricated by CAID Industries, Inc. in Tucson, AZ, who delivered these components to the Mayall. The new top ring mounts to the telescope's Serrurier truss via eight truss members that attach to the ring at four locations, spaced at $90^\circ$ intervals (e.g., see Figure~\ref{fig:overview}). The top ring provides the interface between the Serrurier truss and the prime focus instrumentation, as well as a convenient location to mount the calibration lamps. The prime focus instrumentation is supported by a new prime focus cage that is constructed from low carbon steel and attached to the top ring with four sets of three vanes. The vanes have adjustment features to center the cage on the optical axis of the primary mirror. There is a requirement that the roll of the cage must be less than $8\,\mu$m at the edge of the focal plane during the longest expected DESI exposure times of on order 20\,minutes. To minimize roll the vanes are not connected to the cage at $90^\circ$ increments, but rather are displaced by 18 cm along the circumference of the cage such that pairs of vanes move in opposite directions. Another key requirement is that the vertical sag of the cage's position relative to the top end of the Serrurier truss must be less than 0.9\,mm at zenith so that it is well within the adjustment range of the hexapod. There are also requirements on the sag and tilt at the Horizon that were developed in conjunction with the specifications for the adjustment range of the hexapod.
 
 The prime focus cage has three rings that are held in place with four rails. The ring that is closest to the primary mirror supports the light baffles, the middle ring supports the hexapod system and thus the weight of the corrector and focal plane system, and the top ring supports the mass of the focal plane enclosure. The cage has an open design, which provides easy access to the hexapod and ADC mechanisms. The cage includes covers that protect the barrel from wind and thermal radiation. These components were painted with an optical black Aeroglaze paint, with the exception of precision mating surfaces. The new prime focus cage and instrumentation for DESI is somewhat larger than the previous cage, and it is larger than the central hole of the primary mirror. As a result, the effective area of the telescope is 8.658 m$^2$ on axis. 
  
 \subsection{Hexapod} \label{sec:hexapod}
 
 The hexapod is designed to adjust the barrel in six degrees of freedom to maintain the optical alignment of the prime focus instrumentation with the optical axis of the primary mirror. For a Cartesian coordinate system with the z-axis aligned with the optical axis, the degrees of freedom are x- and y-decenter, tilt about the x- and y-axes, translation along the z-axis (focus), and rotation about the z-axis. The hexapod has six actuators that form three triangles, and the actuators at the base of each triangle are attached to a plate that is directly attached to the cage. The hexapod is controlled by the Active Optics System, which is described in \S\ref{sec:aos}. Figure~\ref{fig:corrbarrel} shows the hexapod attached to the corrector barrel. 
 
 We purchased the hexapod from ADS International, and it met all of our requirements. These included that the ranges (resolutions) are $\pm8$mm ($\pm 10\,\mu$m) for x- and y-decenter, $\pm 10$mm ($\pm 5\,\mu$m) along the z-axis, $\pm 250''$ ($\pm 1.5''$) in tilt around the x- and y-axes, and $\pm 600''$ ($\pm 4''$) in rotation about the z-axis. The DESI hexapod is very similar to the hexapod used for DECam \citep{flaugher15}, with the main difference that the hexapod for DESI has a somewhat larger diameter (1.75\,m vs.\ 1.55\,m). The hexapod was shipped directly to the Mayall telescope and was integrated with the corrector barrel during installation. 
 
 \subsection{Barrel and ADC Mechanism} \label{sec:barreladc}
 
 The corrector barrel was fabricated from carbon steel by Dial Machine in Rockford, IL. It is divided into three main sections to support integration and testing: the Front section includes the C1 and C2 lenses, the Middle section includes the two-element ADC and associated mechanisms, and the Aft section includes the C3 and C4 lenses (e.g. see Figure~\ref{fig:corrbarrel}). The hexapod is mounted to the  cage, and attaches to the Aft section of the barrel. There is also a section called the Shroud that attaches to the Front section. The Shroud extends somewhat beyond C1 and its primary purpose is to protect that lens. It includes a provision for a protective cover, as well as references for alignment. The barrel also includes a light baffle that faces the primary mirror, and an adapter that supports the majority of the mass of the focal plane system. 
 
 The lens cells for C1, C2, C3 and C4 were constructed from a nickel-iron alloy, and the cells for the ADC lenses were constructed from low carbon steel. All of the cells were fabricated by University College London. Each lens is attached to its corresponding cell with axial and radial pads. The thickness of the radial pads for each cell was customized to accommodate the differences in the coefficient of thermal expansion (CTE) between the lens and cell materials, essentially an athermal design. The nickel-iron cells incorporate flexures to allow for the difference in CTE between the cell and steel barrel. The axial pads in the cells both support the lenses over the full range of gravity vectors ($0-60^\circ$ zenith angle) and account for surface irregularities that could produce uneven loading of the lenses.

 The team at Fermilab meticulously measured the barrel and cells prior to their shipment to University College London (UCL). The measurements were to verify the static deflection tolerances and to provide key information in advance of the optical alignment at UCL. The static deflection measurements of the complete barrel were made with a large coordinate measuring machine (CMM) that Fermilab built for this project. This temperature-controlled facility was used to measure the assembled corrector barrel at a range of orientations relative to gravity. A smaller CMM was used to obtain precision measurements of the lens cells, lens spacers, axial and radial pads, and the barrel flanges. These tests also verified that the barrel could be re-assembled to a position accuracy of $\sim 6\,\mu$m by use of a Moglice pinning system. 
 
 The tolerance requirements on the ADC lenses apply to the ADC mechanism as well. Each ADC cell is mounted within a large bearing from Kaydon. This bearing is bonded within a large bull gear and rotated with an SHA servo motor via a REL-230 motor controller from Harmonic Drive. Operation tests prior to shipment included a demonstration that it rotates to the commanded angle within $\pm 0.015^\circ$, which substantially exceeded the requirement. Thermal testing included some operation outdoors during winter in Chicago. 
 
 \subsection{Pre-Ship Assembly, Testing, and Performance}
 
 Once the measurement and alignment of the corrector barrel and cells were complete at Fermilab, all of the components were shipped to UCL. At UCL the lenses were installed in the lens cells with custom room-temperature-vulcanizing (RTV) pads for both axial and radial support. The axial RTV pads were then installed in the cells with a height accuracy of $\pm 25\,\mu$m. This was measured with a Micro-Epsilon laser sensor to ensure the even support of the lens. The cell and lens to be mated were mounted on a rotary table on an independent X-Y-Z support system. The lens was then brought into contact with the axial pads on the cell while ensuring the correct centering of the lens using dial gauges on the cell and lens. Once the lens was supported on the axial pads, the radial RTV pads mounted on the cell inserts were installed in opposite pairs with a layer of RTV to glue each pad to the lens. After the lens installation was complete, we measured the distance from the surface of the lens to the cell mounting flanges with a Faro gauge arm. We used this value to produce spacers to position the lens at the correct axial position in the barrel.

 All of the lenses were installed within their cells by April 2018. The lenses in their cells were next integrated with the barrel sections. The C1, ADC1, and C3 lenses were mounted by lowering their corresponding barrel section close to the cell. The fiducial pins were then inserted and the attachment bolts were put in place and carefully used to raise the cell to the barrel section.

 We checked the alignment of the first lens in each section with a pencil beam laser alignment system installed on the rotary table. This system directs a narrow laser beam along the axis of rotation of the table, and a camera records the beam location as it is deviated by the rotating lenses in the beam path. We measured the position of the return and transmitted beams to calculate the tilt and decenter of the lens. The second lens in its cell was then lowered into the section without disturbing the laser system, and we repeated the laser beam test to determine the position of the second lens. Once both lenses were installed in each of the three barrel sections, we integrated the three corrector sections and mounted the full barrel on the rotary table. The complete corrector was then mounted on the rotary table and the position of the three barrel sections measured and compared to the measurement made at Fermilab. These tests demonstrated that the static alignment tolerances were met. 
 
 Once these measurements were complete, the corrector was disassembled into three sections for shipment to the Mayall. For safety reasons, each section was mounted inside a custom frame, and that frame had vibration isolators to prevent shocks greater than 2\,g in any direction. The crates were designed for shock loads up to $\pm 6$\,g in both the horizontal and vertical directions. The corrector sections were shipped via charter aircraft directly from London to Tucson, AZ. The rotary table and pencil beam alignment system was shipped with the corrector sections to support the realignment and reassembly. The reassembly of the corrector at the Mayall is described in \S\ref{sec:topinstall}. 
 
\section{Focal Plane System} \label{sec:fps}

\begin{figure*}[ht!]
\includegraphics[width=7in]{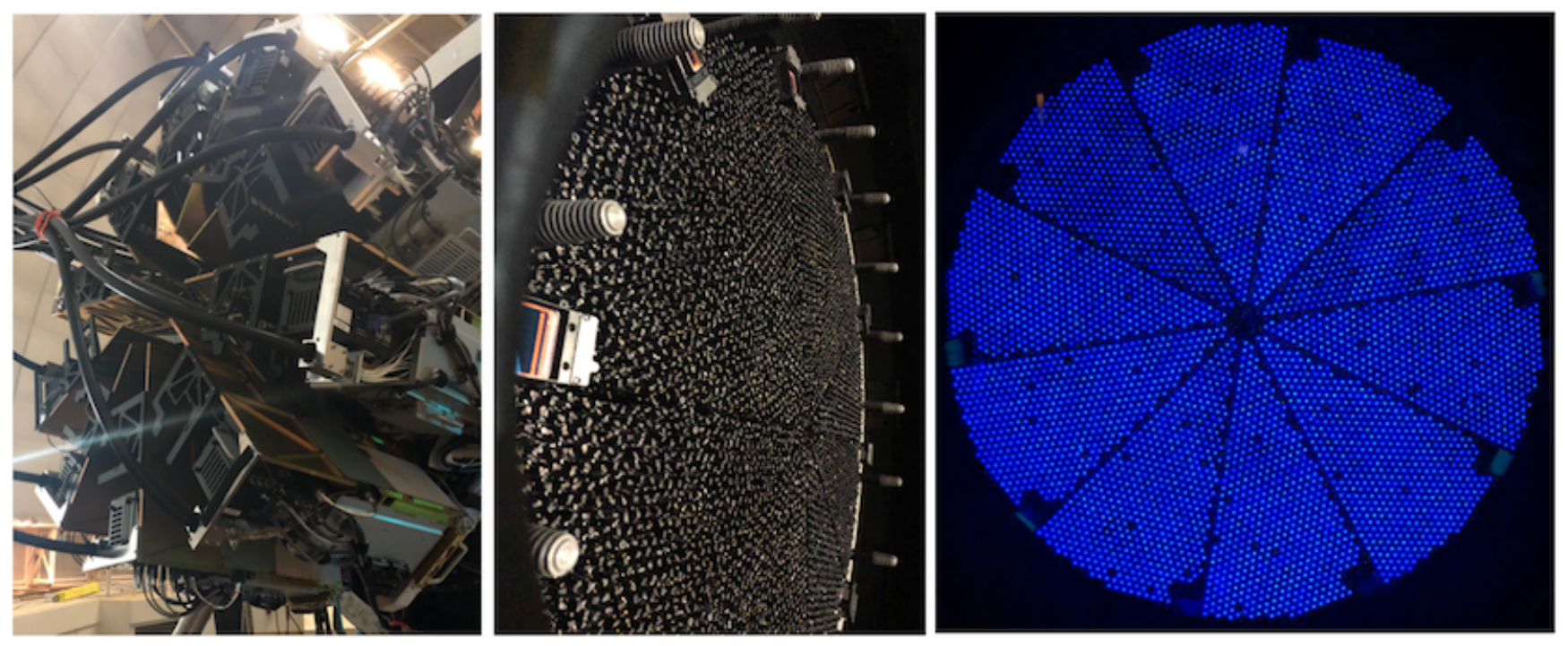}
\caption{DESI Focal Plate Assembly. ({\it Left}) Focal plate assembly after installation of the ten petals, but before installation of the focal plane enclosure. ({\it Middle}) Oblique view of the tips of the fiber positioners, fiducials, and the GFAs on the aspheric focal surface. ({\it Right}) FVC image of the focal surface with the fibers back illuminated and the fiducials off. \label{fig:fps}}
\end{figure*}

 The focal plane system includes both the main Focal Plate Assembly (FPA) and the auxiliary Fiber View Camera (FVC). The FPA is mounted to the corrector at the prime focus and is subdivided into ten identical, $36^\circ$ wedges or ``petals'' that fill the circular, aspheric focal surface of the corrector. Each petal has 500 fibers that feed a single spectrograph, two additional fibers that connect to the sky monitor system described in \S\ref{sec:skymon}, 12 illuminated fiducials\footnote{One petal has 13 fiducials and one petal has 14 to break any symmetries when viewed by the FVC.}, and a Guide/Focus/Alignment (GFA) detector system. Six of the GFAs are configured as guide cameras, with the physical positions of their pixel arrays tied by metrology to two adjacent fiducials. The remaining four are configured to obtain intra- and extra-focal images that we use to maintain the optical alignment between the prime focus instrumentation and the primary mirror. All of these components are sealed within the Focal Plane Enclosure (FPE), a well-insulated control volume into which we plumb liquid coolant lines to regulate the interior air temperature and clean, dry air to suppress the dew point. The FPA is connected to the spectrographs via the fiber system described in \S\ref{sec:fibers}. The FVC is mounted below the primary mirror and points at the FPA through the corrector. We use the FVC to obtain images of the illuminated fiducials and back-illuminated fiber tips to measure their relative locations. The next sections briefly describe each of these components of the focal plane system. More details will be published in Silber et al. (2022), {\it in preparation}. Previous descriptions include \citet{schubnell16}, \citet{leitner18}, \citet{baltay19}, and \citet{fagrelius20}. Figure~\ref{fig:fps} shows the fully assembled FPA, a side view of the focal surface including the tips of most of the 5020 fiber positioners, and an image of the illuminated focal surface obtained with the FVC. 

 \subsection{Technical Requirements}
 
 The key requirements on the focal plane system flow down from three main functions: accurate initial placement of the fiber tips; maintenance of the fiber tips in position for the duration of each exposure; and rapid reconfiguration to set up the next field. The first two functions flow down from the throughput requirement, while the third function flows down from the survey efficiency requirement. 
 
 The accurate placement of the fiber tips on astronomical sources sets requirements on the relative, lateral accuracy of the fiber tips and illuminated fiducials on the aspheric focal surface, precise metrology of the illuminated fiducials relative to the pixels of the guide cameras, and the placement of stars with precise astrometry at specific locations on the guide cameras. The quantitative requirements on the fiber tips are lateral accuracy of $\leq 10\,\mu$m rms of their nominal target position on the focal surface relative to the guide sensors and absolute accuracy of $\leq 35\,\mu$m for each fiber. The fiber tips are also required to be within $\pm100\mu$m of the focal surface to maintain focus, and for the ensemble to be oriented within $0.3^\circ$ mean absolute error perpendicular to the focal surface to limit the fiber injection angle. One of the two key metrology requirements is the measurement of the relative positions of the illuminated fiducials with $10\mu$m accuracy. The second is the measurement of the relative position and orientation of each CCD and the nearest illuminated fiducial with $5\,\mu$m accuracy. In both cases these requirements are on measurements in a plane that is locally parallel to the focal surface. 
 
 The four GFAs configured as wavefront sensors are used to determine and then maintain focus and alignment relative to the primary mirror, while the six configured as guide cameras are used for initial acquisition and subsequently to maintain telescope tracking for the duration of each observation. The guide centroid accuracy requirement is a tracking error signal better than $0.03''$. The GFAs are required to deliver images at a refresh rate of 0.2\,Hz and deliver the images over ethernet for analysis in $\leq 2.5$\,s. This requirement was established based on the known performance of the Mayall telescope. These high-level guiding and wavefront sensing requirements in turn set requirements on the total area, detector properties, and readout that are provided in \S\ref{sec:gfas}. 

 The L3.2.1 requirement on survey area and duration flows down to a requirement on the inter-exposure time of two minutes, and one of the main contributors to this time is the alignment of the fiber positioners for the new field. Given the performance of the positioner design, our expectation was that we would need two iterations of the focal plane in this time. As this reconfiguration applies power to 5020 fiber positioners and the 123 illuminated fiducials, reconfiguration produces on order 5-10\,kW of heat. There are consequently requirements on the cooling of the focal plane system to both minimize thermal gradients that could impact image quality and to avoid damage to equipment due to overheating. These requirements are described in \S\ref{sec:fpe}.  

 \subsection{Positioners} \label{sec:pos}
 
\begin{figure*}[ht!]
\includegraphics[width=7in]{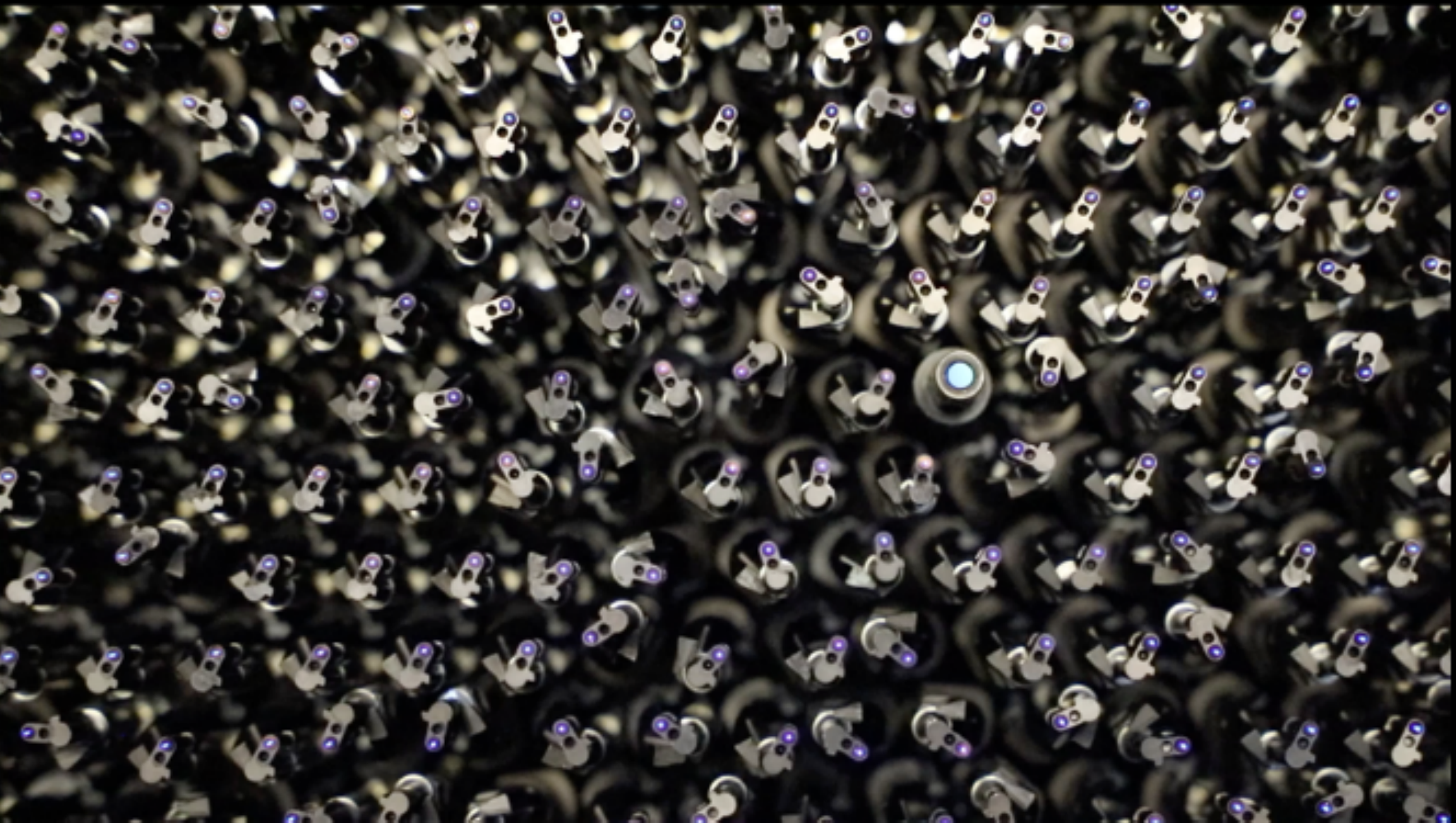}
\caption{Photo of some positioners and a fiducial in the back-illuminated focal plane. The small, blue dot in each positioner is at the location of the fiber tip. This close-up view shows about 150 positioners. One fiducial is visible toward the center right of the photo. \label{fig:pos}}
\end{figure*}

 The fiber positioners have two rotational axes. The rotation of the first (or $\theta$) axis is centered on the positioner. The second (eccentric or $\phi$) axis is centered along an arm nominally 3\,mm from the $\theta$ axis. The fiber tip is located along a second arm nominally 3\,mm from the center of rotation of the $\phi$ axis. The positioner can consequently locate the fiber tip within a nominal 12\,mm diameter patrol region. The positioners were designed to be very compact in order to maximize the number on the focal surface, and may be placed as close as within 10.4\,mm of neighboring units. The patrol regions overlap to maximize coverage of the focal surface, although it does also mean that we need software to prevent collisions between neighboring positioners. Figure~\ref{fig:pos} shows the tips of $\sim 150$ positioners, and illustrates their footprint. 
 
 One of the key technologies that enabled the compact positioner design are high-precision \o 4\,mm brushless DC gearmotors. These were produced for DESI by Namiki Precision Jewel Co. There is one of these per axis per positioner,  for a total of 10,040 motors in the focal plate assembly. These motors do not have encoders, and therefore we use the FVC (see \S\ref{sec:fvc}) and the PlateMaker software (see \S\ref{sec:pm}) to determine if the positioners are in the correct location relative to the fiducials. The motors each have three coils and are driven by pulse-width modulation (PWM) of the current applied to its coils. The PWM frequency is 18\,kHz (55\,$\mu$s), thus for our controller (operating at 72\,MHz) we have 4000 counts of resolution for each 55\,$\mu$s update period. This gives us fine control over the net current delivered to each coil. To drive the motor we apply sinusoidally varying currents to the three coils, phased $120^\circ$ apart. The frequency of this waveform sets the shaft angular speed. The firmware provides a user-facing `duty cycle' scalar parameter, by which the user can scale the pulse-width durations (thus scaling the effective amplitude of the waveform, thus the net current to the coils, thus the torque) by a multiplier between 0\% and 100\%.
 
 The positioners are powered by a 7.5\,V DC supply voltage and a 3.3\,V voltage regulator on the positioner electronics board provides local power. The power draw per positioner with both motors in operation at 100\% duty cycle is 3.23\,W. Only a few dozen motors in the instrument are presently run at 100\% duty cycle in order to overcome internal component misalignment or friction and the remainder are run at 70\% duty cycle. The power draw is 180\,mW with the motors idle and the electronics available for active communications. The positioner electronics are put into a low-power sleep mode most of the time, including during observations. The power draw is 15\,mW per unit in this mode. 

 The shaft position is controlled open-loop by rotating the sum magnetic field of the three coils. For large motions, the magnetic field, and thus the motor shaft, is ramped up rapidly (`cruise speed') to a rotation rate of 9,900\,rpm. This is very fast in practical terms: 176.07\,deg/sec at the output shaft. For smaller motions, such as re-positioning, the final $3^\circ$ ($\sim160\,\mu$m) of the move is done at low speed (`creep speed', 150\,rpm) and very fine resolution ($0.0003^\circ \approx 0.002\,\mu$m at the output shaft). After positioning, the coils are completely de-energized and the position remains stable. We measured the stability by taking repeated centroids with the fiber view camera before and after de-energizing the motor coils. 
 
 There is a positioner electronics board mounted to each positioner that hosts a Cortex microcontroller running custom firmware. This board connects the positioner to power (two wires), a Controller Area Network (CAN) communications bus (two wires), and an independent common logic line (SYNC, one wire) used to coordinate positioner moves. The firmware contains an integral boot loader so that after initial programming, subsequent reprogramming can be performed remotely via the CAN bus command lines. This feature allows full access to the microcontroller while minimizing the wire connection count. 
 
 Before a move, the move commands are uploaded to the individual positioners through the CAN bus. The start of the moves are synchronized at the petal level through the SYNC line on the positioner electronics board. Each positioner is uniquely addressed by an ID number and can be physically placed anywhere on the CAN bus. Groups of positioners are connected in parallel, and in any order, to two-line power and CAN signal rails. There are also a temperature monitor and motor current monitors on the positioner electronics board that can be queried through CAN commands. 
 
 \subsection{Fiducials} \label{sec:fifs}
 
 The illuminated fiducials in the DESI focal plane provide an important dual-purpose reference. They tie the pixel positions of the fibers imaged by the FVC to their physical locations in the focal plane, and tie physical locations in the focal plane to the celestial coordinates of stars imaged by the GFAs. As such, they are crucial to precisely position every fiber on its intended target.  

 The physical shape of each fiducial is an 8\,mm diameter tube that is sized and threaded for interchangeability with the positioners. At the tip of each fiducial there is a small glass block with a ground finish on the inner surface and an opaque blue chrome finish on the outside. Four 10\,$\mu$m diameter pinholes in a standard pattern are etched into the chrome finish. In operation, the pinholes are illuminated by a monochromatic 470\,nm LED inside the support tube. Each fiducial has a microcontroller, the same as for the positioner, for communication on the CAN bus and control of the LED intensity. The tip of one fiducial is visible in Figure~\ref{fig:pos}. 
 
 The fiducials were produced in two types. The vast majority were Field Illuminated Fiducials (FIFs) that are distributed throughout the focal surface. In addition, we produced shorter Guide Illuminated Fiducials (GIFs), two of which were mounted to the GFA camera body, in order to obtain the most precise mapping from the reference frame of the fiducials to the GFA detector pixels. 

 After we installed the fiducials in the petals, we measured the average position of the four pinholes on each fiducial to high precision relative to a petal reference. The resolution and stability of the FVC is sufficient to measure the centroids of the back-lit fibers and the fiducial pinholes to $\sim20$\,millipixel precision. The fiducial centroids provide important information to calibrate the transformation from fiber centroid coordinates to focal plane coordinates. In practice, we refined these measurements during commissioning as part of the dither tests described in \S\ref{sec:dither}. 

 \subsection{Petal and Petal Controller} \label{sec:petal}

 Each physical petal has 514 precision holes for the positioners and fiducials. Each hole is machined at a unique angle to orient the fiber positioner or fiducial along the chief ray at its location on the focal surface. The petals are mounted into the focal plane ring (FPR). Boston University (BU) fabricated both the petals, Boston University (BU) fabricated the petals, and contracted Dial Machine (Rockford, IL) to fabricate the FPR and to have it anodized using an inorganic black process. The FPR was delivered to BU where it was inspected and found to be satisfactory for flatness and for mounting the petals. BU contracted and closely worked with Zeiss in Boston to validate each petal with a custom CMM program. We shimmed the petals and conducted several test alignments with the FPR at LBNL prior to the installation of the positioners to verify the alignment precision, as well as gain experience with integration. The alignment error of each petal and the FPR was $\pm15\,\mu$m rms. \citet{duan18} provide more details about the fabrication, metrology, and alignment of the petals and FPR. 
 
 The petals also include power supplies, control electronics, and substantial cable routing fixtures. The control of the positioners and fiducials is via a small, single-board computer (BeagleBone Black Industrial) which was integrated on a petal control board. This control board is housed in a small box (the `petalbox'), along with circuitry to send and receive external signals, three 12 VDC power supplies, and two commercial 8-channel CAN control boards from Systec for communication with the positioners and fiducials. The petalbox also carries signals for the GFA power supplies, thermoelectric cooler, and fan. The petal control computer in the petalbox communicates with the instrument control system via ethernet (see \S\ref{sec:ics}).

\begin{figure*}[ht!]
\includegraphics[width=7in,angle=0]{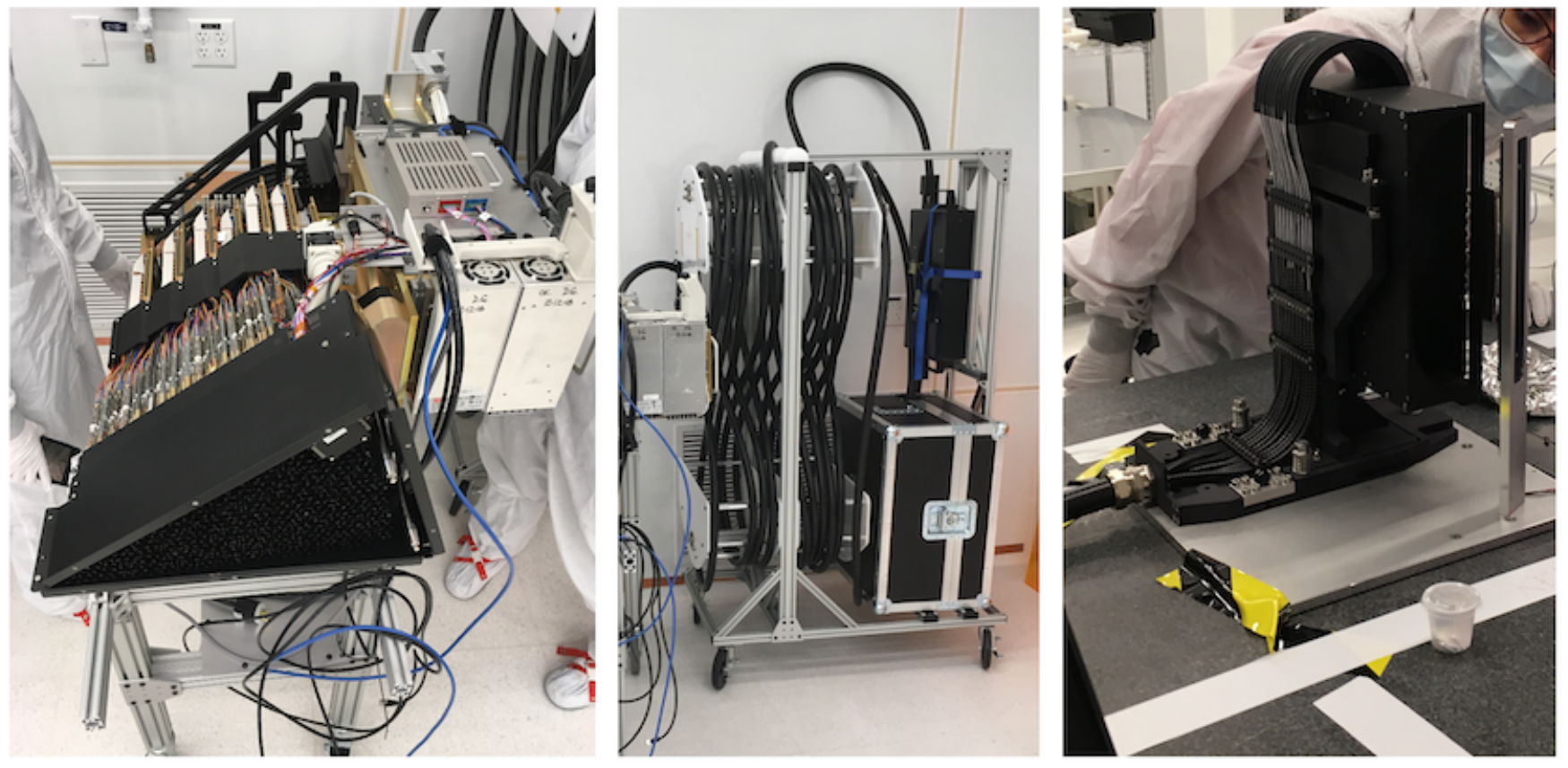}
\caption{Photo of one of the ten DESI petal assemblies on 21 February 2019 in a cleanroom at the LBNL. The petal assembly includes the petal, fiber cable, and slithead. ({\it Left}) Petal with all 502 positioners, 12 fiducials, and the GFA unit, which will be aligned with the focal surface produced by the corrector. The black, triangular structure is a temporary, protective housing around these components. Behind this structure are the FIPOS electronics that drive the fiber positioners and fiducial LEDs, and wire and fiber guides. The petalbox electronics are to the right. ({\it Middle}) The $\sim50$\,m fiber cable that will extend from the petal in the focal plane system to a slithead in one of the ten spectrographs. In this photo the slithead is inside the protective case on the lower right on the cart. ({\it Right}) One of the DESI slitheads. The fiber cable attaches from the left in this photo. The pseudo slit of fibers is pointed at an illumination system on the right. \label{fig:petal}}
\end{figure*}

 There are two 600\,W power supplies (from Mean Well) on each petal. These connect to the positioners via a large printed circuit board that runs radially along the back of the petal. This circuit board also carries the CAN and SYNC bus traces. There are then seven transverse circuit boards that are perpendicular to the large printed circuit board. The seven transverse boards carry the total of ten CAN communication buses for each petal. A photograph of a complete petal is shown in the left panel of Figure~\ref{fig:petal}. We integrated and tested all of the petals at LBNL. In the summer of 2021 we made several upgrades to the petal electronics that are described in \S\ref{sec:problems}. 
 
 \subsection{Positioner Moves and Anti-Collision Software} 
 
 The positioners are moved via open-loop control of the two motor axes. The starting point to calculate a move is the last position measured with the FVC. The move consists of a series of rotations and pauses of the motors that define the path of the positioner to the intended location. Once the move is complete, we measure the position again with the FVC. In order to measure the locations of the fiber tips with the FVC, we back-illuminate the fibers with an illumination system that is mounted to the spectrograph shutters (see \S\ref{sec:mech}) and the locations of the fibers measured with the FVC are converted to the focal plane coordinate system with the PlateMaker software described in \S\ref{sec:pm}. 
 
 During typical nighttime operations, we use two move iterations to acquire targets for a new field. We call the first move the `blind' move because it is based on the known (expected) position, which is typically based on the FVC image obtained before the prior science exposure. The blind move puts fibers within $\sim\,50\,\mu$m\,rms of their target positions, and therefore do not meet our requirements. The second, `correction' move is based on an FVC image obtained after the first move, and this meets our requirements with an rms accuracy of better than $6\,\mu$m with respect to the commanded position. 
 
 The move commands for each positioner have to take into account their neighboring positioners because the patrol regions overlap. This impacts the allowed final positions and the move path (including timing).  Our software evaluates and avoids collisions during moves with a model of each positioner as a pair of 2D polygons. The fiberassign software (see \S\ref{sec:fiberassign}) that matches fiber positioners with targets is also aware of this geometry and avoids targets that would produce collisions. To avoid collisions during the moves to these positions, the typical motion is to first retract the eccentric or $\phi$ arm, then rotate about the central axis $\theta$ arm, and lastly extend the eccentric arm to the final position. This avoids most potential collisions. We spread out these moves in time  so that not all positioners are moving at once, which reduces the instantaneous power draw and further decreases potential collisions. We lastly eliminate any remaining, potential collisions with small adjustments to the move paths. In rare cases the algorithm finds no solution for a given positioner, and that fiber is not moved to a target for that observation. 

 \subsection{Guide/Focus/Alignment Arrays} \label{sec:gfas}
 
 There are ten Guide/Focus/Alignment (GFA) cameras at the periphery of the focal plane. Six of these are configured as guider cameras and four as wavefront sensors. Each unit contains a back-illuminated e2v Model CCD 230-42 detector with $2048 \times 2064 $ pixels that are $15\,\mu$m square. We use the central $1024 \times 2048$ pixels for imaging and the two remaining $512 \times 2048$ regions as a frame store. The centers of the detectors are approximately 398.5 mm from the center of the focal surface. At this distance, the meridional and sagittal plate scales are approximately $76.3\,\mu$m$/''$ and $70.4\,\mu$m$/''$, respectively. The detectors therefore subtend approximately $3.34' \times 7.27'$ on sky and have an average scale of about $0.205''$/pixel. The GFA cameras were built by a consortium of four Spanish institutions: Institut de Física d'Altes Energies (IFAE), Institute of Space Sciences (ICE-CSIC, IEEC), Centro de Investigaciones Energéticas, Medioambientales y Tecnológicas (CIEMAT), and Instituto de Física Teórica (IFT-UAM/CSIC). Their work included custom packaging to minimize the footprint of the GFAs on the focal surface. As part of this design, the voltage supplies are in a separate electronics box that is located behind the petal, and they are cooled by air flow driven in part by fans mounted behind each camera.
 
 All ten GFA cameras have an $r-$band filter directly in front of the detector. The cameras do not have shutters. The filter was chosen to maximize sensitivity to stars. These filters do not transmit the bluer light used by the back-illumination system and the illuminated fiducials. The guide sensors have a 5\,mm thick filter, and each wavefront sensor has half of the active area covered by a 1.625\,mm thick filter, and the other half covered by an 8.375\,mm thick filter. Half of each wavefront sensor is consequently designed to record extra-focal images, and the other half to record intra-focal images. 
 
 The wavefront correction requirements are to measure defocus of the corrector with respect to the primary mirror with $30\,\mu$m rms, decenter with $300\,\mu$m rms, and tip/tilt with $10''$ rms, and report corrections to the hexapod within 5\,s. The software that calculates these corrections is called the Active Optics System (AOS) and is based on the similar system developed by \citet{roodman14} for DECam. One important difference between AOS for DECam and DESI is that the DESI GFAs do not have shutters, and it is consequently possible to calculate corrections prior to the start of science exposures. The guiding and especially the wavefront correction requirements were that the read noise must not exceed $25\,e^-$/pixel. Due to the tracking requirement, the detectors also have requirements on frame rate and data transfer. The central area of the CCD is read into memory in 2.5\,s through four amplifiers. In regular operation we obtain a 5\,s guiding image approximately every 8\,s. While this is somewhat below the refresh rate requirement, in practice this frame rate has worked very well. 
  
 \subsection{Environmental Control and Machine Protection} \label{sec:fpe}
 
 The Focal Plane Enclosure (FPE) maintains the FPR at a constant temperature of $11^\circ$\,C, removes heat generated by the GFAs, positioners, and their associated electronics, monitors the humidity, and maintains the exterior within $\pm1^\circ$\,C of the dome ambient temperature. We sized the FPE insulation to prevent significant surface temperature differentials with respect to the ambient environment, so as not to induce dome turbulence by natural convection. The air supply to the FPE is provided by an external, facility dry air system and cooling is provided by an external chiller that removes heat with 3M Novec 7100 fluid. There are two distinct volumes of air within the FPE, one between the C4 lens and the focal surface, and the other behind the focal surface that contains the remainder of the FPE, including electrical and other utilities. The FPE has two large fans. One drives air via four ducts into the focal plane adapter volume, where it then passes through the positioners, the GFA cameras, and through narrow gaps between petals. The other fan drives air through a heat exchanger. The heat exchanger transfers heat into the Novec coolant, which then transports it out of the FPE control volume. These fans are counter-rotating to minimize vibrations and drive more air through the positioner array and heat exchanger. The FPE Controller (FXC) continually monitors the temperatures and humidity within the FPA and can adjust the temperature of the refrigerant and the speed of the fans.

 There are hardware interlocks within the FPE to protect equipment. One is that the FXC has hardware interlocks that can shut down power to the entire focal plane system if the temperature and/or humidity exceeds threshold values. The FXC is powered by an independent 24 VDC line so that it maintains power in the event the interlock is tripped. The FXC will also issue warnings and alarms in case the measured values approach pre-determined thresholds. The electronics box for each GFA camera monitors the humidity and temperature of the electronics for that camera. If either exceed safe limits, the bias voltages for the CCD are turned off. If the electronics exceed the temperature limit, the GFA power is also shut off. There is also an environmental power interlock for the focal plane on the M floor of the Mayall. If smoke is detected or the dewpoint exceeds $-2^\circ$\,C, power is shut off to the fans, petalboxes, GFAs, positioners, and fiducials. This interlock system also stops the Novec chiller that provides cooling to the focal plane system. If the temperature exceeds $+35^\circ$\,C, then the power is shut off to the positioner power supplies and petalboxes. In addition to these hardware interlocks, there is a fault monitoring system that provides warnings if parameters exceed the nominal range (see \S\ref{sec:icsinfrastructure}). 
  
 \subsection{Fiber View Camera} \label{sec:fvc}

 The Fiber View Camera (FVC) is designed to measure the locations of the back-illuminated fiber tips relative to the illuminated fiducials. The camera is mounted approximately one meter behind the primary mirror and about 12.25 meters from the focal surface. The detector is a Finger Lakes Instrumentation Microline camera body with Kodak KAF-50100 CCD. This CCD has a $6132 \times 8176$ pixel format with $6\,\mu$m pixels, although we just read out the central $6132 \times 6132$ pixels. The key requirements on the FVC system are to return centroid measurements with $<3\,\mu$m accuracy with a $<5$ second cycle time. \citet{baltay19} describe the FVC system in detail.
 
 The FVC system observes the location of the fiber tips and the fiducials through its own lens and the DESI corrector system, and therefore we require accurate knowledge of the distortions of both optical systems to compute the transformation from coordinates on the FVC to the focal plane coordinate system. Through the end of 2019 the FVC had a 600\,mm f/4 Canon telephoto lens described in \citet{baltay19}, although like most telephoto lenses it had significant, non-axisymmetric distortions. We ultimately decided to switch to a simpler, achromatic lens in 2020 to minimize distortion. The new lens is a 25.4\,mm diameter singlet lens with a 600\,mm focal length. There is also a 50\,nm wide filter centered at 475\,nm in front of the lens. The singlet is a plano-convex N-BK7 lens that is part KPX119AR.16 from Newport and the filter is part 86-950 from Edmunds Optics. 
 
 \begin{figure*}[ht!]
\includegraphics[width=7in,angle=0]{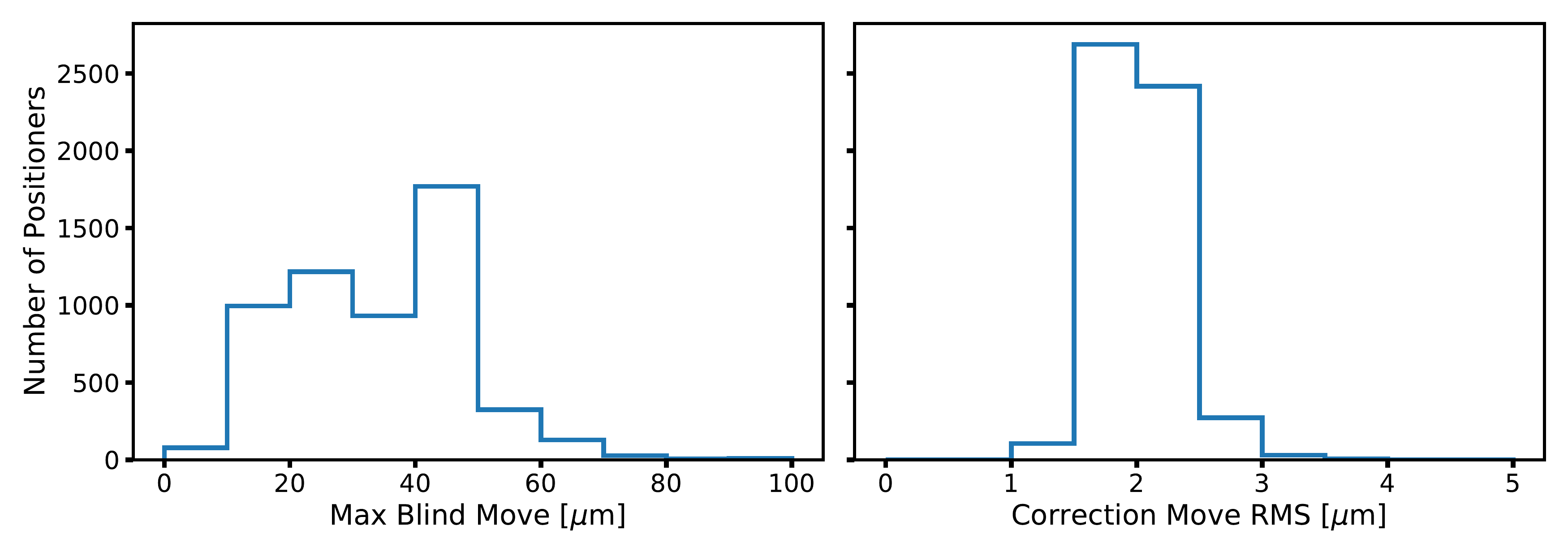}
\caption{Results from pre-ship testing of the positioners in ten petals plus one spare petal. These are measurements of fiber positioner accuracy on a uniform grid of points in a clean room at LBNL. ({\it Left}) Measurement of the maximum positioner error after the initial or `blind' move. The distance from starting point to target for each blind move is typically several mm. ({\it Right}) Measurement of the positioner rms error after the `correction' move to remove the residual errors of the blind move. This move is based on an FVC measurement. The positioners move very precisely over the small distances required to correct for the residual errors after the blind move. \label{fig:posmove}}
\end{figure*}

 \subsection{Pre-Ship Testing and Performance} 
 
 We thoroughly tested each petal prior to shipment. The test suite for each petal included electrical tests, mechanical and optical metrology, characterizations of the GFA noise properties, measurements of the fiber throughput, and fiber positioner performance. The metrology included measurement of the locations of the two GIF fiducials relative to the GFAs, measurement of alignment features on the petal relative to the focal plane ring, and measurements of the relative locations of all of the FIF and GIF fiducials and alignment features. The electrical tests confirmed the safe operation of the petal electrical systems. We characterized the performance of each GFA camera after integration with its petal and measured an average read noise of $18\,e^-$ and an average gain of $3.7\,e^-$ per ADU. This meets the read noise requirement of $25\,e^-$ per pixel. The fiber thoughput measurements are described at the end of the next section. 

 Measurement of fiber positioning speed, accuracy, and precision was a major activity prior to shipment, both to confirm requirements and to further develop the software and procedures for operations. The key measurement was the lateral accuracy of the fiber tips, which we conducted in a clean room at LBNL. The test setup included a system to back-illuminate the fiber tips that was based on the shutter illumination system, a spare FVC to measure the locations, and substantial specialized software. We measured the lateral accuracy\footnote{Lateral accuracy is measured in a plane normal to the central axis of each positioner.} with an `XY Test' in which we moved each positioner to a series of points on an evenly-spaced $5 \times 5$ grid of 24 targets that was approximately 3 mm on a side. There was no central target because we cannot position at the exact center of the patrol field due to small differences in the arm lengths. For each position, we first execute a blind move of all positioners and measure the errors in their locations with the FVC. We then used the error vectors from that FVC measurement to calculate and send correction moves. Finally, we measured the errors in the correction moves with a second FVC image. The distribution of the uncertainties in the blind and correction moves are shown in Figure~\ref{fig:posmove}. The blind move typically achieved a median accuracy of $29\,\mu$m in the lab. The median accuracy was $2.2\,\mu$m after the first correction move, which met our requirement of $\leq 10\,\mu$m rms lateral accuracy. Note that this measurement did not include several important effects that impact performance on sky, including astrometric transformations, optical distortions, dome turbulence, and uncertainties in the metrology for the complete focal plane system after assembly at the Mayall. We obtained on-sky measurements that included these effects during commissioning and describe those results in \S\ref{sec:results}.

\section{Fiber System} \label{sec:fibers}

 The fiber system extends a continuous run of optical cable from the tips of the fiber positioners to the slitheads in the spectrographs with high throughput and minimal degradation of the output beam. Through a combination of careful attention to stresses on the fibers, optimised termination, fusion splicing, and anti-reflection coating, we have achieved $\geq$90\% throughput in the fiber system. The three main components of the fiber system are the Positioner Fiber Assemblies (PFAs), the fiber cable bundles, and the slitheads. The PFAs are approximately 3\,m lengths of AR-coated fiber cable, and there is one in each positioner. The PFAs were produced by LBNL. The fiber cable bundles are 47.5\,m long and each contain over 500 strands of fiber optic cable that are wound and reinforced following similar construction techniques to under-sea fiber cables. There are ten of these fiber cable bundles, one per spectrograph. The slitheads are a linear array of 500 fibers on a convex surface that is matched to the spectrograph axial focal prescription. There are ten slitheads, one in each spectrograph. The slitheads and cable assemblies were built by Durham University, UK and shipped to LBNL who fusion-spliced the fibers to the PFAs after the robotic fiber positioners were inserted in the petals. Earlier descriptions of the fiber system were published by \citet{poppett18} and \citet{poppett20}. Poppett et al. (2022), {\it in preparation} will present a more detailed description of the fiber system than the brief summary in this section. 

 \subsection{Technical Requirements}

 The key requirements on the fiber system relate to the throughput. The throughput of this system is impacted in three ways: 1) Fresnel losses; 2) attenuation of light as it propagates through the fiber; 3) losses due to the degradation of the output beam. The throughput requirement from propagation alone is $>99$\% over nearly the entire wavelength range, except for percent-level decreases below 375\,nm and above 900\,nm. This requirement is imposed as both an average over 5000 fibers, and that no fiber may be below $85$\% of the nominal value at any wavelength. The degradation of the output beam is an increase in angle of the output beam compared to the input beam, which is commonly referred to as focal ratio degradation (FRD). This characterizes how much energy is within some output cone that is accepted by the spectrograph, and impacts the fraction of the total light within the PSF. Our requirement is that the encircled energy should be $\geq90$\% within an $f/3.57$ output beam when illuminated by a uniform $f/3.9$ input beam that mimics the corrector, including a scaled obscuration at prime focus. There are also requirements on the throughput stability, the PSF stability, and the fraction of intact fibers at the time of integration with the telescope. 

 \subsection{Fiber Cables and Slit Heads} \label{sec:cables}
 
 The DESI fiber system has ten, identical fiber cable bundles, one between each petal and spectrograph. Each cable bundle includes about 530 fibers, more than the 502 positioners per petal so that some spares are available in case of breakage before integration with the petal and slithead. The individual fibers are Polymicro ``FBP'' broad spectrum optical fiber that have a core diameter of $107 \pm 3\,\mu$m. Inside of the fiber cable bundle there are 11 protection tubes from Miniflex, ten with 53 fibers each and one with three sky fibers. We used a boron nitride lubricant and a fiber control guide to avoid tangles that might induce stress when inserting the fibers into these tubes. The 11 tubes were then helically wound around a tensile element that limits stresses due to axial loads and temperature changes, and then inserted into a protective conduit from Adaptaflex. The middle panel of Figure~\ref{fig:petal} shows one of these fiber cable bundles. \citet{schmoll18} present more details about the fiber cable production.

\begin{figure*}[ht!]
\includegraphics[width=7in,angle=0]{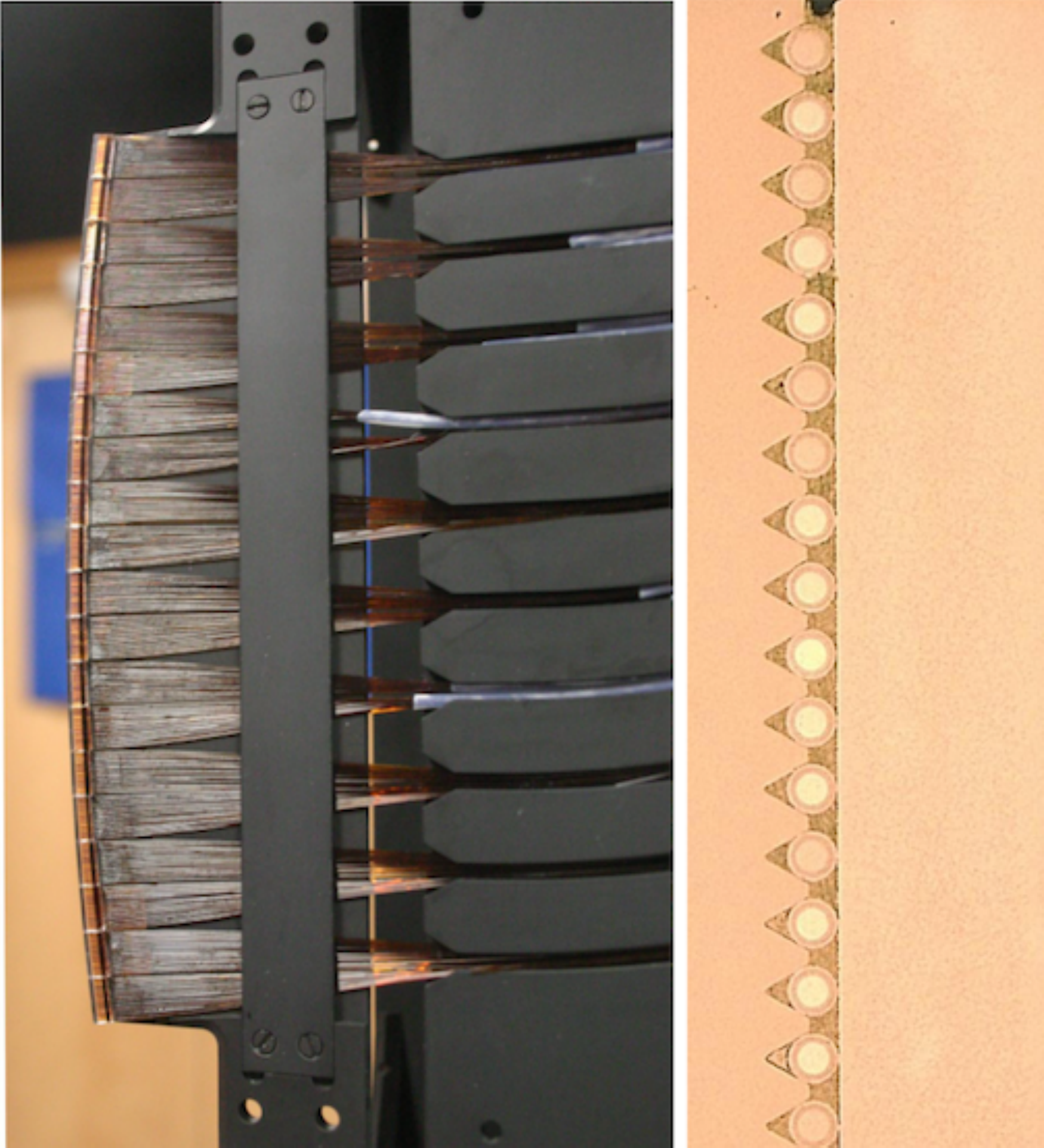}
\caption{Photos of the interface between the fibers and the slithead. ({\it Left}) Side view of the termination of one of the fiber cables in a slithead. The slithead is an array of 500 fibers that form a pseudoslit with curvature along the optical axis. The 500 fibers are divided into 20 blocks of 25 fibers each. ({\it Right}) Magnified view of the end of one of the blocks of 25 fibers. The 25 fibers tips are each bonded into a separate V-groove, sandwiched between the V-groove and a coverplate, and the fiber tips have been co-polished with the block. \label{fig:slithead}}
\end{figure*}
 
 One end of each fiber cable bundle terminates at the slithead \citep{tyas18}, which consists of 500 fibers in a linear array that is grouped into 20 blocks of 25 fibers each. Each block has 25 parallel V-grooves machined in fused silica that hold the fiber ends, and consequently each block points at a slightly different angle along the linear slit. To assemble each block, we bonded the individual fibers between the V-grooves and a cover plate, and  co-polished the fiber ends and the block. We then bonded a slit cover window with a UV-transparent glue (EPO-TEK 301-2) onto each fiber block. The slit cover windows were AR-coated by Infinite Optics. A side view of one slithead is shown in the left panel of Figure~\ref{fig:slithead}, and a view along the optical axis of one fiber block is shown in the right panel. The other end of each fiber cable bundle initially terminated in a fiber spool with sufficient excess fiber for splicing to the Positioner Fiber Assemblies. This process is described in the next subsection.  
 
 The FRD of the fiber cables was characterized by collimated (or ring) tests. In this test, a collimated light beam at 625nm is injected into a fiber, which azimuthally scrambles the light to form a ring illumination pattern. The diameter of the ring is a direct measurement of the incidence angle of the input light, and the thickness of the ring in the radial direction is a measure of the FRD. While not as quantitative as a measurement of the angular diffusion of a filled light cone matched to the nominal $f$-ratio, the collimated FRD test is very simple to perform and interpret. We performed both cone and collimated FRD tests on a subset of fibers and used the results to calibrate the collimated FRD tests that were applied to a larger sample. Upon the completion and verification of each fiber cable bundle and slithead, we shipped the unit from Durham to LBNL for integration with the PFAs and the focal plane system. 
 
  \subsection{Positioner Fiber Assemblies} 
 
  A Positioner Fiber Assembly (PFA) is a single, $\sim 3$\,m strand of the same fiber employed in the fiber cable bundles and slithead. These 3\,m strands exist because the positioners were designed to be loaded into the petal from the front, and we did not want to pull the entire $\sim 50$\,m fiber associated with each positioner through the petal before wrapping the fiber cable. The assembly of the PFAs was carefully optimized to minimize degradation of the optical performance. One end of each fiber is precision-cleaved with a diamond cleave blade while under tension in order to produce a high-quality optical surface. We optimized the tension for the cleave with feedback provided from FRD measurements. We chose to optimize the precision cleave to produce a high-quality surface that did not require additional polishing because this eased requirements on the hardness of the glue required to bond the fiber into the ferrule. Since we did not polish the fiber ends, we bonded the ferrule with a low stress adhesive that cured very quickly under UV illumination.  We then bonded the fiber into a fused silica ferrule to minimize stress when it is in the fiber positioner. There is also a polyimide tube bonded to the back of the ferrule to provide strain relief and reduce stress when the positioner moves. The assembly of each PFA took only three minutes per fiber, in large part due to the short cure time, and this helped us assemble and test these units fairly quickly. 
  
  Once the PFA was complete, we had the input face of each PFA AR-coated by Infinite Optics. We loaded sets of 320 PFAs into custom coating jigs that could be placed directly into the coating chamber, and therefore minimize handling by the vendor. After coating, we inserted each PFA into a positioner and used optical metrology to check the fiber tip location with respect to the focus direction against the $\pm 100\,\mu$m requirement on all fiber tips relative to an ideal focal surface. Finally, the positioner with the PFA was installed into the petal from the side of the petal that faces the focal surface. 
  
  \begin{figure*}[ht!]
  \includegraphics[width=7in,angle=0]{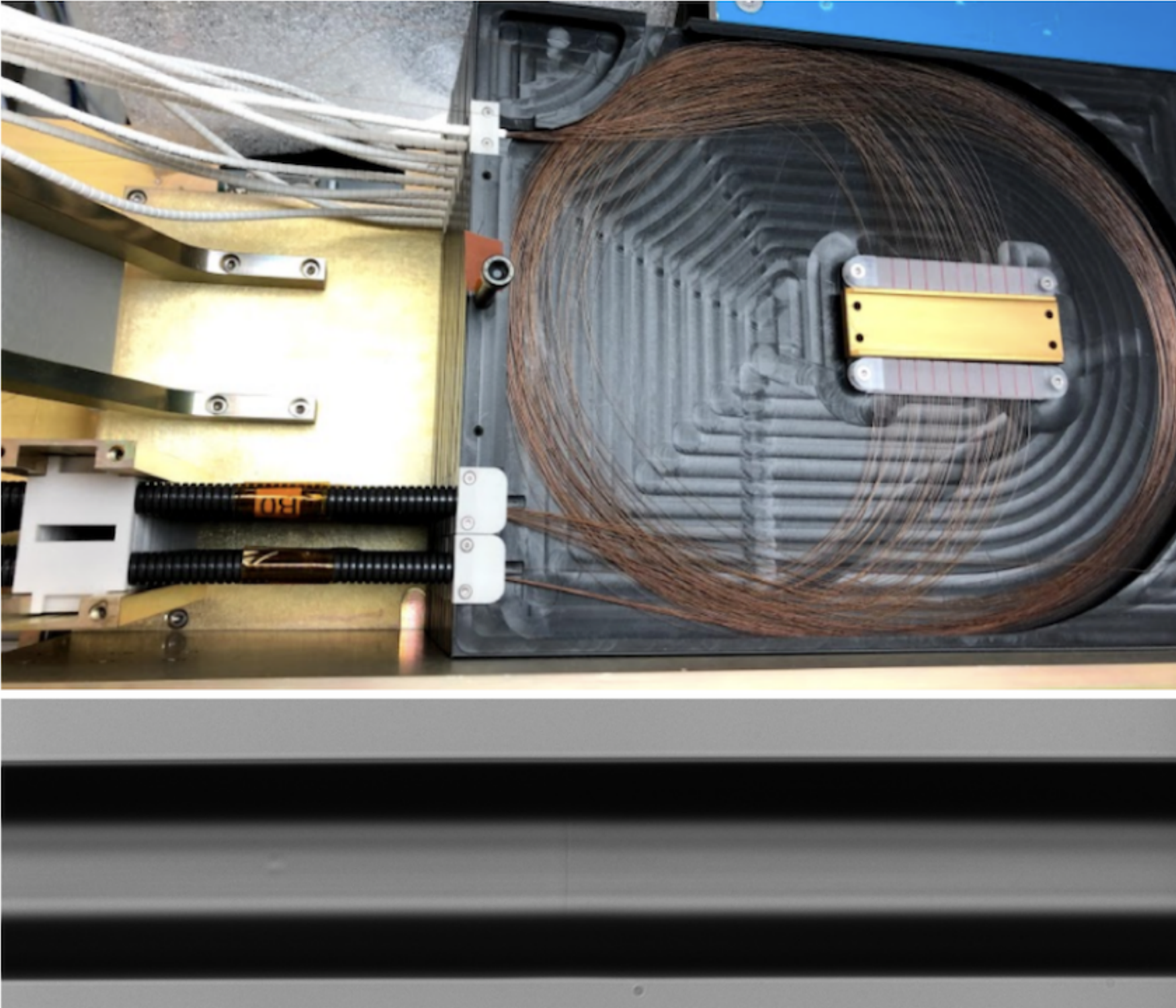}
  \caption{Photos of the interior of a Spool Box ({\it top}) and a single spliced fiber ({\it bottom}). The top panel shows the inside of a Spool Box, which contains the excess fiber required for the splicing process. The spliced region of each fiber is underneath the cover plate at the center of the Spool Box. The bottom panel shows an image of a single fiber after it has been spliced together. The location of the splice is barely visible as a narrow, vertical line at the approximate center of the image. \label{fig:fibersplice}}
  \end{figure*}

  After the 500 positioners were installed in a petal, the PFAs were connected to the Spool Box at the end of the fiber cable bundle, which contained a sufficient excess of fiber for the splicing process. The fibers were connected with fusion splicing, which proved to provide superior performance to other splicing and mechanical connector technologies \citep{fagrelius16}. To employ this method, we carefully aligned the two uncoated ends of fibers, melted them with an electric filament, and pushed them together to create a solid bond. This technique is used in the telecommunications and medical industries, and several commercial machines are available for this procedure. We experimented with various fusion splicing parameters and optimized our choice of parameters based on feedback from the collimated FRD performance. 
  
  \subsection{Pre-Ship Testing and Performance}
  
\begin{figure*}[ht!]
\includegraphics[width=7in,angle=0]{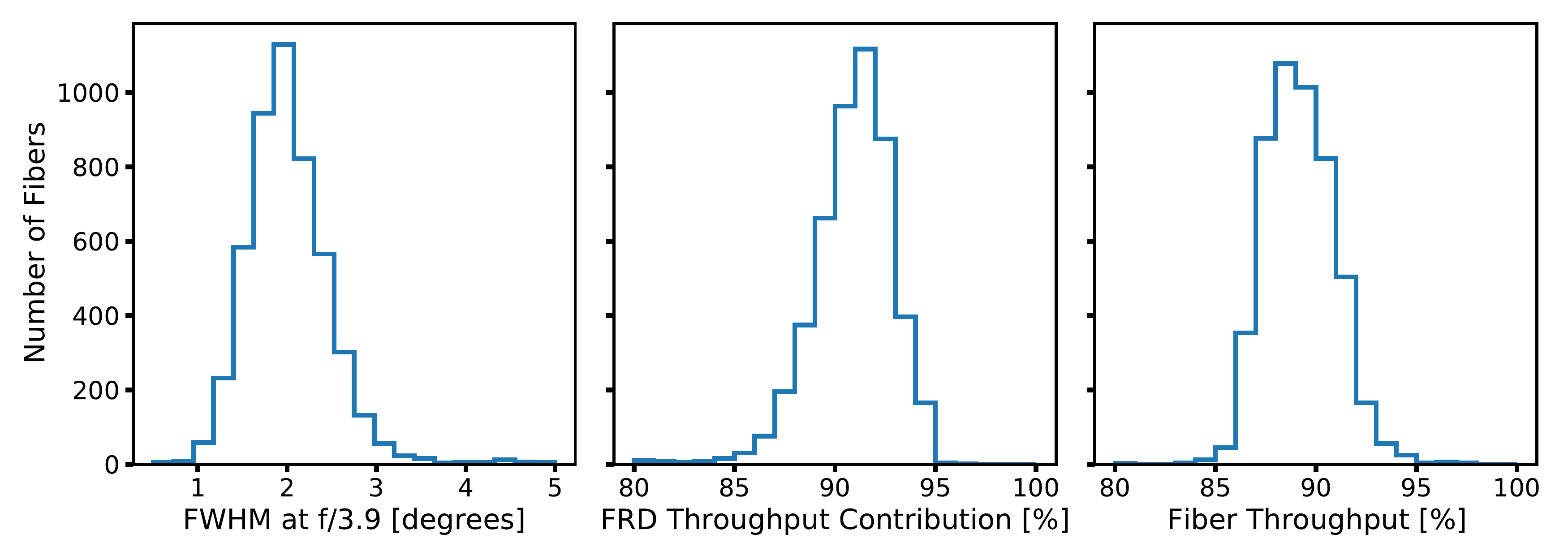}
\caption{Results from pre-ship testing of the fiber system for all ten petals. {\it (Left)} Measured FWHM of the output ring in degrees when illuminated by a collimated red LED at $7.3^\circ$ relative to the optical axis of the fiber. Broader FWHM corresponds to greater FRD, and consequently less flux within the nominal f/3.57 output beam. {\it (Middle)} Estimate of the impact of FRD on the fiber throughput. {\it (Right)} Distribution of the total fiber throughput. The total fiber throughput is the combination of the flux throughput and the FRD contribution. The flux throughput is the fraction of the input light transmitted by the fiber, and the FRD throughput contribution accounts for loss due to light that is out of the f/3.57 output beam. \label{fig:fiberthru}}
\end{figure*}
 
  We performed FRD and absolute flux throughput measurements of the complete fiber system for each petal after it was assembled. The FRD measurements used the collimated (or ring) illumination test described in \S\ref{sec:cables}. The average FWHM of the output ring was $1.92^\circ$, and the distribution is shown in Figure~\ref{fig:fiberthru}. We empirically determined the relationship between the measured FWHM and the inferred FRD throughput with measurements of multiple fibers with both full cone illumination and the collimated FRD test \citep{poppett18}. The FRD throughput contribution corresponds to the encircled energy within an f/3.57 output beam. The average inferred FRD contribution to the throughput is $90.3$\%, which met the requirement of 90\%. The other contribution to the fiber throughput is the flux transmission, which includes both surface losses and attenuation in the fiber. We measured the flux transmission of each fiber with measurements relative to a well-calibrated test fiber. The fiber throughput is the combination of the FRD contribution and the flux transmission. The average fiber throughput is $89.1$\%. Figure~\ref{fig:fiberthru} also shows the distribution of the FRD throughput contribution and total fiber throughput for all ten petals. 
  
 \section{Spectrograph System} \label{sec:spec}
 
\begin{figure*}[ht!]
\includegraphics[width=7in,angle=0]{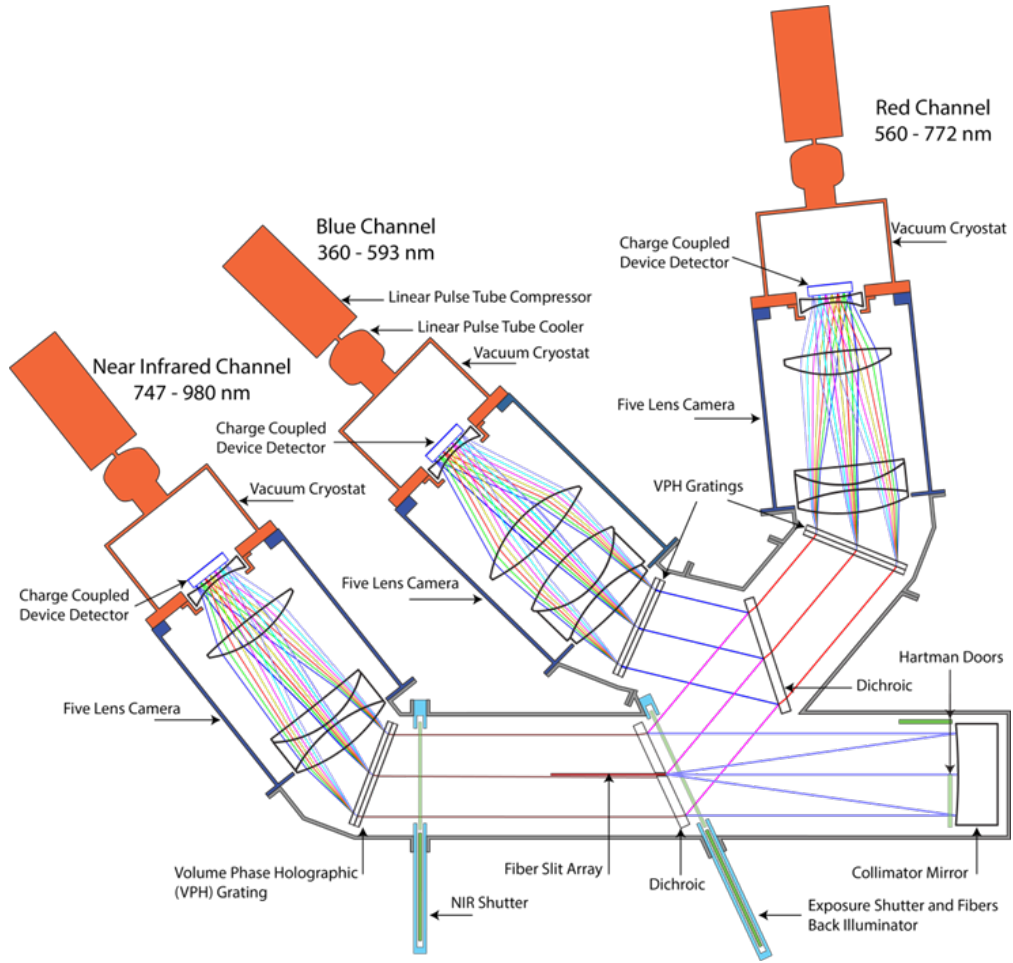}
\caption{Schematic of one spectrograph that shows the optical design and mechanical structures in each spectrograph. Light enters the spectrograph at the fiber slit array and is directed toward the collimator mirror through a slot in the NIR dichroic, and the collimated light is reflected directly back toward the fiber slit array. Light in the NIR channel ($747-980$nm) passes through the NIR dichroic, the NIR VPH grating, and enters the NIR camera. Shorter wavelength light is reflected off the NIR dichroic and encounters the red dichroic. Blue light ($360-593$nm) reflects off the red dichroic, is dispersed by the blue VPH grating, and enters the blue camera. Red light ($560-772$\,nm) passes through the red dichroic, is dispersered by the red VPH Grating, and enters the red camera. The exposure shutter is immediately after the NIR dichroic in the optical path. There is a second (NIR) shutter before the NIR VPH Grating, which is closed when the fiber illumination system is in use. All of the spectrographs are identical.  \label{fig:speclayout}}
\end{figure*}

\begin{figure*}[ht!]
\includegraphics[width=7in,angle=0]{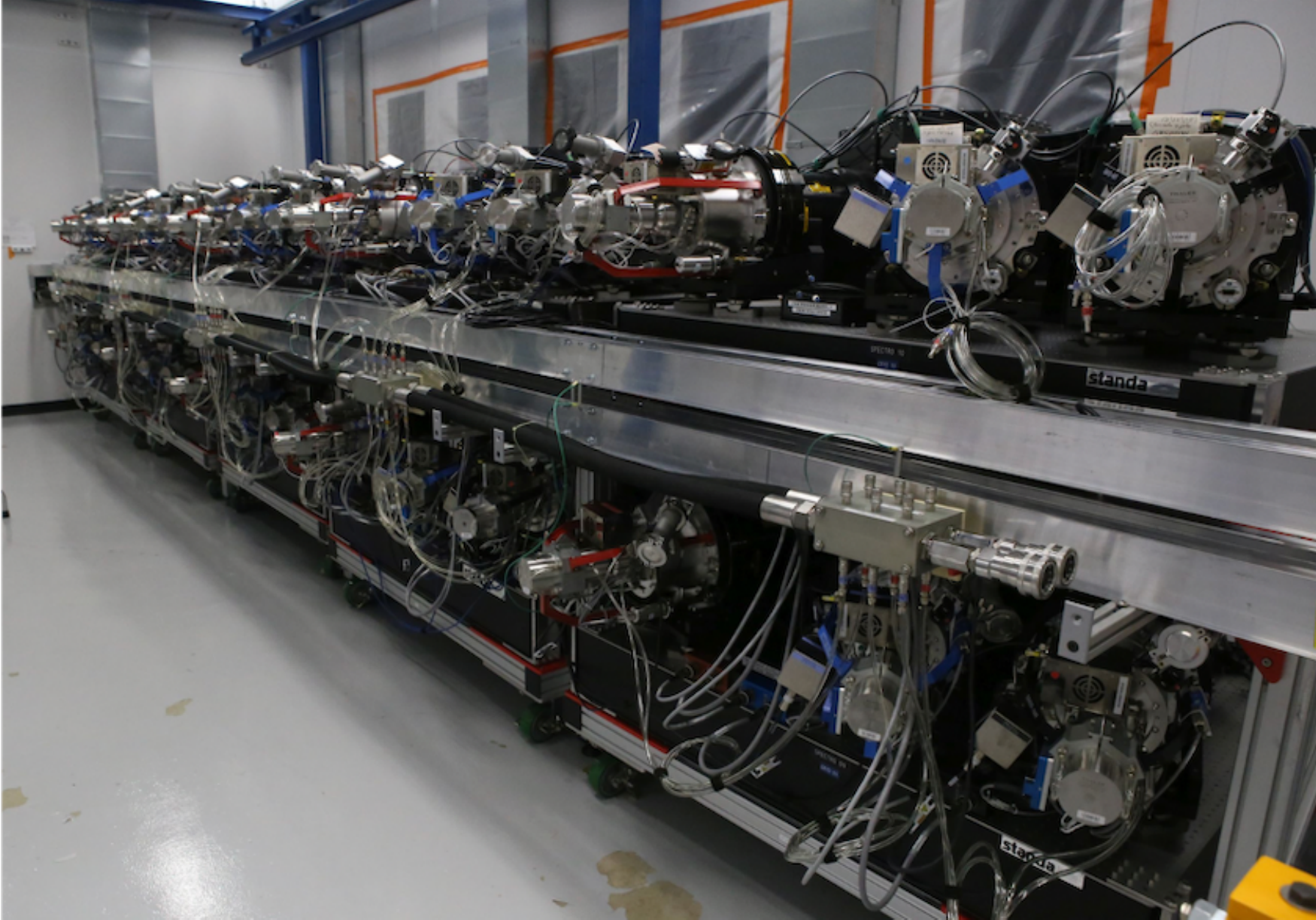}
\caption{The ten, completed spectrographs installed on the Rack in the Shack at the Mayall. Five spectrographs are along the top, fixed structure called the Upper Rack. The other five are mounted on low-profile carts called the Lower Rack and may be rolled out for servicing. The Shack includes an overhead, gantry-style crane that may lift any spectrograph off of the Rack. \label{fig:spectros}}
\end{figure*}

 The DESI spectrograph system includes ten identical, bench-mounted spectrographs, their support structure and environmental enclosure, the calibration system, and the sky monitor. The spectrographs have two dichroics that split the light into three wavelength channels that are referred to as Blue, Red, and Near-Infrared (NIR). This split helps to optimize throughput, increase spectral coverage, and allows for a distinct spectral resolution for each channel. Figure~\ref{fig:speclayout} shows an overview of the spectrograph optical design and Figure~\ref{fig:spectros} is a photo of the ten spectrographs after installation. All ten spectrographs are mounted on a support system referred to as the Rack, and these are maintained at constant temperature in a clean, environmental enclosure referred to as the Shack. The spectrograph calibration system is comprised of four identical calibration lamp units mounted to the top ring of the telescope, and the light from these lamps reflects off a new dome flat field screen. The sky monitor is a pair of cameras that monitor the night sky continuum brightness through dedicated sky fibers, and are one of the inputs to the Dynamic Exposure Time Calculator (\S\ref{sec:etc}). Some earlier descriptions of the spectrographs include a report from \citet{edelstein18}, measurements of the gratings by \citet{ishikawa18}, and the pre-ship testing of the spectrographs by \citet{perruchot20}. Jelinsky et al. (2022) {\it in preparation} will present a more detailed description of the spectrograph system than the brief summary in this section. 

 \subsection{Technical Requirements} \label{sec:specreq}
 
 A series of Level 4 technical requirements were imposed upon the spectrograph design, and these were used to also generate Level 5 specifications for the cryostats, detectors, Rack, and Shack. One class of Level 4 requirements stipulated conformance with the design adopted for other subsystems, namely the fiber diameter of $107\,\mu$m, the input light cone of $f/3.57$, the spectral bandpass from $360 - 980$nm, and accommodation of 5000 total fibers. The other class of requirements specified the resolution, throughput, PSF stability, and noise. Those key Level 4 requirements and Level 5 specifications are summarized in the relevant subsections. 
 
 \subsection{Optomechanical Design} \label{sec:optomech}
 
 The spectrograph design has a reflective, $f/3.57$ spherical Schmidt collimator and three refractive, $f/1.7$ cameras. This produces a magnification of 0.48 of the $107\,\mu$m fiber diameter onto the $15\,\mu$m pixels in the three wavelength channels. Light enters the spectrograph through a 121\,mm pseudo-slit composed of 500 fibers from one of the petals in the FPA. This pseudo-slit protrudes through a slot in the NIR dichroic such that it is immediately behind the exposure shutter. Collimated light is reflected directly back toward the slit. Light for the NIR channel passes through the NIR pass dichroic, the NIR VPH grating, and into the NIR camera. The slithead vignettes approximately 1\% of the light that enters the NIR channel. Light in the red and blue channel wavelength range is reflected off the NIR pass dichroic and the blue light reflects a second time off of the red pass dichroic, and then is dispersed by the blue VPH grating and enters the blue camera. Light in the red channel wavelength range passes through the red dichroic, is dispersed by the red VPH grating, and enters the red camera. 
 
 The collimator mirror was fabricated by Winlight Optical Systems and coated by Infinite Optics with an enhanced silver coating. While the spectrographs are maintained at a constant temperature, the collimator mount design includes athermalization to further maintain stability in the event of small temperature fluctuations. The mount also includes adjustment features for tip/tilt and focus, and the Hartmann door mechanisms described in \S\ref{sec:mech}. 

 The NIR pass dichroic transmission requirements were $>95$\% from 772 to 980\,nm, and reflectance $> 95$\% from 360 to 757\,nm. The average incidence angle is $25.4^\circ$. The red pass dichroic requirements were $>95$\% transmission from 593 to 772\,nm, and $>95$\% reflectance from 360 to 566\,nm, and the average incidence angle is $32.1^\circ$. The substrates for both types of dichroic are fused silica, and all of the dichroics were produced by Materion. The mount design includes several degrees of tip/tilt adjustment to aid the optical alignment. 

 There is a distinct grating design for each wavelength channel. All three gratings have physical dimensions of $180 \times 160$\,mm$^2$, are 16\,mm thick, and the incident beam diameters are approximately 140\,mm. The VPH grating fringe frequencies and average incidence angles are 1103.5 l/mm and $11.48^\circ$ (Blue), 1157.4\,l/mm and $18.73^\circ$ (Red), and 992.5 and $21.19^\circ$ (NIR). The fringe angles for all three were $3^\circ$ to avoid Littrow ghosts, and in all cases the diffraction angle is $<16^\circ$. There is also an AR coating on the exterior faces of the gratings. All of the gratings were produced by Kaiser Optical Systems, Inc. and mounted to the cameras by Winlight. The resolutions of the blue, red, and NIR channels are approximately 2000 -- 3000, 3500 -- 4500, and 4000 -- 5500, respectively. These values meet the resolution requirements of FWHM ($\lambda/\Delta\lambda$) $\geq 1500$ from $360 - 555$\,nm, $\geq 3000$ from $555-656$\,nm, and $\geq 4000$ from $656 - 980$\,nm. 
 
 There are different optical designs for the blue, red, and NIR cameras. The optical designs for the cameras each use a triplet, a singlet, and a field lens that also serves as the window of the cryostat. These designs produce excellent image quality, and achieve the throughput requirement. The average rms radius is better than $6\,\mu$m and the maximum is approximately $10\,\mu$m. The blue camera has two large ($\sim 160-$mm diameter), convex lenses fabricated from Calcium Fluoride crystal that sandwich a convex N-BAK2 lens. Due to the mismatch in CTE between these materials, and the fragility of CAF2, Winlight used a fluid-coupled solution for this triplet that was based on the solution designed by \citet{obrien14} for the C/KOSMOS instruments \citep{martini14}. The triplets in the red and NIR cameras were fabricated from optical glasses with similar CTE (S-BSM4, S-TIM22, and S-LAL12) and cemented. All three field lenses were constructed from Fused Silica. All of the lenses were polished by Winlight and coated by Oerlikon Balzers. Winlight built and aligned the cameras. 

 \subsection{Mechanisms} \label{sec:mech}
 
 There are four mechanisms in each spectrograph: two shutters and two Hartmann doors. The exposure shutter is the first element in the light path after the slithead, and the operation of this shutter sets the length of exposures. The exposure shutter also contains the back-illumination system for the focal plane system. The second shutter is located between the slithead and the NIR camera. Both shutters have pneumatic seals that produce an extremely light-tight seal around the shutter blade when inflated. When the back illumination system is in use, both of the shutters are closed and the seals are inflated; otherwise just the shutter doors are closed. The control system includes interlocks to prevent the activation of the back illumination system when the shutter is not closed and the seal is not inflated. The two Hartmann doors are integrated with the collimator mirror mount. The locations of these mechanisms are shown on Figure~\ref{fig:speclayout}. 
 
 The main specifications on the exposure shutter were that it move from completely open to completely closed in less than 0.5\,seconds, and that the actuation time be uniform to 0.1\,seconds. These specifications minimize uncertainties in the true length of each exposure. The illumination system included specifications for total flux within the $400-550$\,nm bandpass to ensure adequate SNR and minimize chromatic aberrations in the centroids measured by the FVC, and to avoid contamination through the $r-$band filters mounted in front of the GFA detectors. Lastly, the shutters were required to transmit $<10^{-7}$ of the flux at the fiber tip into the red and blue channels so the back illumination system could be used when the detectors are still reading out the previous exposure. \citet{derwent16} present a detailed description of the exposure shutter design. The NIR shutter is identical to the exposure shutter, except it does not have a back illumination system, and it has a somewhat smaller aperture on the shutter housing due to the smaller beamsize at its location. The two Hartmann doors are used to measure the focus of the three optical channels through a classic Hartmann test, and consequently each door is actuated by a separate motor. 
 
 All control electronics for the spectrograph mechanisms are housed in an electronics box called the spectrograph controller, and there is one per spectrograph. The control box includes a digital stepper motor controller for each of the four mechanisms, a NUC PC that handles communications with the ICS, an air cylinder for the shutter seals, solenoid valves that control the air to the seals, and other control electronics to read various sensors. Each motor controller has a distinct command set to perform specific actions, such as to open or close a shutter, as well as to provide low-level equipment safety. More details of the control system were presented in \citet{coker16}. 
 
 \subsection{Detectors and Cryostats}
 
 All three spectrograph channels have back-illuminated CCDs with $15\,\mu$m pixels, four readout channels, and a readout rate of 100\,kHz. The blue devices are  $4096 \times 4096$ STA4150 CCDs from Semiconductor Technology Associates (STA) that were processed and packaged by the University of Arizona Imaging Technology Laboratory (ITL) to meet high efficiency and low noise requirements. The quantum efficiency requirements included better than 75\% from 360 -- 400\,nm and $\geq 85$\% from 450 -- 600\,nm. The read noise requirement was better than $3\,e^-$ rms and the dark current requirement was less than $10\,e^-\, {\rm pix}^{-1}\,{\rm hr}^{-1}$. There were also numerous other requirements on quantities such as charge transfer efficiency and cosmetics. All of the devices were tested at ITL prior to shipment and then retested with the DESI electronics. ITL delivered 13 science grade devices and we mounted the best ten plus one spare into cryostats. 
 
 The devices for the red and NIR channels were especially developed for DESI by LBL with wafers fabricated by Dalsa. These are thick, high resistivity, fully-depleted $p-$channel  $4114 \times 4128$ CCDs that provide very high quantum efficiency in the near-infrared. The quantum efficiency requirements included better than 85\% from 600 -- 900\,nm and better than 60\% from 900 -- 980\,nm. The read noise requirement was better than $3\,e^-$ rms and the dark current requirement was less than $10 e^-\,{\rm pix}^{-1}\,{\rm hr}^{-1}$. These devices achieved higher quantum efficiency in the near-infrared and lower noise than the previous generation, such as those for DES \citep{flaugher15}, through two main developments. First, improvements to the AR coating through the addition of higher-index, ZrO$_2$ to the multi-layer coating increased the quantum efficiency. Second, the development of a buried-contact technology combined with optimizations to the transistor size substantially reduced the read noise \citep{holland14,bebek15}. We mounted the best twenty devices plus two spares into cryostats.

 The cryostat system has one self-contained vacuum vessel for each spectrograph channel. The requirements on the cryostats include precise $\pm1$ K and stable $\pm0.1$ K temperature control, and the ability to accommodate the different operational temperatures of the devices (163\,K for the blue channel, and 140\,K for the red and NIR channels). The cryostats maintain temperature control with Thales linear pulse tube cryocoolers. An ion pump maintains the pressure below $3 \times 10^{-7}$\,mBar for all cryostats (well below the $<10^{-3}$\,mBar requirement). The cryostats are monitored with a slow control system that includes setting and regulation of the CCD temperature, setting the pulse tube frequency and electrical power, and control of the warmup and cooldown process. The field lenses for the spectrograph cameras form the windows of the cryostats. The detectors and field lenses were integrated with the cryostats at CEA Saclay, tested, and then delivered to Winlight for integration with the spectrographs. CEA/Saclay also reintegrated the cryostats with the spectrographs and tested them after they arrived at Kitt Peak.

 \subsection{Rack and Shack} \label{sec:shack}
 
 The ten spectrographs are supported in two rows of five on a structure called the Rack, which is shown on  Figure~\ref{fig:spectros}. The main design requirements on the Rack were to place the spectrographs as close as practical to the entry point of the fiber cables into the Large Coud\'e Room in order to minimize the total fiber length and to minimize vibration frequencies. The Rack is approximately 8.5\,m long and 1\,m deep and constructed with an extruded aluminum frame with steel braces that provide both structural rigidity and the mounting surface for the spectrographs. The Upper Rack is a fixed, interconnected structure that supports five spectrographs. The upper spectrographs were installed directly into position on the Upper Rack via an overhead crane. The Lower Rack is comprised of five, low-profile wheeled carts. The lower spectrographs were installed onto these carts with the overhead crane and then wheeled into position under the Upper Rack. The Upper Rack includes cable trays along its length on both sides, which support all utility lines such as communications, power, and air lines. The Upper Rack also has shelves to support the Slitheads when they are removed for special calibrations. As several spectrographs had not yet been installed after installation of the focal plane system, those Slitheads were placed on the shelves with an illuminator similar to the one on the exposure shutter. This stand-alone fiber illumination system was used to test the focal plane system before all of the spectrographs arrived. 
 
 The ten spectrographs are installed in a clean, thermal enclosure called the Shack. The Shack is a commercial lab enclosure that we purchased as a kit from Western Environmental Corporation (WEC) to our specifications and then assembled in the Large Coud\'e Room. The key specifications for the Shack included thermal stability of $20\pm2^\circ$, relative humidity of $<50\%$, and the ability to provide a clean room environment better than class 10,000 when the optics are exposed for $>12$ hour service operations. The thermal specification optimizes the spectrograph optical design performance and the humidity specification is to prevent condensation on the Dewar windows. The exterior dimensions of the main room of the Shack are $10.7 \times 4.7$ meters and the exterior height is nearly 4\,m. The interior height is 3.5\,m, which includes space for the operation of an overhead gantry crane. There is also an Annex that is $4.4 \times 2.5$ meters with a lower ceiling height (the crane does not extend into the Annex) that was intended to support major maintenance activities. The dimensions of the Large Coud\'e Room were the most significant limitations on the size of the Shack, and this area was carefully surveyed before the start of the detailed design work. 
 
 There are two industrial control electronics boxes for the Shack. One was supplied by WEC, which controls the thermal environment, as well as reads telemetry data from all of the components supplied by WEC. These include temperature, humidity, and pressure, as well as information about the HVAC system. The second control box is called the shack controller and it provides remote power control to the ten spectrograph controllers, monitors the air pressure of the shutter seals, operates the stand-alone fiber illumination system, and reads the telemetry data from the WEC control box. The ICS regularly queries and stores these telemetry data in the project's database. 

 \subsection{Calibration System} \label{sec:cals}
 
 The calibration system for the spectrographs includes a new dome screen and new calibration lamps. The key specifications included continuum and spectral line coverage across the DESI bandpass, azimuthally averaged pupil uniformity to 20\%, field uniformity over the $3.2^\circ$ diameter FOV to 5\%, and a specification on the flatness of the continuum emission. The pupil uniformity requirement is designed to limit PSF inhomogeneities, while the field uniformity is relative to the sky, and is designed to minimize sky subtraction residuals. The spectral line coverage is motivated by the wavelength calibration requirements needed to achieve the Level 3 requirements on systematic redshift uncertainties, as well as the relative wavelength calibration between fibers to minimize sky subtraction residuals. 
 
\begin{figure*}[ht!]
\includegraphics[width=7in,angle=0]{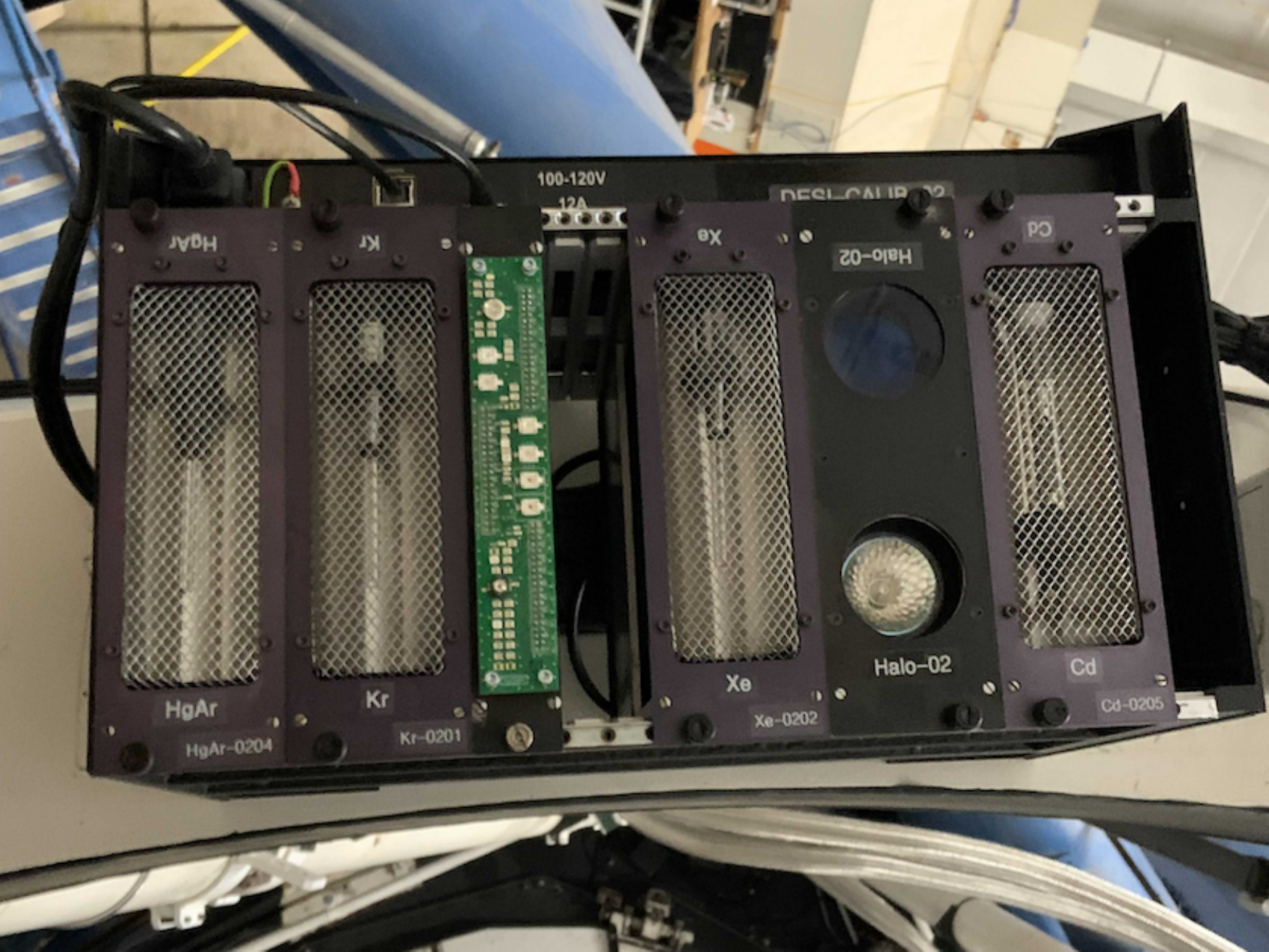}
\caption{Photo of one of the four calibration lamp boxes installed on the top ring. Each calibration lamp box has six drawers (vertical structures, four of which are protected by mesh screens). From left to right these drawers contain HgAr, Kr, a collection of LEDs, Xe, two halogen lamps, and Cd. The two halogen lamps are continuum lamps, where one has a blue filter and the other is unfiltered.  \label{fig:callamp}}
\end{figure*}

 The new dome screen is a replacement of the previous screen, which was too small for DESI. It has a usable diameter of 5173 mm, which corresponds to a FOV of $3.4^\circ$, and this allows for a lateral placement error of $\pm 230$ mm. The screen is coated with Permaflect (sold by Labsphere), and has reflectivity $>90$\% over the entire DESI bandpass that is independent of incidence angle. The new dome screen was completed in 2017. The calibration lamps are mounted in four boxes that are attached to the top ring. One calibration lamp box is shown in Figure~\ref{fig:callamp}. Each calibration lamp box has slots for six drawer units that contain lamps. The first drawer of each box contains a set of LEDs that together produce a reasonably flat, broad-based SED, and the second drawer contains two halogen lamps, one with and one without a blue filter. The remaining drawers have Cd, Xe (2/4 boxes) or Ne (2/4 boxes), Kr, and HgAr wavelength calibration lamps. There is a Raritan Power Distribution Unit (PDU) for each lamp box, and the lamps are activated through the PDU. The PDUs also have temperature and humidity sensors to protect each lamp box. 
 
 \subsection{Sky Monitor} \label{sec:skymon}
 
 DESI uses the Dynamic ETC to actively estimate the spectroscopic SNR while exposures are in progress, and adjusts the exposure time as needed to ensure the SNR requirements are met. The inputs to the Dynamic ETC include real time measurements of the image quality, transparency, and the sky continuum brightness. The first two quantities are measured with the guider CCDs, and the sky continuum brightness is measured with the sky monitor. 

 The sky monitor was designed to take input from 20 dedicated sky fibers mounted in fiber positioners on the periphery of the focal plane, with two per petal. The sky fibers are part of the fiber bundle from the focal plane system to the Shack, although they terminate at SMA connectors on the side of each fiber spool box, rather than in the slithead in each spectrograph. Short segments of DESI fiber with SMA connectors link the spool boxes to one of two dedicated camera systems that sit below the spool boxes on the exterior of the Shack inside the Large Coud\'e Room. Each unit has an SBIG STXL-6303e CCD camera, an eight-position FW8S-STXL filter wheel, a Nikon 50 mm $f/1.2$ lens that reimages the fiber tips onto the CCD, and a back-illumination system for the fiber tips. \citet{tie20} describe the sky monitor in further detail. 
 
 \subsection{Pre-Ship Testing and Performance} 
 
\begin{figure*}[ht!]
\includegraphics[width=7in,angle=0]{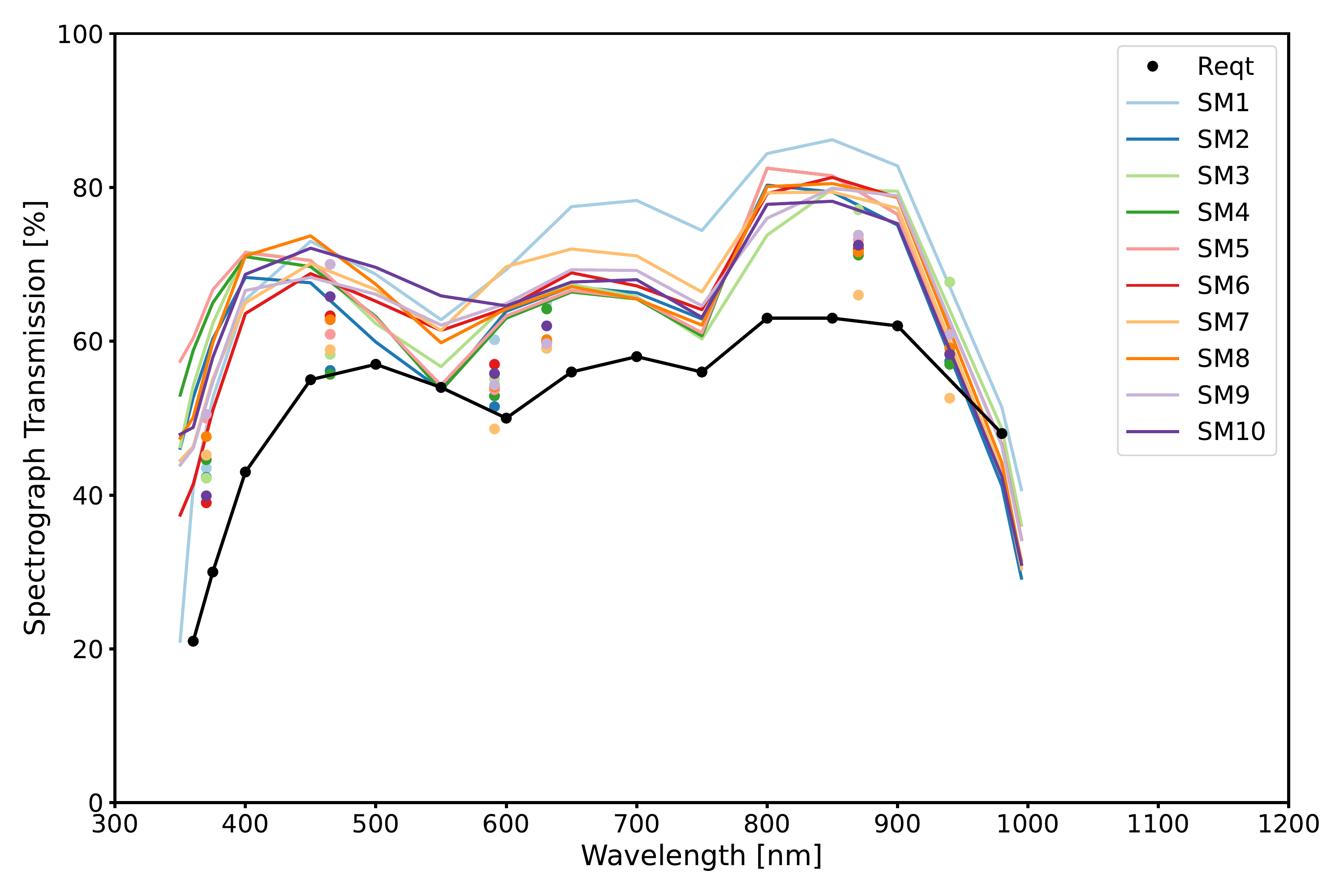}
\caption{Lab throughput measurements for the ten spectrographs from 350 -- 995\,nm. The throughput for each spectrograph ({\it solid lines}) is based on the measured reflectivity of the collimator, the  reflectivity and tranmission data for the dichroics, the efficiency of the gratings, acceptance reports of each camera, and quantum efficiency measurements for the CCDs. Integrated throughput measurements from lab tests are shown as filled points with the same color as the solid line for a given spectrograph. The spectrographs exceed the requirement ({\it black points and line}) over the entire wavelength range (see \S\ref{sec:spec}) with the exception of 980nm. \label{fig:specthru}}
\end{figure*}

 Each spectrograph was extensively tested and characterized before shipment to the Mayall. These tests included focus, optical performance, CCD properties, throughput measurements for each channel, and stray light measurements and functional checks of the mechanisms. This lab system included a range of LEDs, arc calibration lamps, and a white light source that illuminated an integrating sphere. An Offner relay then reimaged the output into DESI-like fibers that comprised a sparse test slit with just 21 fibers, rather than 500. We performed the focus tests with the Hartmann doors, and then used arc calibration lamps to quantify the image quality, resolution, and dispersion. We also used the sparse test slit, in conjunction with a calibrated photodiode, to measure the throughput of each spectrograph as a function of wavelength. Finally, we measured the CCD readout noise and gain from data with a separate, flat field slit which was illuminated with a tungsten lamp. These tests verified that the spectrographs met their requirements. 

 The calibration system and sky monitor underwent numerous pre-ship tests. The key requirements for the calibration system were the bandpass coverage, pupil and field uniformity, spectral line coverage, and continuum flatness. Most of these quantities were measured with lab tests prior to shipment, with the exception that the pupil uniformity was verified by analysis. The continuum flatness required a maximum variation of no more than a factor of three in any spectrograph channel, whereas we only achieve a factor of five with many LEDs. In practice, this variation is acceptable as we simply take more exposures and coadd the results to obtain high SNR flat fields across the bandpasses. For the sky monitor, we verified the readout cadence, and heat dissipation prior to shipment. The remaining requirements were verified after installation as part of commissioning the Dynamic ETC (see \S\ref{sec:etc}). The verification of the temperature stability, humidity control, and cleanliness of the Shack environment after installation is described in \S\ref{sec:specinstall}.  

\section{Instrument Control System} \label{sec:ics}

 The Instrument Control System (ICS) manages observations and other instrument activity via active control of all of the instrument components and the telescope. These many components have an extensive range of potential configurations to update and track, as well as provide a substantial amount of telemetry. The ICS controls all of these components with a distributed, multi-processor system with about 80 nodes that range from embedded controllers to server class computers, and many more applications and process. The ICS has a database system that maintains the configuration and status information, and stores all telemetry. These data form the basis of an integrated warning and alarm system. Observers interact with the ICS via its series of web-based, graphical user interfaces (GUIs). These are a key element to support remote observing and instrument operations for DESI.
 
 Observations are initiated with exposure requests to the ICS Observation Control System (OCS). A typical exposure request is a pre-established design for a DESI field or tile that includes the telescope pointing, the assignment of individual science targets to fiber positioners, and pre-determined guide stars. In regular survey operations, the OCS receives each exposure request from the Next Tile Selector (NTS). The ICS uses this information to orchestrate a complex sequence that includes slewing the telescope, adjustments of the ADC and hexapod, field star acquisition and astrometry, control of the guider, the start of regular exposures with the four focus/alignment GFAs and the sky monitor system, iterations of the fiber positioner sequence via back-illumination and the FVC, and the start of spectroscopic observations. Once an exposure begins, the ICS actively controls the exposure times of spectroscopic observations through the Dynamic ETC. This application combines data from the guide cameras and sky monitor to adjust the duration of exposures while they are in progress to ensure the spectra meet our SNR requirements. The active optics system (AOS) analyzes the wavefront from images obtained with the focus/alignment GFAs during the spectroscopic exposure.  We allow focus updates during exposures. Upon the completion of each exposure, the AOS determines the best focus and sends any correction to the hexapod. During the spectrograph exposure we compensate for field rotation by changing the hexapod rotation once per minute.
 
 The ICS has been actively developed over many years, including early versions that were used for the Legacy Surveys \citep{dey19} and the ProtoDESI campaign \citep{fagrelius18a}. This development included support for a number of different observation modes to support component testing, and this remains a key functionality for testing and to provide support for specialized observations, such as the regular acquisition of arc, flat, and other spectroscopic calibration data. Previous descriptions of the ICS during development were presented in \citet{honscheid16} and \citet{honscheid18}. The next subsections describe the state of the ICS upon the completion of commissioning and the start of survey operations. 
 
 \subsection{Technical Requirements}
 
 The ICS requirements flow from the Level 3 requirements on survey efficiency and duration. The overall requirement is that the ICS must perform all control and monitor functions to enable successful operation of the instrument and the completion of the science program. One of the main performance requirements is that the time between successive exposures, called the inter-exposure time, must be less than two minutes\footnote{When moving to a neighboring tile with telescope slew angles under 10 degrees.}. Another is that the system up time must be greater than 97\% by the end of the first year of the survey. There are several monitoring requirements that include storage of instrument telemetry, provision for interfaces to the telemetry data, and a warning and alarm system. The warning and alarm system is informational by design, as ultimately hardware protection was a requirement on each subsystem. The instrument control requirements include a user interface, connection to the hardware subsystems, and the Dynamic ETC. Lastly, the ICS has to provide computing hardware to perform nighttime survey operations, as well as sufficient storage for at least three nights of observations should there be a loss of off-mountain data links. The off-mountain data transfer, along with data processing, is the responsibility of the DESI data systems group. That work will be described in a separate series of papers.

 \subsection{Software Architecture}
 
\begin{figure*}[ht!]
\begin{center}
\includegraphics[width=0.9\textwidth]{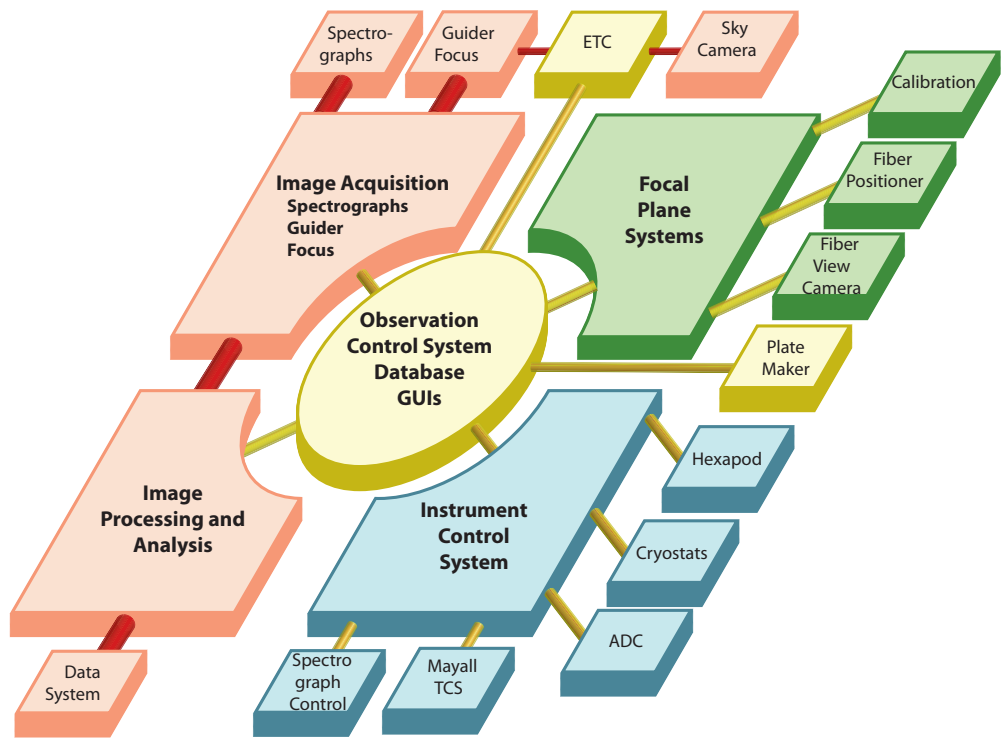}
\end{center}
\caption{Schematic view of the DESI readout and control system. The observer console and other user interfaces are not shown.}
\label{fig:daqoverview}
\end{figure*}

 The design of the DESI online system is modelled on the readout and control system architecture developed for the Dark Energy Camera \citep{diehl15}, which has been successfully used for both the Dark Energy Survey and other DECam observing programs. There are detailed descriptions of the DECam data acquisition system in both \citet{SPIEhonscheid2012} and \citet{SPIEhonscheid2014}. A schematic view of the overall architecture for DESI is shown in Figure~\ref{fig:daqoverview}.
 
The core of the system is the observation control system (OCS), which manages the complex DESI exposure sequence. The OCS interacts closely with the focal plane system, which includes the fiber positioners, the FVC, and the GFA cameras. Image acquisition originates with the GFA and spectrograph CCD controllers, and then flows through the image processing and analysis stage to a storage device. At that point the images are transferred and further processed by the data systems group.  The ICS connects to hardware components, including the spectrographs, hexapod, cryostats, and ADC, as well as the Mayall Telescope Control System (TCS) and telemetry database. 

The readout system is tightly integrated with the DESI instrument monitoring and control system. This system monitors every component of the instrument and includes detailed information about the instrument status, operating conditions, and performance. All such data are archived in the DESI operations database. The monitoring and control system provides all components of the instrument with access to the telemetry and alarm history databases, and can do so even if the rest of the OCS and the readout system are offline.

\subsubsection{Infrastructure Software} \label{sec:icsinfrastructure}

The DESI readout and control system is implemented as a distributed multi-processor system. Approximately a hundred processes are used to operate the DESI instrument.
Due to the nature of this architecture, inter-process communication was a major consideration in the design of the DESI infrastructure software. Our solution employs the Python Remote Objects (PYRO) software package\footnote{https://pyro4.readthedocs.io} and draws on extensive heritage from the very successful DECam online system. PYRO is a free, advanced, and powerful distributed object technology system written entirely in Python. It allows objects to interact just like normal Python objects even when they are spread over different computers on the network. PYRO handles the network communication transparently. PYRO provides an object-oriented form of remote procedure calls similar to Java's Remote Method Invocation. A name server supports dynamic object location, which renders (network) configuration files obsolete. Using the name server, DESI processes can locate and establish communication links with each other irrespective of the underlying hardware architecture. This is an important feature for any multi-processor communication system and a key element to support fast reconfiguration in case a component fails in the middle of the night.

The DESI Online Software (DOS) communication distinguishes between command messages and telemetry data. Commands are used to request information from a remote application or to activate a remote action. The command or message passing system is implemented using a Client-Server design pattern with a thin custom software layer (Program Macro Language or PML) on top of PYRO. PML introduces the concepts of roles and devices to provide a uniform naming scheme for all DESI online applications. The telemetry system is based on the publish-subscribe design pattern using ideas similar to data distribution services, a standard for publish-subscribe middleware for distributed systems. We developed a concept called shared variables that is built on the core functionality provided by PYRO, although adapted to the DESI online system. This concept includes a client stub library and a central server built on top of a Redis\footnote{https://redis.io} in-memory data structure store, which allows user applications to publish information such as temperature readings or readout status to a virtual data space. Other applications can subscribe to information placed in this virtual data space and receive updates whenever a publisher submits a new value. The shared variable system supports asynchronous callbacks, guaranteed delivery, multiple publishers of the same shared variable, and group subscriptions.

DOS is based entirely on open source software. The default operating system is RedHat Enterprise Linux 7/Centos 7, but since the software is written entirely in Python~3, other systems like Mac~OS and various Linux distributions used by embedded controllers are supported as well. The inter-process communication system is built on top of PYRO4 and Redis. The operations database uses PostgreSQL and the web-based graphical user interfaces are based on the SproutCore toolkit. All code resides in an svn repository hosted at LBNL. Our release management tools are based on the eUPS software originally developed for SDSS and now maintained by the LSST project.

\paragraph{Configuration}

Initialization and configuration of a complex distributed system such as the DESI online system is a multi-step process.  The complete system has about 60 nodes that range from embedded controllers to server class computers, and an even larger number of applications and processes. Some device applications will already be running and need to be integrated with the newly started processes. We developed a flexible startup system called the Architect to start the online processes in the correct order, on the correct computer, and with the correct arguments with a single command. This Architect connects all compute nodes in the chosen configuration and initializes DOS system services before starting all user applications. Initialization files in standard Windows .ini notation are used to describe the configuration. The .ini files identify the participating nodes and specify which applications should be started on each node. They also specify the calling arguments and other configuration information for each process. Once all applications and services are running, the Architect continuously monitors all processes at the operating system level.

\paragraph{Application Framework}

All DESI online applications are based on a common software model. This application framework serves as a base class and provides a unified interface to all DESI services such as the configuration system (Architect), the DOS data cloud (shared variables), the message passing and remote procedure system (PML), alarms, logging, database access, and the interlock and monitoring system. The application framework manages all resources centrally, which allows it to release all allocated resources when the application is about to exit. Proper exit handling is critical for the ability to stop and restart individual processes without the need to end the entire instance. Additional functionality provided by the application framework includes a heartbeat that can be used to monitor the overall state of the system and a standardized management interface with process control functions and access to an application state variable. This design of the DOS application framework significantly reduced the complexity of the development. The most important characteristics are that it is lightweight and written in pure Python, and the later feature makes it very  easy to port to different host environments including all of the embedded controllers used by the spectrograph and focal plane systems. The application framework also supports stand alone operation for devices that operate all the time and not just when the rest of the online system is running.

 \paragraph{Database}
 
At the heart of the DESI online system is the database system that hosts all databases for mountain-top operations. The system consists of a number of different components,  each with its own schema within the database. This structure keeps the different sets of tables organized while still permitting queries that join together tables in different schema. This would not be possible if we kept the different components in separate databases. The database server is PostgreSQL\footnote{http://www.postgresql.org}. The Python interface to the database uses the SQLAlchemy\footnote{http://www.sqlalchemy.org} module with  psycopg2\footnote{http://initd.org/psycopg} as the underlying database driver. The web viewers use the mod\_wsgi package to connect with the Apache server and the Grafana\footnote{https://grafana.com} PostgreSQL datasource \citep[e.g.,][]{abareshi18}.

\paragraph{Notifications and Alarms} 

Every DESI component includes an independent machine protection and fault monitoring system that is responsible for protecting that hardware from physical damage. Built on top of this is the ICS alarm and notification system that alerts the observers -- both on site and remote -- if a problematic condition has been detected. Critical alarms trigger acoustic signals, as well as automatic email and/or text messages to experts. All alarm messages are logged and archived in the operations database. In addition to alarms, the ICS provides a software interlock facility that can be used by the DOS application to monitor the status of other components, such as the spectrograph or petal controllers.

 \subsection{ICS Computer Hardware}
 
The DESI mountain top computer system consists of four server class machines for the database and the software and user home areas (including two hot standby servers for these services) and 17 rack-mounted Dell PowerEdge R540 workstations. Each workstation is configured with two Intel Xeon Gold 6126 processors running at 2.6\,GHz and with 96\,GB of memory. Dual network interfaces and dual hot-plug redundant power supplies increase the overall system reliability. The workstations and servers are connected by a 10G-base-T network using Cisco network switches. NFS-shared 44\,TB disk arrays in a RAID-10 configuration attached to each of the server machines provide ample storage for observer data, ICS software, and observation and other telemetry data in case the network connection off the mountain is down for an extended period. Intel NUC computers are used to control the FVC, the spectrograph mechanisms, and the spectrograph shack environmental system. The fiber positioners in each of the ten focal plane petals are controlled by BeagleBone Black micro controllers. Custom FPGA-based controllers with an ARM processor core are used to readout the spectrograph CCDs and the GFAs. Apple iMac Pro computers, each equipped with three high resolution 5k Retina displays, are used as observer consoles in the control room at the Mayall telescope. 

 \subsection{Observer Console and User Interfaces}
 
\begin{figure*}[!hbt]
\centering
\includegraphics[width=0.9\textwidth]{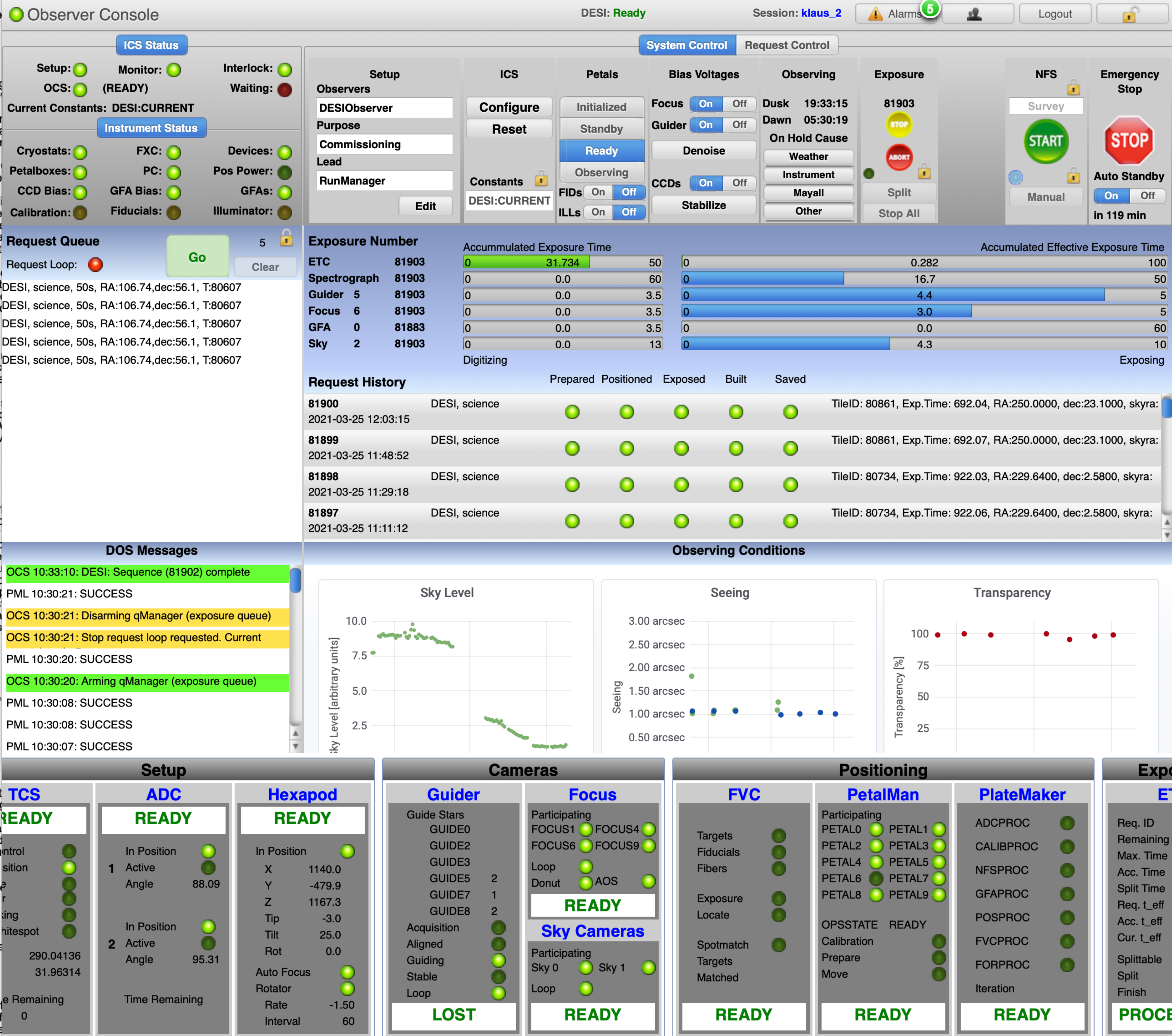}
\caption{Screenshot of the DESI Observer Console running in a web browser window. The top section provides status information (ICS Status and Instrument Status) and elements to control the instrument (System Control and Request Control). The center section shows the observing queue (Request Queue), the ongoing exposure (Exposure Number), and the exposure history (Request History). This is followed by information on the observing conditions (Sky Level, Seeing, Transparency). The bottom section provides more detailed status information on key components of the instrument (e.g., Cameras, Positioning). Additional GUIs not shown here provide more detailed information about the instrument status and ongoing information about the most recent observation.}
\label{fig:observerconsole}
\end{figure*}

 The DESI observer console acts as the primary observer interface for day-to-day operations. This set of graphical user interfaces includes elements for system and exposure control, alarm displays, and telemetry monitors. Figure~\ref{fig:observerconsole} shows the DESI observer console GUI. Combined, they present the user with the most commonly used information and everything that is needed to operate the DESI instrument. Near real-time performance is needed to provide quick access to more specific information and to achieve the responsiveness expected from a modern system. The DESI GUI architecture follows the Model-View-Controller pattern that is based on the realization that all applications are essentially interfaces that manipulate data. On the ICS server side, we have implemented the controller component as a web server so that we can use standard web browsers to render the views. Web-based user interfaces provide many desirable features such as platform independence, remote access, a large number of 3rd party tools, and also a certain level of security. Thanks to developments that have enabled faster rendering and JavaScript execution, as well as standards such as HTML5 and the Web-Socket API that allow for advanced functionality, this approach now provides equal performance to desktop GUI toolkits.  The client side (web browser) code for the DESI user interface is based on HTML/CSS, JavaScript and SproutCore,an HTML5 application framework.  Websockets, defined as part of the HTML5 standard, are used to provide  bi-directional information exchange and a responsive user experience.

\subsection{Exposure Sequence} \label{sec:exposure}

Exposure sequences contain the information necessary to setup the instrument and point the telescope for an observation. The ICS was structured to allow both direct control of exposure sequences and nearly autonomous operation through an application called the Next Tile Selector (NTS). The direct control capability was designed to support instrument testing, commissioning, and other specialized observations, and the NTS is used to initiate each exposure sequence during survey operations. In either case, the exposure sequences are loaded into a request queue and managed by the OCS. The architecture of the OCS is pipelined such that the setup of multiple components will proceed concurrently as long as there are no conflicts accessing instrument hardware or certain software components.

\begin{figure*}[!hbt]
\centering
\includegraphics[width=7in]{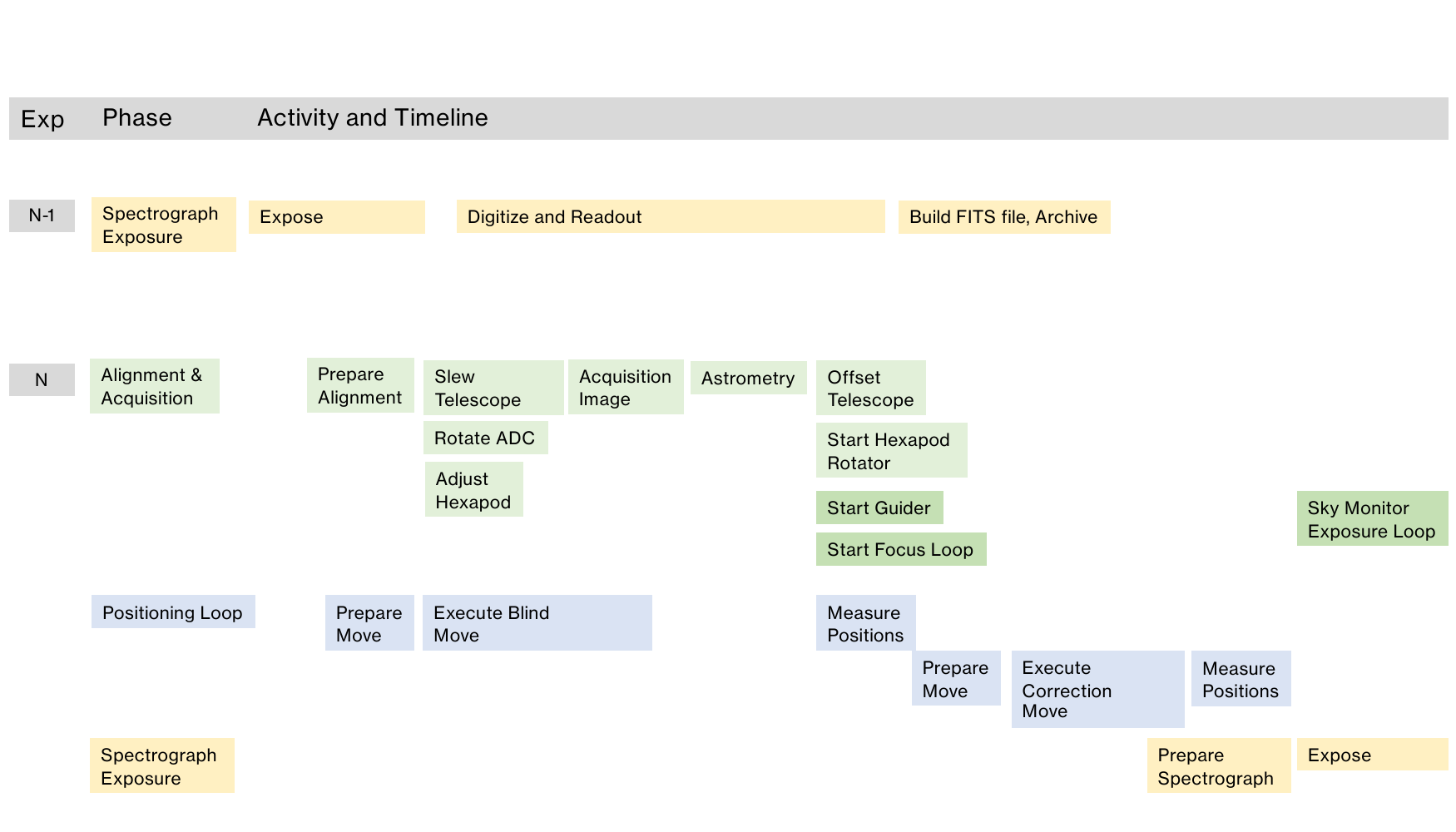}
\caption{Schematic view of the DESI inter-exposure sequence to align the instrument and the fiber positioners with a new target field. The pipelined architecture of the ICS allows interleaving processing of sequential exposures. The location of the boxes in the figure reflects the timeline with time increasing toward the right. See the \S\ref{sec:exposure} for further discussion.}
\label{fig:inter_exp}
\end{figure*}

Processing of the next exposure sequence begins with a notification from the Dynamic ETC that the current exposure is about to end within the next minute. The activities required to setup for the next exposure are shown schematically in Figure~\ref{fig:inter_exp}. The sequence starts by processing the information provided in the exposure request by the NTS. We determine the telescope, ADC and hexapod settings and then transform the sky coordinates of the targets to physical focal plane coordinates. These coordinates are then sent to the petal software to prepare for the next positioner move, which also runs the anti-collision algorithm. When the spectrograph shutters close to end the previous exposure, the OCS instructs the telescope to slew to the new field and adjusts the ADC and the hexapod and commands the positioners to execute the first or blind move. Once the telescope, dome, ADC, and hexapod are ready, we obtain a 15\,s acquisition image with the six guide GFA cameras that is then processed by the astrometry module (GFAPROC) of the PlateMaker application.  Each tile design includes pre-established guide stars that must be placed on specific pixels on the guide cameras to align the targets with fibers. The GFAPROC outputs are used to offset the telescope pointing to be exactly on the field center and to initialize the guider with the expected guide star positions in pixel coordinates for each of the GFA cameras. 

The next steps are performed concurrently: begin guiding, start observations with the four focus and alignment GFAs, and complete the first iteration of the fiber positioning loop by taking and processing an FVC image. The fiber positioning loop uses real-time feedback from an FVC image of the back-illuminated focal plane to measure the locations of the fiber positioners (see \S\ref{sec:fpi}). The PlateMaker FVCPROC module converts these measured positions in FVC pixels into focal plane coordinates, determines the corrections needed to place each fiber on target, and those correction moves are sent to the fiber positioning system to prepare and execute the second iteration (correction move) of the fiber positioning loop. The correction move begins as soon as the analysis of the FVC image taken after the blind move is complete. Upon completion of the correction move, we automatically obtain another FVC image to record the final positions of the fiber positioners and fiducials. The fiber back illumination and fiducials are then turned off while in parallel the spectrograph CCDs are prepared for the next exposure. Once the illumination is off, the spectrograph shutters open and the exposure begins. The analysis of the final fiber positions occurs in parallel with the start of the spectrograph exposure. The exposure loop for the sky monitor cameras starts when the spectrograph shutters open. Optimization of these steps continued into the start of operations, and we present the latest time estimates below in \S\ref{sec:inter}. 

The spectrograph exposure time is determined by the Dynamic ETC, which uses information from the sky monitor system and full frame images from the guide GFAs to analyze the observing conditions in real time and to calculate the accumulated effective exposure time of the spectroscopic exposures. Once the requested effective exposure time has been reached, the Dynamic ETC triggers the OCS to close the spectrograph shutters to end the exposure. In order to control the amount of cosmic ray contamination on the spectrograph CCDs, the Dynamic ETC can trigger the OCS to initiate an exposure split. This is a spectrograph readout followed by the restart of the exposure at the same location with only minimal adjustments to fiber positions to correct for field rotation and similar effects. At the end of an exposure, the OCS initiates the readout and digitization of the CCDs. While exposing, the AOS analyzes the images obtained with the focus and alignment GFAs. The focus and alignment loop takes 60\,s exposures and each frame is analyzed by the AOS to determine any alignment corrections. Focus corrections are sent to the hexapod and applied while exposures are in progress.

 \subsubsection{Next Tile Selector} \label{sec:nts}
 
 The Next Tile Selector (NTS) determines which tile to observe at a given time of night and loads the exposure request for this tile into the exposure request queue when queried by the ICS. The priorities for each night are determined in the afternoon through an automated process based on which tiles have been completed, and general considerations about the design of the survey.  The chief concerns are finding the optimal Local Sidereal Time (LST) at which to observe each tile to maximize survey efficiency, and to maximize the amount of full-depth area obtained early in the survey.  The NTS also considers other parameters, such as the positions of the moon and sun.  During the night, the ICS will request a new tile from the NTS when informed by the Dynamic ETC that the ongoing exposure is about to complete. The NTS takes into account the current observing and environmental conditions, the target LSTs of potential tiles, and their relative priorities before selecting the best available tile to be observed next.  The NTS provides an initial exposure time based on conditions and specifies the maximum exposure time for the tile. The data returned by the NTS also include the requested effective exposure time that is used by the Dynamic ETC to control the actual length of the exposure.

 \subsubsection{Focal Plane Imaging and Analysis} \label{sec:fpi}
 
Processing of the images taken for the fiber positioning loop by the FVC occurs in multiple steps. First, centroids of spots detected in the FVC image are computed by fitting a 2D Gaussian to each significant local maximum after filtering out common noise artifacts such as cosmics and hot pixels. The computations are performed in parallel on the computer that receives the images directly from the FVC. Because the aperture of the FVC lens is intentionally reduced to yield diffraction-limited spot profiles ($\sim f/22$), and because the incident flux from the illuminated fiducials and fibers is very high, the centroids are measured to $\sim 20$\,millipixel precision ($3\,\mu$m\,projected on the DESI focal surface). In practice, the precision is $\sim 6\,\mu$m\, in focal plane coordinates due to dome air turbulence.

A crucial task of the focal plane imaging and analysis is to match each fiber spot measured by the FVC with the expected location of the fiber (see next section). This is a trivial procedure when the fibers are positioned accurately and the transformation from DESI focal plane to FVC pixel coordinates is well determined, that is the closest spot to the intended position is always the correct match. However, inadvertent collisions between fibers and erratic positioner motions happen on occasion and can place the fiber well away from the expected position. In some cases, the spot can only be reached by one positioner, so a match is still straightforward. However, it is possible that two spots will appear in the overlap of the patrol regions of neighboring positioners. In this case, the two positioners, and perhaps some neighbors, are not moved until we can employ a special procedure to resolve the ambiguity. 

 \subsubsection{PlateMaker} \label{sec:pm}
 
The PlateMaker code calculates the astrometric transformations between celestial coordinates and focal plane coordinates and the corrections to fiber positioners that are necessary to align them with their targets \citep{kent16,kent18}. These two actions are organized into two operation phases. In the first phase, the PlateMaker processes images from the guide GFAs and computes the astrometric solution for the current telescope pointing. In the second phase, the PlateMaker processes the fiber spot locations measured with the FVC and determines the correction moves required to position the fibers on target. We find that only one correction move is necessary to achieve our desired positioning accuracy, although multiple iterations are possible. 

The first phase begins once the coordinates and exposure time of the next field are known. The PlateMaker determines the appropriate ADC angles based on the anticipated start time and duration of the exposure, and transforms the celestial coordinates of the targets into focal plane coordinates based on the field elevation and the distortion induced by the ADC and other optical elements. Once the positioners have moved, PlateMaker estimates the approximate pixel coordinates of the positioners and fiducials in the next FVC image in order to aid spot identification. At this point the guide GFAs take an acquisition exposure. This exposure is processed by a star-detection algorithm that is fast and robust against variable dark current and the occasional pattern noise that is present in the GFA images. The stars are matched with an astrometric catalog \citep[currently Gaia DR2,][]{gaia16,gaia18} to determine the astrometric solution. This solution treats the guide GFAs as a single, rigid focal plane with relative physical positions determined from a set of exposures taken close to zenith. This astrometric solution from the GFAs is tied to the focal plane with the special guide illuminated fiducials (GIFs) that are mounted on the GFA assembly blocks and have the most accurate laboratory metrology with respect to the GFA CCD pixels. This astrometric solution is used to offset the telescope to center on the target field as well as to identified pixel locations on the guide GFAs for the set of pre-determined guide stars specified in the exposure request. Because the fiber assignment code assumes the focal plane maintains a fixed position angle in celestial coordinates, any field rotation determined by the astrometric solution is compensated for by a rotation of the hexapod. In order to minimize residual field rotation effects, we update the hexapod rotation once per minute during each exposure.  

In the second phase, we turn on the fiducials and back illumination system and take an FVC image. We use the predicted pixel coordinates for the fiducials and fibers to aid in matching the spots in the FVC image to the positioners and fiducials. The singlet lens of the FVC can be treated as a pinhole camera such that the sky is assumed to map linearly onto the FVC CCD with a tangent plane projection. We calculate the astrometric coordinates of the GIFs and use their spot locations to astrometrically calibrate the FVC CCD.  We then use the tangent plane projection and corresponding FVC coordinates of each fiber target to compare the desired and actual coordinates of the positioners and calculate correction moves. The fiducials are used to determine the transformation from the FVC pixels coordinates to the focal plane physical coordinates. This transformation relies on knowledge of the corrector distortion independent of any distortion from the FVC lens; the latter was intentionally minimized in the FVC lens design. The distortion model includes both radial and non-axi-symmetric terms, the latter changing as a function of ADC lens angles. During commissioning, we used special dither sequences (see \S\ref{sec:dither}) to quantify the accuracy of the coordinate transformation from FVC pixels to celestial coordinates. We also identified small, static patterns in the residuals that are applied as corrections to observing sequences.

Two iterations of the fiber positioning loop are sufficient to meet the required fiber positioning accuracy of better than $10\,\mu$m (or 0.15$''$) positioning accuracy in focal plane coordinates. The largest residual contribution to the positioning error comes from dome seeing and turbulence effects. We mitigate this effect by using the known and measured positions of all stationary devices in the focal plane (fiducials, non-functional fiber positioners) to fit large scale turbulence modes in every FVC image and to derive corrections at each fiber position. Under good observing conditions this approach has yielded fiber positioning accuracies as good as $2.7\,\mu$m rms between the coordinates determined by PlateMaker and the fiber position measured by the FVC after the correction move.

 \subsubsection{Guider}
 
The DESI guider system provides correction signals to the Mayall TCS to maintain a stable and accurate position during science exposures. Based on the acquisition image taken with the six guide GFAs at the start of every DESI exposure, the PlateMaker application finds the astrometric solution for the current pointing and determines the location of guide stars taken from the GAIA catalog on each of the guide GFAs. For each GFA, we select two guide stars based on SNR and isolation criteria and load their locations in GFA CCD pixel coordinates to the guider software. Once initialized, the ICS starts the guider loop and 5\,s GFA exposures are taken in a $\approx9$\,s cadence. For each guide star and every GFA frame, $50 \times 50$ pixel regions of interest are extracted from the full frame images. The Guider application receives these postage stamps and uses star finding and centroiding algorithms to measure the star locations relative to the pre-assigned positions. ICS combines the results of these algorithms for all guide stars to compute a correction signal that is sent to the TCS. The TCS employs a filtering algorithm to process these correction signals.

The DESI guider algorithm is robust against small disturbances, low SNR stars, and occasional lost signals, for example due to passing clouds. Studies during commissioning showed that a relatively long exposure time of 5\,s is needed to obtain guide star images of sufficient SNR when observing conditions are poor. While a higher cadence is preferable, we determined that this rate is sufficient to maintain the required level of telescope tracking accuracy.

For testing and performance evaluation, the DESI guider application supports a special mode that automatically locates suitable guide stars in the full frame GFA images and then reconfigures the software to cut out the same $50 \times 50$ pixel regions of interest around these guide stars. When operated in this mode, no star catalogs are required and the guider runs independently of other ICS components such as PlateMaker. Early guiding and tracking studies during DESI commissioning relied heavily on this mode.

 \subsubsection{Active Optics System} \label{sec:aos}
 
 The DESI Active Optics System (AOS) maintains the alignment of the prime focus instrumentation with the primary mirror through corrections applied with the hexapod. The AOS software calculates the alignment corrections with intra- and extra-focal images obtained with the four GFA cameras that are configured for wavefront sensing, as described in \S\ref{sec:gfas}. We fit the out-of-focus stars with a pupil plane Zernike expansion and use an optical model of the corrector to calculate hexapod corrections for five degrees of freedom: tip and tilt of the focal surface, x and y decenter of the focal surface, and defocus. The AOS does not correct for rotation. We use a separate, empirical model to correct for field rotation during exposures.
 
 The AOS is an adaptation of the similar system for DECam that was developed by \citet{roodman14}. The key difference is that the DESI GFA wavefront sensors are not behind the shutter used for science exposures, and therefore may operate independent of the science exposures. The advantage of this design is that we may use the AOS to check the alignment prior to the start of each science exposure. In practice, we have found that a static look-up table provides sufficiently good values for the correction to adjust the hexapod decenter, tip and tilt settings as a function of telescope position. We use the results of the wavefront calculation to set the focus and to compensate for temperature effects. The AOS system is in routine operation and works reliably. 
 
 \subsubsection{Dynamic Exposure Time Calculator} \label{sec:etc}
 
 The purpose of the Dynamic Exposure Time Calculator (ETC) is to achieve uniform depth in all spectroscopic exposures and therefore uniform redshift success rates. The Dynamic ETC achieves this with real-time estimates of the accumulated effective exposure time of each science exposure while it is in progress and provides feedback via the ICS to adjust the length of that exposure if observing conditions change. The inputs to the Dynamic ETC calculation are the images from the sky monitor cameras and the same images from the guider GFAs that are used for guiding. Both sky monitor and guider GFA frames are collected continuously during each spectrograph exposure. The Dynamic ETC uses these image data to determine the seeing, transparency, and the night sky continuum brightness. This summary information is also provided to the observers and archived in the telemetry database. 
 
 The Dynamic ETC algorithm uses the full-frame guider GFA images to measure both the seeing from stellar profiles and the transparency with stellar photometry relative to their known brightness from Legacy Survey photometry. We calibrated the photometric performance of the GFAs with observations obtained under the best conditions, and regularly verify the stability of this calibration. The guide exposures have a typical duration of 5\,s and there are a little over 3\,s between exposures. The noise in the spectrograph data is dominated by Poisson fluctuations in the sky level. The sky monitor (\S\ref{sec:skymon}) measures the sky continuum flux in real time through dedicated broad-band filters. We have used numerous spectroscopic observations from fibers that observed ``blank sky'' under a wide range of conditions to calibrate the relationship between the continuum observed by the sky monitor and the sky recorded by the spectrographs. This calibration includes smaller SNR contributions due to finite galaxy sizes and the spectrograph CCD read noise. The sky monitor measures the sky flux every 60\,s, as this time is needed to obtain a sufficient detection of the sky continuum brightness under the darkest conditions (new moon with the telescope pointed at the zenith).
 
 The Dynamic ETC combines these data to determine if sufficient effective exposure time has been accumulated to end the exposure. In some cases the observing conditions require exposure times that would accumulate too many cosmic rays or incur a loss of throughput due to changes in atmospheric refraction with airmass. In these cases, the Dynamic ETC sends a trigger to the ICS to read out the spectrographs and start a new exposure at the same location. The only overhead associated with this split is the 60\,s spectrograph readout time and about 4\,s to prepare the CCDs and the spectrograph controllers. The Dynamic ETC can also automatically stop an exposure based on twilight or airmass limits, or when conditions deteriorate sufficiently that the accumulated SNR starts to decrease. Kirkby et al. (2022), {\it in preparation} will provide a more complete description of the Dynamic ETC.

\subsection{Integration, Testing and Performance}

The ICS is connected to every component of the instrument as well as the Mayall telescope. In order to manage the large number of interfaces, we developed interface control documents (ICDs) that detail the hardware connection and define deliverables and responsibilities. These documents were reviewed by the project teams and were placed under change control. While the hardware components were under development, we created software simulators for each interface to allow early testing and validation. We adopted a spiral development approach with increasingly complex systems and included a number of functional verification tests throughout the construction project. Some of these tests were coordinated with the project to support other activities, for example spectrograph testing and evaluation. The four key DOS system tests were: ICS--Mayall TCS integration; single spectrograph support; ProtoDESI described in  \S\ref{sec:protodesi}; and a full system test using the emulated hardware components. 

The ICS has been in routine operation since the start of the DESI pre-commissioning activities described in \S\ref{sec:precmx}. During DESI commissioning, a steady stream of improvements, bug fixes and feature enhancements were applied to the ICS. The requirements to provide telemetry storage, data mining, and alarm and error handling infrastructure were demonstrated before the start of commissioning. Since the start of DESI operations, the robustness and ICS availability have surpassed the required uptime of 97\% and the typical inter-exposure times are close to the two minute design goal.

\section{Data Systems} \label{sec:datasystems}

Data systems provides the software and infrastructure to plan the DESI survey, code to implement that survey plan within the ICS environment, and software to process the resulting data into calibrated spectra and redshifts for analysis by the DESI collaboration.  Its key components are: 1) Target Selection; 2) Survey Planning; 3) Fiber Assignment; 4) Data Transfer; 5) Spectroscopic Pipeline; 6) Nightwatch. Target Selection uses data from the Legacy Imaging Surveys \citep{dey19}, WISE \citep{wright10}, and Gaia \citep{gaia16,gaia18} to select which astronomical objects should be observed by DESI. Survey planning selects a strategy for observing a tiling pattern of overlapping telescope pointings that can be completed within the 5-year DESI program while achieving the required density of target observations. Fiber Assignment selects which targets should be assigned to which fibers for each telescope pointing. Data Transfer copies raw data from KPNO to the National Energy Sciences Computing Center (NERSC)\footnote{\url{https://nersc.gov}} in near-realtime for processing and makes backups to tape at NERSC and a geographically separated mirror at NOIRLab\footnote{\url{https://noirlab.edu}}. The Spectroscopic Pipeline processes those raw data into sky-subtracted flux-calibrated spectra, classifications, and redshifts. Lastly, Nightwatch provides a simplified and faster processing of the data, capable of running on a single 24-core computer at KPNO, in order to provide data quality feedback to observers within minutes of each exposure. The next subsections briefly describe each of these components. Further details will be provided in future papers on the Target Selection Pipeline (Myers et al.\ 2022), Survey Operations and Simulations (Schlafly et al.\ 2022), Fiber Assignment (Raichoor et al.\ 2022), the Spectroscopic Pipeline, Data Transfer, and Nightwatch (Guy et al\ 2022), and Spectral Classification and Redshift Fitting (Bailey et al.\ 2022). 

\subsection{Technical Requirements}

Data systems developed and implemented the target selection, survey plan,  fiber assignment, and spectroscopic data processing capable of producing a dataset meeting the science requirements in Tables~\ref{tab:level12} and~\ref{tab:level3} given the raw data from the DESI instrument. The derived technical requirements for data systems include the assignment of 10 standard stars and 40 sky fibers per petal for calibration, the need to transfer and backup data within 12 hours of sunrise, and to provide fully processed data to the collaboration within 48 hours of observation. The detailed performance of data systems relative to the requirements will be provided in the papers cited above.

\subsection{Major Components}

\subsubsection{Target Selection}

Algorithms to select science targets from imaging catalogs were developed by the Target Selection Working group of the DESI science collaboration. Data systems provided a parallelized implementation of those selections, including Quality Assurance (QA) and infrastructure to track input provenance and ensure repeatability. Target Selection also provided selection algorithms for: standard stars for flux calibration; identification of blank sky locations for modeling and subtracting the sky; GFA targets for guiding, focus, and dynamic exposure time decisions; and coordination of targets for commissioning, survey validation, and secondary programs. Most selection algorithms are simple cuts in observed parameters (e.g.\ colors of objects), though QSO selection used a random forest algorithm. Targeting bit masks track which objects pass which selection criteria, thus providing the flexibility for the same object to pass multiple selections.

\subsubsection{Survey Plan}

The survey plan selected a 14,000 deg$^2$ footprint observable from the Mayall Telescope and designed a tiling pattern of 15.6k overlapping tiles with an average of 5.2 overlapping visits per location on the sky for the dark program, and 3.0 for the bright program. In practice, these values are somewhat lower by an amount that scales with the percentage of non-functional positioners. The survey strategy prioritizes what tiles should be observed when given the current observing conditions, tile airmass, galactic dust extinction, and the distribution of remaining observable hours vs.\ LST. This survey strategy is algorithmically encoded in the Next Tile Selector described in section \ref{sec:nts}.  A survey simulation using 10\,years of Mayall weather history and a model of instrument overheads and downtime tested the survey plan to ensure that it is viable to complete the full footprint within a 5\,year survey.

\subsubsection{Fiber Assignment} \label{sec:fiberassign}

\begin{figure*}[!hbt]
\centering
\includegraphics[width=0.9\textwidth]{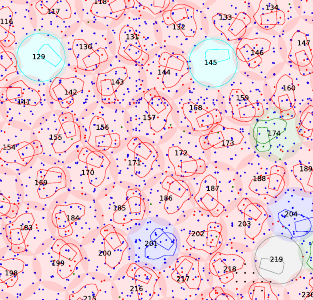}
\caption{A view of $\sim$1\% of the DESI focal surface for a particular set of fiber assignments. The coverage regions of each positioner are shown in overlapping circles with red for those assigned to science targets, green for standard stars, blue for sky locations, cyan for disabled positioners serendipitously pointing at a blank sky location, and gray for disabled positioners not assigned to anything. The irregular polygons show the keepout regions for the individual positioners. The smaller polygon is the outer $\phi$ arm, and the larger polygon is the inner $\theta$ arm that spins about the central axis of the positioner. Targets that are reachable by non-disabled positioners are shown in small dots with science targets in red, sky locations in blue, and standard stars in green. Location 160 in the upper right shows a case where it was necessary to avoid a collision with a neighboring positioner (174) even though the assigned targets are well away from each other.} \label{fig:fiberassign}
\end{figure*}

Fiber Assignment selects which targets should be assigned to which fibers for each tile, while also assigning fibers to sky locations and standard stars for calibrations. Lower density targets such as QSOs are given higher priority and assigned first, followed by higher-density lower-priority targets such as LRGs and ELGs.  Standard stars and sky locations are initially assigned to positioners that cannot reach any science targets. Any disabled positioners that are serendipitously pointing at blank sky locations are also used as sky calibration fibers. If needed, low priority science targets are replaced with standard star and sky assignments if required to achieve the minimum calibration target density (10 standard stars and 40 sky fibers per petal). Sky fibers are also distributed such that groups of 50 fibers on the focal plane will have at least one sky assignment to ensure uniform coverage over both the field of view and across the spectrograph CCDs. Figure~\ref{fig:fiberassign} shows an illustration of fiber assignment for a very small subset of the focal surface. 

Since neighboring positioners have overlapping reach, fiber assignment also checks for and excludes target positions in which the mechanical components of neighbors would interfere. Since pre-planned fiber assignments cannot perfectly know the exact transforms between sky coordinates and focal plane coordinates (e.g., due to the exact airmass of observation), fiber assignment includes a buffer zone of $200\,\mu$m in total reach and an extra $50\,\mu$m around each positioner (in addition to the anti-collision safety buffer zone used in operations) so that small variations between the expected and actual transforms do not introduce new collisions. Although fiber assignment does not purposefully make assignments that would collide, this is additionally checked by the online code using the final as-observed conditions and colliding assignment requests are not fulfilled.

\subsubsection{Data Transfer, Archive, and Distribution}

Data are transferred from KPNO to the National Energy Research Scientific Computing Center (NERSC)\footnote{\url{https://nersc.gov}} every 10 minutes using rsync. A long running process on a data transfer node at NERSC initiates the transfer. This design ensures that data are transferred frequently, while also being robust to individual transfers taking longer than the cadence for checking for new data. A cronjob checks every minute and restarts the process if needed, providing rapid recovery robustness to unexpected events such as a reboot of the transfer node. Once a day, the previous night's raw data are archived to tape storage at NERSC (also leaving a copy on disk), and copies of the raw and processed data are synched to NOIRlab in Tucson to provide a geographically separated backup. Although other technologies such as Globus\footnote{\url{https://globus.org}} could have provided better throughput, they also come with additional configuration and maintenance burdens and the simple rsync method is sufficient to keep up with the data in realtime to meet the requirements.

\subsubsection{Spectroscopic Pipeline} \label{sec:specpipe}

The Spectroscopic Pipeline converts the raw data into sky-subtracted flux-calibrated spectra, classifications (star, galaxy, QSO), and redshifts. Similar to the raw data transfers, a long running process monitors the NERSC disk for the arrival of new data and launches jobs into a realtime queue for immediate processing \citep{kremin20}.  This job launcher process is purposefully separated from data transfer so that either can be paused or restarted as needed without requiring direct coordination.  For example, if the compute queue is down for maintenance, the data transfers can still continue and the processing jobs will catch up when they can.

The spectroscopic pipeline uses the ``spectroperfectionism'' formalism from \citet{bolton10} to extract spectra using a 2D model of the spectrograph PSF as measured from arc lamp spectra each afternoon.  This method provides spectra with uncorrelated flux bins on a uniform wavelength grid plus a model of the non-Gaussian line spead function for each fiber at each wavelength (the ``Resolution Matrix'').  These components are used throughout the data processing steps, for example to model how an input sky spectrum appears slightly differently on each of the different fibers with their unique resolutions.

The DESI spectroscopic pipeline keeps up in real time with DESI exposures with 10 nodes (32 Intel Haswell cores/node) of the NERSC realtime queue such that it delivers fully calibrated spectra and redshifts each morning after observations. These are used to make decisions on the quality of the data for individual targets and these decisions inform the observing plan and fiber assignments for the following night. The spectroscopic pipeline meets all science requirements in Tables~\ref{tab:level12} and~\ref{tab:level3}.  Further performance details will be provided in Guy et al.\ (2022), {\it in preparation}.

\subsubsection{Nightwatch}

Nightwatch provides a simpler, faster version of the spectroscopic pipeline for the purposes of semi-realtime data QA.  It uses row-by-row boxcar spectral extractions instead of the full 2D PSF modeling of spectroperfectionism; it does not use the per-fiber resolution information when performing sky subtraction; it only applies an average flux calibration corrected for airmass and galactic dust extinction instead of performing a per-exposure fit to standard stars; and it does not fit classifications and redshifts.
 
A summary webpage per-exposure shows the CCD readnoise per amplifier (30 cameras $\times$ 4 amplifiers/camera) and their expected ranges; the integrated sky-subtracted flux per fiber in each of the 30 cameras; and five randomly selected spectra. This page alone is sufficient to identify many problems, e.g.\ with CCD performance or configuration, stuck spectrograph shutters, or light contamination in the dome or spectrograph room. Additional pages provide details on per-CCD amplifier QA metrics (overscan levels, cosmic ray rate); raw data CCD images; offsets of the spectral traces compared to expected locations; integrated fluxes, signal-to-noise, and positioning accuracy vs.~fiber number and vs.~(x,y) location on the focal plane; guiding offsets vs.~time during the exposure; and the ability to view individual spectra.
 
One copy of Nightwatch runs on a dedicated $24$-core machine at KPNO, providing results within about three minutes of the end of the spectrograph readout.  This copy provides the fastest turnaround, but is only available to observers at KPNO or with VPN access, so a second copy runs at NERSC.  This has additional data transfer delays of up to 10 minutes, but provides a semi-realtime view of the data QA to any collaborator without requiring VPN access to the KPNO network.

\subsection{Parallelism and Code Design}

The data systems code was designed from the beginning to run in parallel at a high performance computing (HPC) center such as NERSC.  This ability has been critical for rapid turnaround of nightly data for survey operations and timely processing of internal data assemblies that process all data using a tagged software release.  DESI uses Message Passing Interface (MPI) as its primary mode of parallelism due to its performance and ubiquity at HPC centers; some algorithmic steps also use Python multiprocessing.  The largest production runs so far have used 40800 Intel KNL cores simultaneously.  Work is in progress to adapt the slowest pieces of the DESI pipeline to GPUs in preparation for the next-generation Perlmutter machine at NERSC, which will have dramatic performance improvements \citep{margala21}.

While designing for HPC deployment from the start, data systems also made some pragmatic design choices to maximize the efficiency of human contributors over raw computing performance.  Most DESI code is written in Python, leveraging compiled code ``under the hood'' via numpy\footnote{\url{https://numpy.org}}, scipy\footnote{\url{https://scipy.org}}, and numba\footnote{\url{https://numba.pydata.org}}. All algorithmic steps can be run on a Linux or Mac laptop using files for input and output, without requiring MPI, centralized databases, GPUs, or other HPC or pipeline infrastructure.  This use of Python and separation of algorithms from pipelining and parallelism has enabled non-computing-expert collaborators to make significant contributions, and greatly facilitated the speed with which new algorithm ideas can be implemented and tested.

\subsection{Testing and Integration}

All data systems code repositories are open source at \url{https://github.com/desihub}, with significant contributions from both DESI project staff and from members of the science collaboration.  Developers make branches of the code, implement their features, push the branch to GitHub and open a pull request for review before integration into the main branch.  Each repository includes unit tests that run automatically when pull requests are opened.

Additionally, each night a cronjob updates the main copy of each repository at NERSC and re-performs the unit tests to ensure cross-repository consistency.  This cronjob also performs an integration test by simulating a small amount of pixel-level raw data, processes it with the spectroscopic pipeline as if it were real data, and verifies the resulting redshifts.  Any integration test failures generate email alerts.

Quarterly software releases of version-tagged repositories include additional testing on real data and end-to-end tests of target selection, fiber assignment, survey operations (via simulation), and the spectroscopic pipeline.  These software releases are used for interim internal data releases, processing all available data to date.

During the DESI construction project, extensive simulations guided the code development.  These included pixel-level simulations of raw data, higher level (faster but less detailed) simulations of raw extracted spectra (including sky and not flux calibrated), and rapid simulations of fully sky-subtracted flux-calibrated spectra with added noise based upon the instrument model.  Any of these levels of simulation could be inserted into the spectroscopic pipeline for processing as if it was real data.

This extensive simulation and testing program enabled the code to work well from the very beginning of operations. Although we continue to make improvements, the core pipeline met the science requirements and scaling performance needs from the start of commissioning.

\section{Mayall Upgrades} \label{sec:mayall}

We performed numerous upgrades to the Mayall telescope infrastructure to support the installation and operation of DESI. The following subsections summarize the technical requirements on the upgrades, improvements to the telescope control system, infrastructure upgrades, and other facility improvements. 

 \subsection{Technical Requirements}
 
The key technical requirement on the telescope upgrades was that they do not degrade the delivered image quality relative to the baseline adopted to plan the DESI survey. That baseline was obtained from the study by \citet{dey14}, who measured the delivered image quality for the Mayall from thousands of images obtained with the MOSAIC prime focus imager between 2006 and 2011. Those data cover a range of wavelength, airmass, temperature differences between the mirror and ambient environment, and exposure time. \citet{dey14} report values of $1.2 - 1.3''$ for the median of the nightly median values in images obtained in the reddest filters ($i, I, z$), and that the median of the best values from each night range from $0.93 - 1''$. 

There were also requirements on improvements to electrical, cooling, and communications infrastructure, as well as facility improvements to support installation. The electrical upgrades included additional capacity in the prime focus cage for new electronics, in the computer room for additional computing and storage capacity, in the Large Coud\'e Room for the thermal enclosure and spectrographs, at the top end to support the calibration lamps, and in the facility uninterruptible power supply (UPS). Cooling capacity was added to both the Large Coud\'e Room and the top end for the spectrographs and focal plane system, respectively. The TCS upgrades included the capability to interface with the DESI ICS. Other facility improvements and additions included a new, larger control room on the ``utility'' or U floor. 

 \subsection{Telescope Control System} \label{sec:tcs}
 
The Mayall TCS was modernized as part of the preparations for DESI installation \citep{abareshi16, sprayberry16}. Key motivations for this project were to minimize the time between successive exposures and to guide accurately during the $\sim 20$ minutes planned for the longest science exposures. The guiding enhancements included an interface between the DESI guider system and the TCS. 

The modernization project drew substantially from the TCS modernization project at the 4\,m Blanco Telescope that prepared for the Dark Energy Survey \citep{warner12}. The main changes were the replacement of the old VxWorks TCS computer, new tape encoders on the hour angle and declination axes, and new digital servo controllers for the telescope drives. We also replaced the old 2D polynomial pointing model with the commercial TPOINT software. These updates occurred during 2014, and regular operations shifted over to the new TCS at the start of the 2015B semester. 

Pointing map measurements with the new TCS showed rms errors of $2 - 3''$ at that time, which is substantially better than the $10'' - 20''$ errors typical of the previous pointing map and the old, mechanical encoders. The unguided or open-loop tracking errors after the upgrades also exhibited substantial improvement. The best model yielded equivalent open-loop tracking errors of $1.6''$/hr in RA and $0.1''$/hr in DEC, far better than the $30''$/hr performance typical of the previous polynomial pointing model with the old mechanical encoders and analog servo controllers.

We evaluated the impact of the TCS upgrade on the image quality with 329 nights of data obtained between December 2015 and February 2018 for the Mayall z-band Legacy Survey \citep{dey19}, which was the largest program on the Mayall telescope since the \citet{dey14} study. The seeing in these images was characterized in a similar manner as the \citet{dey14} study, namely we first calculated the median seeing and the best seeing for each night, and then calculated the median of the median values and the median of the best values. The median of the median values of each night was $1.03''$, and the median of the best values per night was $0.84''$. Both metrics indicate that the upgrades prior to installation somewhat improved the delivered image quality. 

 \subsection{Facility Improvements}
 
 Some significant improvements for DESI included increases in the capacity of the building’s electrical wiring and battery backup, cooling system enhancements, and bandwidth upgrades to the building’s ethernet system \citep{sprayberry14}. Updates to the facility electrical system and battery backup capability were needed to support the greater power draw from DESI components, and especially to maintain and protect them against interruptions that would impact survey efficiency. The updates to the cooling system were needed to support the stable thermal environment for the spectrographs, the linear pulse tubes for the spectrograph cryostats, and the additional processing and data storage in the computer room. The DESI focal plane system has a separate cooling system described in \S\ref{sec:fpe}.
 
 One major change was the surveying and clearing of the Large Coud\'e Room. This room is located on the ``main'' or M floor, the bottom floor of the dome enclosure of the telescope. The Large Coud\'e Room was originally intended for a large spectrograph, and was used for a number of instrument setups, although over the years the room also became a convenient storage area. This room was carefully surveyed in 2014 to support the design of the thermal enclosure or Shack for the ten spectrographs (see \S\ref{sec:shack}), and then cleared of its venerable equipment collection and the heavy fixtures intended to support the original large Coud\'e spectrograph in order to prepare for the assembly of the Shack. 
 
 The DESI project also needed infrastructure improvements to the computer room, which is located on the Mezzanine level just above the M floor.  Mayall staff cleared a number of obsolescent computer racks and tape storage files out of this room and increased the cooling capacity of this climate-controlled room to support the expanded computer processing and storage requirements of the DESI project. Much of this work was completed by 2016 to support the early integration efforts of the ICS with the TCS. 
 
 The DESI installation plan included reassembly and testing of the new optical corrector on site prior to integration with the new top ring and subsequent integration with the telescope. The reassembly and testing of the corrector occurred on the ground floor of the telescope building. This area has access to the exterior via a large garage door, high ceilings, and is connected to the telescope level with a pair of cranes (5-ton and 50-ton capacity) that may extend through a hatch in the M floor. Mayall staff cleared the ground floor area to support the delivery of the many large crates that contained the corrector components, as well as numerous other components that moved through this space. They also erected a large clean tent (4\,m $\times$ 4\,m footprint, and high enough to support a crane with a 6\,m hook height) to support the re-assembly and testing of the corrector. 
 
 The operations model for DESI is to have two personnel on site every night: a highly trained expert observer (called the Lead Observer) and the staff Observing Assistant (OA), who is responsible for telescope operations and personnel safety. Observers drawn from the collaboration membership also connect remotely every night and assist with observations (Observing Scientist) and data quality assurance (Data Quality Scientist). Pre-COVID, the operations model had been for the Observing Scientist and Data Quality Scientist to be on site as well. 
 
 Many more people were expected to be on site during the installation, functional verification, and commissioning periods.  As the original control room on the “console” or C floor could not accommodate more than five people, the Observatory built a new, larger control room on the “utility” or U floor, which is several floors below the M floor. The new control room accommodates up to 15 people, is more easily and safely accessible than the original control room, and reduces the thermal load on the dome environment.

\section{Pre-Commissioning Activities} \label{sec:precmx}

 The development of DESI included two technology demonstration and risk reduction activities with on-sky observations prior to the start of commissioning of the complete instrument. The first of these employed an instrument called ProtoDESI whose goal was to reduce technical risks associated with the alignment and stability of the robotic fiber positioners. ProtoDESI was used for observations from August 14 to September 2016. The first subsection provides a brief description of ProtoDESI, with an emphasis on how it influenced subsequent development of the project and planning for commissioning. ProtoDESI is described in detail in \citet{fagrelius18a}. The second subsection describes the Commissioning Instrument (CI). We developed the CI to take advantage of a schedule gap between the installation of the corrector system and the start of the installation of the FPA so that we could begin commissioning tasks with the new corrector system. The CI was used for observations in April and May of 2019. \citet{ross18} describe the CI, and \citet{coles18} describe the critical metrology program for the CI. The description below emphasizes the completion of the CI, the observing campaign, and how this work influenced commissioning of the complete instrument.  

 \subsection{ProtoDESI} \label{sec:protodesi}
 
 The main purpose of ProtoDESI was to test the fiber positioners, including both the  physical mechanisms and the acquisition procedure. It was also useful to test a GFA prototype and improve the guiding algorithm. The instrument included three fiber positioners, 16 fiducials, one GFA camera, a back-illumination system for the fibers, and a backup finder scope camera in the event of difficulties with the fiber alignment. The three fiber positioners each had a 3\,m length of fiber that was reimaged on an SBIG STF-8300M camera. ProtoDESI was mounted to the $f/3.2$ refractive Mosaic corrector at the prime focus. We operated ProtoDESI with an early version of the FVC and the ICS and therefore we were able to test key software components several years in advance of commissioning. 
 
 We used ProtoDESI to test the fiber positioner technology in conjunction with an early test of the fiber positioning software PlateMaker, which calculates the coordinate transforms necessary to place fibers on targets (see \S\ref{sec:pm}). The alignment of fibers with targets required that ICS coordinate the activation of the back illumination system with exposures from the FVC, use the measurements from the FVC to determine the projected location of the fiber tips on the focal plane, and combine these data with astrometry from the GFA to project the fiber tip locations into sky coordinates. The observations with ProtoDESI were extremely valuable to develop and test all of these software tools, as well as to develop tools and procedures for the eventual calibration and commissioning of the complete FPA. In addition to fiber positioning, we used ProtoDESI to characterize the GFA detectors, such as to demonstrate that the GFA CCD met the quantum efficiency requirement. We also used these observations to develop more optimal GFA exposure times to minimize errors due to atmospheric scintillation, decrease heating of the GFA unit, and improve the guiding algorithms. 
 
 The ProtoDESI experience was invaluable to help plan DESI commissioning. One impact was that we more thoroughly defined the purpose, conditions, data requirements, analysis plan, criteria for success, time estimate, and one or more responsible parties for each commissioning task. Another was that we identified a number of new tasks. These included the development of robust tools for automatic focus sequences, use of the GFAs to refine fiducial positions, telescope tracking and guiding tests, and positioner dither tests. We also identified multiple pre-commissioning tasks that better prepared us to analyze data once we went on sky. These tasks included a more complete development of data models and interfaces, a more thorough plan for the metrology of the FPA, enhanced lab testing to minimize the number of systems that would be tested on sky for the first time, and motivated us to substantially expand the core group of experts who would contribute to commissioning. 
 
 \subsection{Commissioning Instrument} \label{sec:ci}
 
 \begin{figure*}[t]
 \includegraphics[width=7in,angle=0]{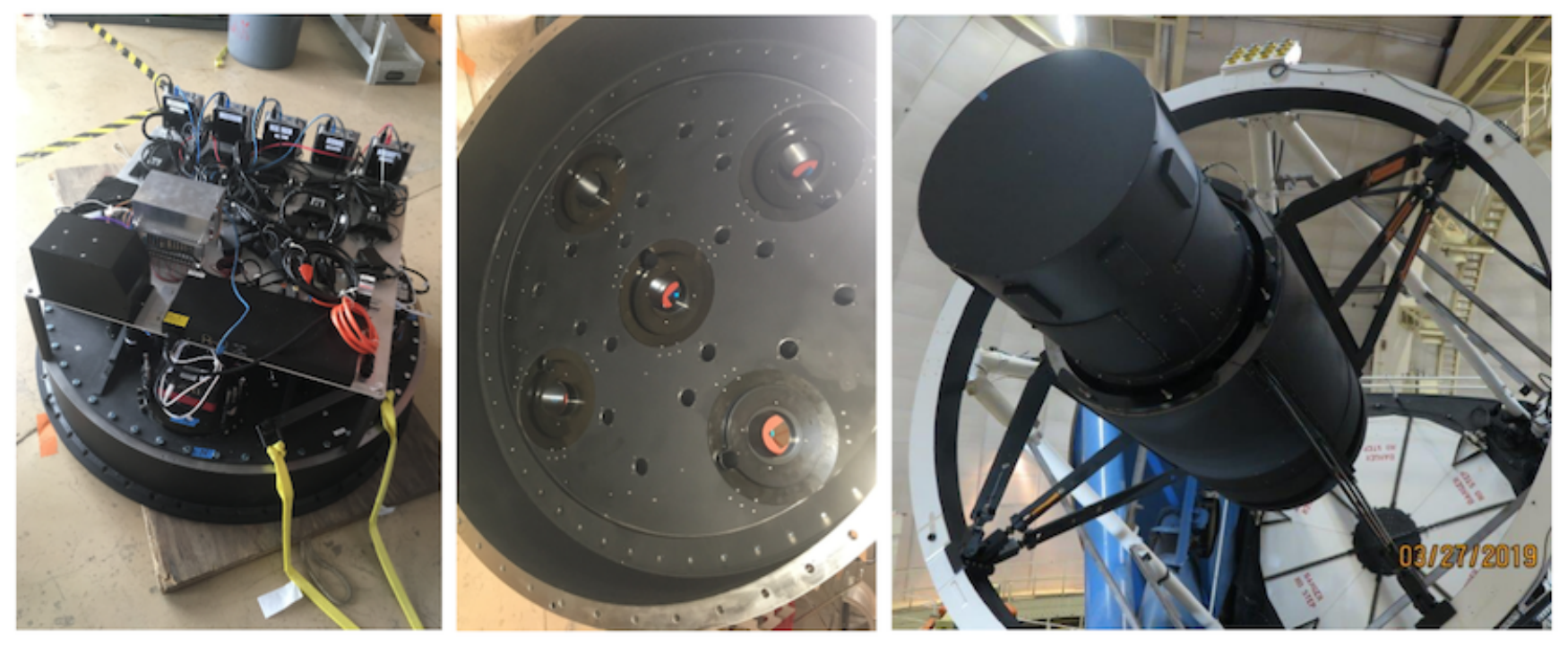}
 \caption{Three views of parts of the Commissioning Instrument (CI). ({\it Left}) Top view of the CI focal plate assembly. This side shows most of the electronics and faces away from the corrector . The electronics are mounted on a platform above the cameras and illuminated fiducials. ({\it Middle}) View of the CI that faces the corrector. The five larger apertures are for the five SBIG cameras, with one at the center of the field of view and the others near the periphery in the four cardinal directions. There are also 22 smaller apertures, each of which houses one of the illuminated fiducials. ({\it Right}) View of the CI after installation on the new corrector in late March 2019. \label{fig:ci}}
\end{figure*}

 Development of the Commissioning Instrument (CI) began in 2017 in order to take advantage of a forecast two-month schedule gap between the installation of the new corrector system and the readiness of the FPA for installation. The CI was designed to realize an early start on commissioning tasks related to the telescope, the performance of the corrector and corrector system, guiding algorithms, the active optics system, the FVC, and the ICS. We determined that tasks from these categories could readily fill the expected two-month period with a fairly simple arrangement of imaging cameras combined with illuminated fiducials. For this reason, and due to the earlier success of ProtoDESI, we decided that the CI would not include fiber positioners, nor a fiber feed to a spectrograph. To obtain good value from the CI, we used the GFA requirements to establish similar requirements on the location, tilt, FOV, noise, pixel size, and readout time on the imaging cameras that would be part of the CI. 

 The main components of the CI were five commercial SBIG STXL-6303 cameras and 22 illuminated fiducials. These SBIG cameras were chosen because their CCD array and pixel size ($3072 \times 2048$, $9\,\mu$m pixels) produced a 28 square arcminute FOV that was comparable to the 25 square arcminute FOV of the GFAs, and with somewhat higher sampling ($0.13''$/pixel vs.\ $0.2''$/pixel) and better noise properties, although somewhat lower efficiency. The fiducials were identical to those produced by Yale \citep{baltay19} for the FPA. The cameras and fiducials were supported by a short, steel cylinder called the CI focal plate assembly. Four of the five commercial cameras were located at the same off-axis distance as the GFAs on the focal plane system, while the fifth provided unique, on-axis imaging through the corrector. Behind the cameras was an electronics rack that contained a dedicated NUC PC for each of the five SBIG cameras, a simplified DESI petal controller electronics box supplied by the University of Michigan to operate the illuminated fiducials, temperature sensors, an ethernet switch, and a Raritan PDU for remote power control. These electronics are shown in the left panel of Figure~\ref{fig:ci}. Each camera had an SDSS $r-$band filter mounted immediately in front of its shutter to mimic the bandpass of the GFA cameras. Most of the cameras and their filters are visible in the middle panel of Figure~\ref{fig:ci}. One of the requirements of the CI focal plate assembly was that it would match the 815\,kg mass and moment expected for the FPA in order to assess the expected pointing and tracking performance of the telescope with the FPA. We met this requirement by surrounding the CI focal plate assembly with a larger, more massive steel cylinder. An additional 221\,kg of counterweights were attached to the corrector  cage to mimic the mass of the focal plane enclosure. 
 
 The greatest challenges in the CI development were the strict alignment and metrology requirements. The five cameras and 22 fiducials needed to be aligned to the aspheric focal surface. The alignment requirements were that the full area of each CCD and each fiducial had to be within $50\,\mu$m of the focal surface, and that each CCD had a tilt of no more than $0.05^\circ$ relative to the best-fit local tangent. The metrology requirements were to measure the x,y positions of each fiducial with $\pm10\,\mu$m accuracy, and the orientation of each CCD and the nearest fiducial with $\pm5\,\mu$m accuracy. We met these requirements with a careful metrology program that was based on a combination of CMM touch probe metrology and a custom microscope with an integrated projection system \citep{coles18}.

 The two-month CI campaign was very successful because it retired multiple risks and identified many areas for further work with sufficient time to address them (approximately six months) before the start of commissioning. One major accomplishment on the first night was the demonstration that the new corrector system could produce subarcsecond images across the entire FOV, including at the center where the focal plane system does not have an imaging capability (see Figures~\ref{fig:firstlight} and \ref{fig:corrpsf}). One important discovery was that the historical value for the radius of curvature of the primary mirror was incorrect by 18\,mm, which corresponds to a shift in the position of best focus by 9\,mm. Fortunately the hexapod had a focus range of $\pm 10$\,mm and was able to compensate for the difference, although in the midst of the CI campaign we used this measurement to adjust the position of the cage to place the center of the hexapod range at the best focus. Another measurement with the CI was the amount of field rotation during exposures as a function of position on the sky due to a combination of polar axis misalignment and variations in atmospheric dispersion. We used these data to develop a model for field rotation that we subsequently refined during commissioning of the full focal plane system. We also made substantial progress on many commissioning tasks, including a new pointing model for the telescope, an initial assessment of the throughput of the corrector system, further development of guiding algorithms, measurement of the performance of the ADC, and development of ICS, PlateMaker, and the active optics system, including mapping the hexapod settings to keep the corrector aligned to the primary mirror over the full range of telescope pointing for the survey. Lastly, we obtained substantial experience with scheduling personnel, documentation, and organization of commissioning activities that helped the subsequent commissioning proceed more smoothly. 
 
\section{Acceptance, Integration, and Testing} \label{sec:ait}

The DESI subsystems went through extensive reviews during the design and construction phases. These included an R\&D review in 2012, reviews mandated by DOE at CD-1 (2014), CD-2 (2015), CD-3 (2016), an annual review in 2017, two reviews in 2018, and at CD-4 (2020). In advance of each DOE review, LBNL organized a ``Director's Review" to assess readiness for the corresponding DOE review. The project also organized Preliminary Design Reviews of all subsystems in advance of CD-2 (Performance Baseline), Final Design Reviews of all subsystems in advance of CD-3 (Start of Construction), and numerous more focused reviews on individual instrument components, manufacturing readiness for major procurements, safety, installation, commissioning, and operations planning.  

The Level 2 managers and cognizant scientists were responsible for representing their subsystems at the relevant reviews, as well as shepherding their subsystems through formal subsystem acceptance by the Project. The subsystem acceptance process is described in the first subsection and usually included a pre-ship review with verification of all requirements that could be met prior to installation, as well as an installation plan. Subsystem acceptance was only complete upon delivery to the Mayall, demonstration that no damage occurred during shipment, and delivery of an installation plan and the documentation required for long-term support. 

Upon acceptance, each subsystem  became the responsibility of the installation team. The installation of each subsystem is described in the second subsection. Upon the completion of installation, each subsystem went through functional verification. This phase included system-level integration and testing activities, although did not re-verify requirements. Functional verification included essentially all testing that could be done with the dome closed, and especially focused on integration with the ICS. 

 \subsection{Subsystem Acceptance}
 
 The completion of the subsystem acceptance process marked the formal hand off of each subsystem to the installation team. In most cases the subsystems were provided by collaboration member institutions (or partners), including several international collaborators, and not directly by LBNL. As a consequence, the acceptance process often represented the formal delivery of equipment provided as a contribution to membership, and in some cases marked the point when the scientists and engineers who developed a subsystem would complete their work with DESI and move to other projects. As a result, the acceptance process was used to ensure we had sufficient information to complete installation and integration and maintain the subsystem during operations. We consequently created a Subsystem Acceptance Board to review the requirements on each subsystem, verify that we had sufficient documentation, and to make a recommendation on acceptance to the Project Manager. The Subsystem Acceptance Board was chaired by the Instrument Scientist and also included the two Project Scientists, the Systems Engineer, and the Acceptance, Integration, \& Testing (AI\&T) Scientist. 
 
 Each hardware subsystem had a review prior to shipping. In practice, this was a dry run for the formal, post-delivery acceptance review. The pre-ship review included documentation for all requirements that could be verified prior to integration with other subsystems. These include laboratory tests such as precision metrology, analyses such as the performance of the as-built lenses in an optical model, and documentation of requirements met by design, for example that the petals had the correct number of  fibers for the sky monitor. The review ensured that the system was in compliance with the relevant ICDs, that the ICD documentation was mature and complete, and that the documentation described the procedures for installation and long-term maintenance in sufficient detail. The pre-ship review frequently helped identify missing documentation well in advance of the final subsystem acceptance review. 
 
 We anticipated that not all subsystems would meet all requirements and maintained margin against this eventuality. We also created a Nonconformance Review Board to review any instance where a subsystem failed to meet requirements. This Board was initially comprised of the AI\&T Manager, L2 Manager, Project Manager, System Engineer, and Instrument Scientist, and we typically added one or more other experts from within the collaboration. The Board performed an initial judgement of the scope of any cost and/or schedule impact of the nonconformance. If the scope was judged to be major, the Board was expanded to include the Project Scientists and Director. Some conditions that led to nonconformances were that a subsystem failed to meet performance, technical, or material requirements or that it failed to meet a requirement in ICDs or other design documents. If the nonconformance could not be easily addressed, the Board conducted an analysis of the potential implications for higher-level requirements and drafted a waiver request. Some examples of waiver requests include that the efficiency of the blue gratings in three spectrographs was somewhat below the requirement and that the ADC rotator exhibited some vibrations at the required speed. In the case of the blue gratings, the throughput of other spectrograph components exceeded their specifications, so the net throughput of the spectrographs were compliant. In the case of the ADC rotator, further study showed that the vibration was minimized at 50\% nominal speed, and that this speed was sufficiently fast relative to the telescope and dome that it would not affect the length of the inter-exposure sequence. Waiver requests were submitted to the Subsystem Acceptance Board as part of the documentation for subsystem acceptance. 

 Subsystem acceptance occurred after post-delivery inspection. The rationale for this timing was to place the responsibility for shipment on the partner who would prepare it for shipment. The partner would also retain responsibility for the subsystem through post-delivery inspection and testing, as the partner would be best equipped to address any issues that arose during shipment. The degree of post-delivery inspection and testing varied substantially between subsystems. In some cases, such as the spectrograph cryostats, it was not practical to conduct on-site testing until after the cryostats were installed on the spectrographs in their final location, so the post-delivery inspection was simply a visual inspection for shipping damage. In other cases, such as the corrector system, there were substantial verification tests during the re-assembly of the corrector prior to installation. Once the post-delivery inspection and testing were complete, the Subsystem Acceptance Board reviewed the subsystem. In all cases, the Board recommended acceptance to the Project Manager.

 \subsection{Integration} \label{sec:install}

\begin{figure*}[ht!]
\includegraphics[width=7in]{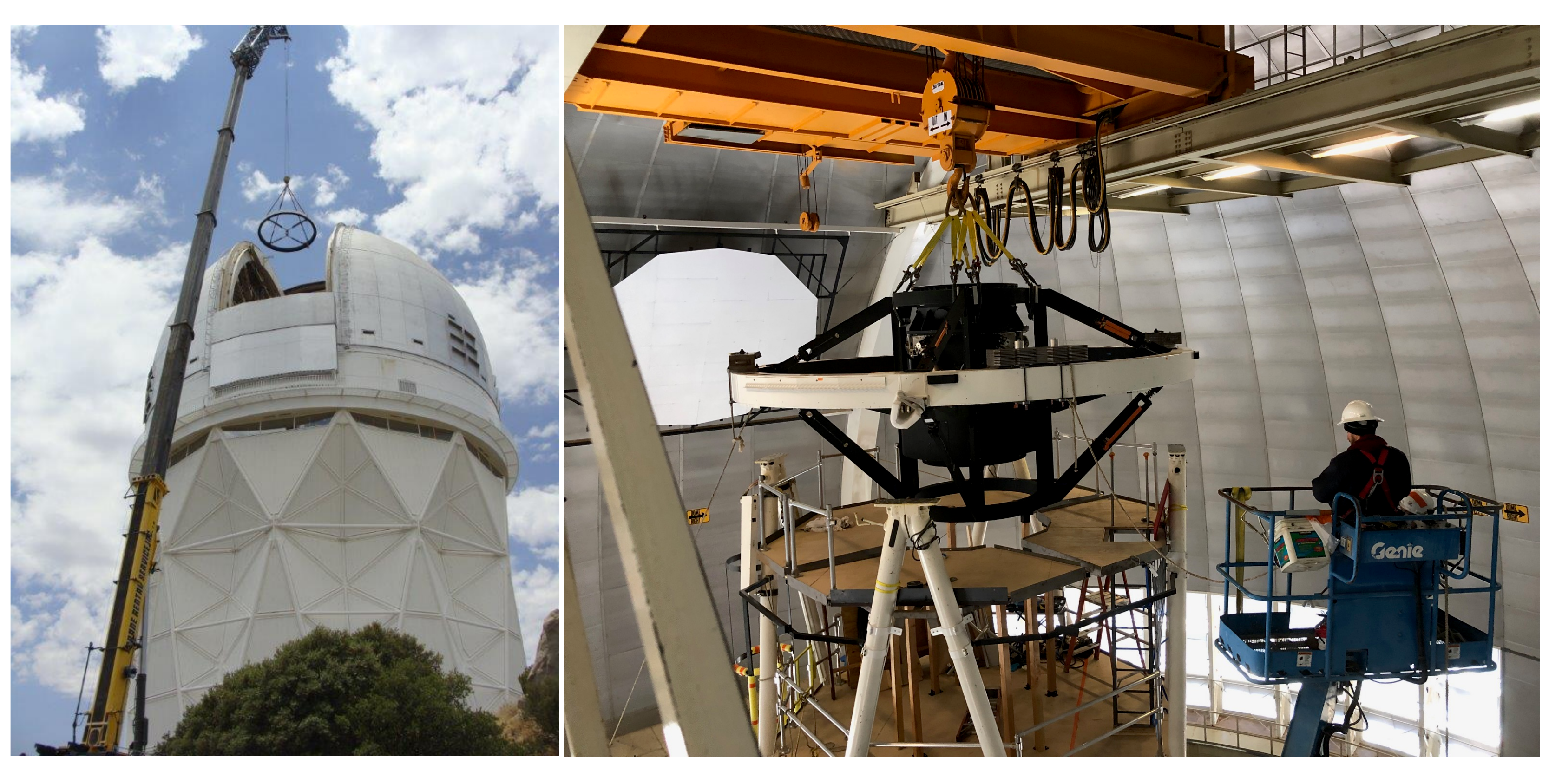}
\caption{Installation phases of the new top ring. ({\it Left}) Transfer of the new top ring from the ground level to the dome floor via the dome slit with a 450 ton crane on 14 June 2018. The same crane was used to remove the old top end. ({\it Right}) Installation of the fully assembled, new top end with cage, hexapod, and corrector barrel to the Mayall's Serrurier truss on 28 November 2018. \label{fig:topend}}
\end{figure*}

 The planning and execution of DESI integration with the Mayall occurred over many years, starting with the upgrades to the TCS and other facility improvements described in \S\ref{sec:mayall}. In this subsection, we describe the integration stages of the new instrumentation. These include the removal and replacement of the top end and the installation of the new corrector, modifications to the bottom end, construction of the thermal enclosure and support structure for the spectrographs, installation of the new focal plane instrumentation, and routing of the fiber cables and other utilities. In addition to the CD reviews, there was a dedicated DOE review of the installation planning  in April 2017. This planning included careful attention to personal and equipment safety procedures, as well as the detailed installation and lift procedures for all components. The peak of the installation activity occurred during shutdowns of the telescope for the corrector system installation from February 2018 to March 2019 and for the focal plane system installation from May through September 2019. During both of these periods, the telescope was locked into position. The interlude between these two phases corresponds to the Commissioning Instrument campaign, which was described in \S\ref{sec:ci}. \citet{besuner20} present a more detailed description of the entire installation process. 
 
 \subsubsection{Top End and Corrector} \label{sec:topinstall}

 Preparations for the removal of the top end began with the removal of the KOSMOS \citep{martini14} instrument from the Cassegrain focus so that we could center an alignment telescope on the rotator axis with the aid of a cross-hair marked on the $f/8$ secondary mirror. The next steps were the removal of the Mosaic prime focus imager \citep{dey16} and corrector optics \citep{sawyer10}, the $f/8$ secondary mirror, and associated utilities. We then disconnected or severed the air, data, and electrical connections running to the top ring assembly and added temporary counterweights to account for the removal of these items. At this point locked the telescope into position at the zenith. The next step was the disassembly of the bottom end, including the Cassegrain guider and rotator and the central baffle assembly. We then removed the primary mirror, which was re-aluminized and stored safely off the telescope until the completion of the work on the top end; however, we did re-install the empty primary mirror cell and the Cassegrain rotator to facilitate later alignment of the new DESI top ring. As the Serrurier truss would be unstable without the upper ring in place, we braced the truss to prevent any relaxation or rebounding of the truss upon removal of the old ring. We also constructed a two-level work platform (visible in Figure~\ref{fig:topend}) to provide personnel access to the upper assembly. These access platforms were useful for the removal of the remaining hardware on the top rings in preparation for the removal of the ring assembly. We designed and built a temporary hydraulic jack system to aid in the separation of the ring from the truss. 

 We hired a 450-ton, 300-foot crane to remove the old Mayall top ring and transfer the new ring from the ground level into the dome. To remove the old ring, we first used the jack system attached to the truss members to slightly raise the ring off of the truss pads. This was more straightforward than feared, as fortunately the ring had not become rusted in place over the preceding 45 years. On June 14, 2018 the exterior crane lifted the old top end, maneuvered it through the dome slit, and placed it on a flatbed truck that carried it to its retirement home in the parking lot by the Kitt Peak Visitor Center. The crane then lifted the new top ring from the ground floor into the dome. Both of these operations were completed with an exterior crane, rather than the 50-ton dome crane, because the top ring is too large to fit through the hatch between the dome floor and the ground floor of the Mayall unless the ring were rotated such that the longest dimension was aligned with the diagonal of the hatch. The exterior crane was a less risky and less complex solution. 
 
 The new ring, vanes, and cage for the top end arrived on April 19, 2018, before the removal of the old top end assembly; they were stored in the ground floor garage until the old assembly was removed. The first step of their installation was to match-drill the locations of the alignment pins and bolts that would be used to fasten it to the truss. As the ring experiences stress and some deformation with the mass of the corrector installed in the cage, we put the ring into a similar state of stress and tension in order to accurately mark the locations for the pins and bolts. We reproduced the expected deformation (a few millimeters) with an assembly that replicated how the vanes attached to the cage and provided a comparable amount of tension.

 The next stage of installation was to match drill the pre-tensioned upper ring. We lifted the ring with the  pre-tensioning system to the top of the truss with the 5-ton dome crane, used the alignment telescope previously centered on the Cassegrain rotator axis to align the ring to the axis, and match-marked the holes for the alignment pins and bolts. The ring was then lowered to the M floor, drilled and tapped for the pins and bolts, and lifted to the C floor for integration with the cage and vanes. Prior to the start of this final assembly, we added strain gauges to the vanes to help balance the forces on them. We used a laser metrology system to center the prime focus cage within the top ring.
 
 The new corrector and ancillary alignment equipment arrived on July 13, 2018 by charter aircraft from the UK. We carefully unpacked each of the three corrector segments, inspected them on a rotary table, and reassembled and tested the complete barrel. We then integrated the corrector with the hexapod, which required the temporary separation of the forward section of the corrector from the other two sections. All of this work occurred on the ground floor of the Mayall. In early November 2018 we used the 50-ton dome crane to lift the corrector plus hexapod to the dome floor, where we integrated it with the new prime focus cage in the new top ring. Finally, on November 28, 2018 we used the 50-ton dome crane to install the 10,700 kg top end assembly onto the top of the Serrurier truss. 

 \subsubsection{Spectrograph System} \label{sec:specinstall}
  
 The kit of components for the thermal enclosure or Shack arrived in mid-May 2018 and we completed most of the assembly by the end of June, although some residual work on utilities and other details continued for the following several months. A Western Environmental Corporation representative verified that the Shack met temperature, humidity, and cleanliness requirements in mid-October 2018. The certification tests showed that the temperature stability was better than $0.1^\circ$C, the humidity was within range, and the air cleanliness met the requirements of a class 10,000 clean facility. Later in October we assembled the Rack inside the Shack to prepare it for the installation of the spectrographs. 
 
 In mid-December 2018 we installed and tested the shack controller, spectrograph purge system, and stand-alone fiber illumination system. This included mounting the Shack electronics box on the south wall of the Shack, running electrical lines, ethernet, and air hoses through the rear cable tray of the Rack, and testing the stand-alone fiber illumination system and the dry air purge system for the cryostat windows. We installed the cryostat electronics in early 2019. This equipment included three computer racks that were placed outside of the Shack, and numerous electrical and coolant cables that connect to the spectrographs. Once the Shack and associated utilities were ready, we moved the first spectrograph from storage at the Kitt Peak Instrument Handling Facility into the Shack and onto the Rack. We then attached the cryostats to the spectrograph, connected various utilities, and began functional verification tests (see \S\ref{sec:fv}).

  \subsubsection{Focal Plane and Fiber System} \label{sec:fpsinstall}

 \begin{figure*}[ht!]
 \includegraphics[width=7in]{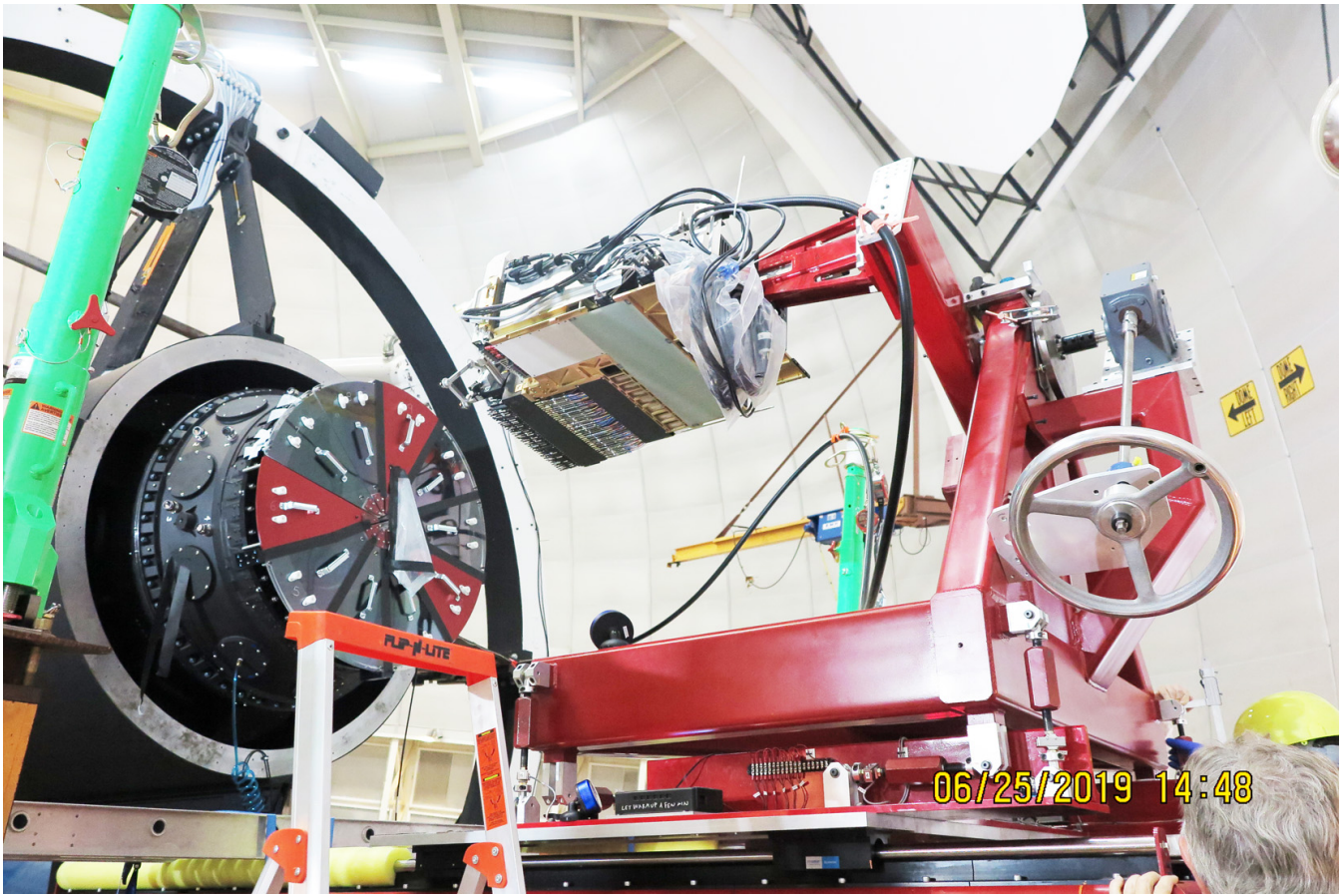}
 \caption{Installation of the first petal of the focal plane system on 25 June 2019. This photo shows the telescope ({\it left}) parked at the Southeast Annex maintenance position. The focal plane adapter assembly is the black steel cone with multiple side port holes that is bolted to the rear flange of the corrector barrel and extends from the end of the cage. The ten petal positions are temporarily covered by ten triangular plates that protect the C4 lens from accidental contact. The red structure in the center of the photo is the Petal Mount Adapter (PMA), which holds the petal at the end of its rotation arm. The hand wheel at right drives the arm via a right-angle gearbox. Several struts are visible connecting the PMA to its base plate. One of the Sled's linear rails is visible at the bottom of the photo. See \S\ref{sec:fpsinstall} for more details.  \label{fig:petalinstall}}
 \end{figure*}
 
 We began preparations for the assembly of the focal plane system at the Mayall during the Commissioning Instrument campaign in April and May of 2019. As described in \S\ref{sec:fps}, we chose to ship individual petals from LBNL and integrate the system on the Mayall after review of the relative merits of integration on the telescope compared to shipment of the fully integrated focal plane system to the Mayall. The greatest challenge in the on-site integration was the insertion of the ten petals, as each petal is a large, complex, delicate piece of equipment with only 0.6\,mm of clearance relative to its neighbors. Each petal had to be inserted at a unique angle (increments of $36^\circ$ about the corrector axis) relative to its neighbors. In addition, the insertion direction was tilted by $\sim 7.2^\circ$ relative to the horizon, as that corresponds to the orientation of the telescope when locked into place at the Southeast Annex maintenance position.
 
 To overcome these challenges, we built a petal insertion structure called the ``Sled'' (see Figure~\ref{fig:petalinstall}). The Sled consists of two large, welded steel frames connected to each other by a set of six adjustable-length struts that control six degrees of freedom. The lower frame of the Sled is rigidly mounted to the Southeast Platform of the Mayall facility. The upper frame supports a pair of linear motion guide rails and a central lead screw. Prior to petal insertion, we adjusted the struts to align the rails parallel to the axis of the telescope when it is parked in the maintenance position. To install a petal, the petal mount adapter (PMA) is driven along the Sled's linear rails, and the base plate of the PMA has linear bearings and a coupling to the Sled's lead screw. Six adjustable-length struts connect this base plate to a roughly tetragonal steel frame above it, and mounted to this frame is a steel rotation arm. We attach a petal to one end of the PMA arm and a counterweight to the other. The rotation axis of the arm is nominally colinear with the corrector axis. This allows us to clock the petal to its designated angular position in the focal plane. We then use the PMA struts to finely align the petal to the focal plate adapter ring. At this point the petal is ready for insertion by driving the PMA down the Sled rails. All driving torques during the petal insertion are provided via handwheels, with significant gear ratios between the operator and the load. Each strut has an inline load cell, so that we can quantitatively monitor the net reaction forces between the telescope and the petal during insertion and load transfer. The cells have a resolution better than $\sim10$\,N.

 One key tool for the alignment process was a Faro VantageE Laser Tracker, combined with Spherical Mount Retroreflectors (SMRs) that were attached to both the focal plane adapter ring on the telescope and to each petal. The SMRs are mounted in ``nests'' that magnetically hold them against repeatable contact points, and the locations of the contact points were measured at LBNL with a CMM. We measured the relative locations of the SMRs on the focal plane adapter ring and on each petal during installation. Typically these measurements had uncertainties of less than $100\,\mu$m, and better than $50\mu$m was achieved under good conditions. \citet{shourt20} provide a more extensive description of the alignment and integration of the FPA with a laser tracker. 
 
 While the Commissioning Instrument campaign was in progress, we conducted a mock installation exercise on the ground floor of the Mayall building. This included a custom corrector simulator with an identical interface flange to the end of the corrector, oriented at the same $\sim 7.2^\circ$ angle relative to the ground, as well as used dummy petals with the same envelope, mounting features, mass, and center of mass. This mock installation led to a number of improvements in the installation procedure.
 
 As soon as the Commissioning Instrument was removed from the telescope, we moved the installation fixtures for the focal plane system from the ground floor to the Southeast Annex. The first petal arrived on May 8, 2019, and the remainder followed at intervals throughout June and July. Before installation, each petal was briefly powered on, back illuminated, and tested to determine if any damage had occurred during shipment. We installed the first petal on June 25, 2019. The installation of the remaining petals proceeded smoothly, and we installed the final petal on July 23, 2019. Figure~\ref{fig:petalinstall} is a photo of the first petal ready for installation. 
 
 One unforeseen challenge in this process was a small drift between the telescope and the Southeast Annex platform with time. This drift occurs because the telescope is mounted to a rigid, concrete pier that is mechanically isolated from the rest of the Mayall building in order to maximize its stability. During the petal installation, we observed differential motion as large as a few tenths of a millimeter in tens of minutes, which we attribute to differential thermal effects on the Mayall building and the pier. We adjusted our procedure to minimize the time between the final measurement and adjustment and the insertion of the petal. Throughout this process we enclosed the workspace as much as practical with plastic sheeting and applied a slight overpressure to maintain a clean and moth-free environment. 
 
 Once the FPA was complete, we installed the FPE and routed the fiber cables to the spectrographs. The FPE is mounted directly to the cage, rather than the corrector, to reduce the total mass and moment attached to the hexapod. The FPE thermally isolates the FPA at a constant temperature, as well as provides a clean and low-humidity environment. The frame of the FPE includes four rigid ``tower'' assemblies that are parallel to the corrector axis and extend the full length of the FPE. The fiber cables, electrical cables, and hoses route through these towers. From the FPE, the fiber cables route to the upper ring along the Southeast and Northeast upper spider vanes, down the telescope truss, and through the hour angle and declination axes. NOIRLab staff designed and built a series of cable carriers for the fibers that avoid twists and enforce their minimum bend radius of 200\,mm. The fiber cables then extend through custom penetrations in the walls of the Large Coud\'e Room and into the Shack. Once we completed the fiber cable installation, we installed the four calibration lamp boxes on the Mayall's top ring.
 
 \subsection{Functional Verification} \label{sec:fv}

The two goals of the Functional Verification (FV) phase were to verify that subsystems were not damaged during installation and to integrate them with the other subsystems. The main focus of this integration was software, as hardware integration occurred during installation. We chose to not spend time on the re-verification of requirements during this phase. Instead, we emphasized the start of on-sky commissioning as soon as practical, as that phase provides more and better opportunities for the verification of the instrumentation. 

The emphasis on software integration meant in practice that FV for each subsystem was complete upon the successful integration of that subsystem with the ICS. As a consequence, ICS was the first subsystem to begin FV and the last to finish. Functional verification for ICS began early in the project with the TCS integration in advance of ProtoDESI. FV for the other subsystems occurred as soon as they were installed. In practice this meant that the Shack was the next subsystem to go through the FV process, as it was needed to install the spectrographs and cryostats. The Shack was followed by the corrector, as it was needed before the installation of the Commissioning Instrument, and then the FPA. Spectrograph FV occurred after each spectrograph arrived and was installed in the Shack.

The main step for the Shack FV was the ingestion of telemetry data (e.g. temperature, humidity) from the WEC control computer to the ICS, which we achieved with a dedicated communications link from the telemetry available from the control computer to our shack controller electronics box. The shack controller also reads the pressures of the air supplies for the spectrograph seals and the spectrograph purge system, and it has the capability to power cycle any of the spectrograph controller electronics boxes. During FV of the focal plane, the shack controller was also used to control the stand-alone fiber illumination system. 

\begin{figure*}[ht!]
 \includegraphics[width=7in]{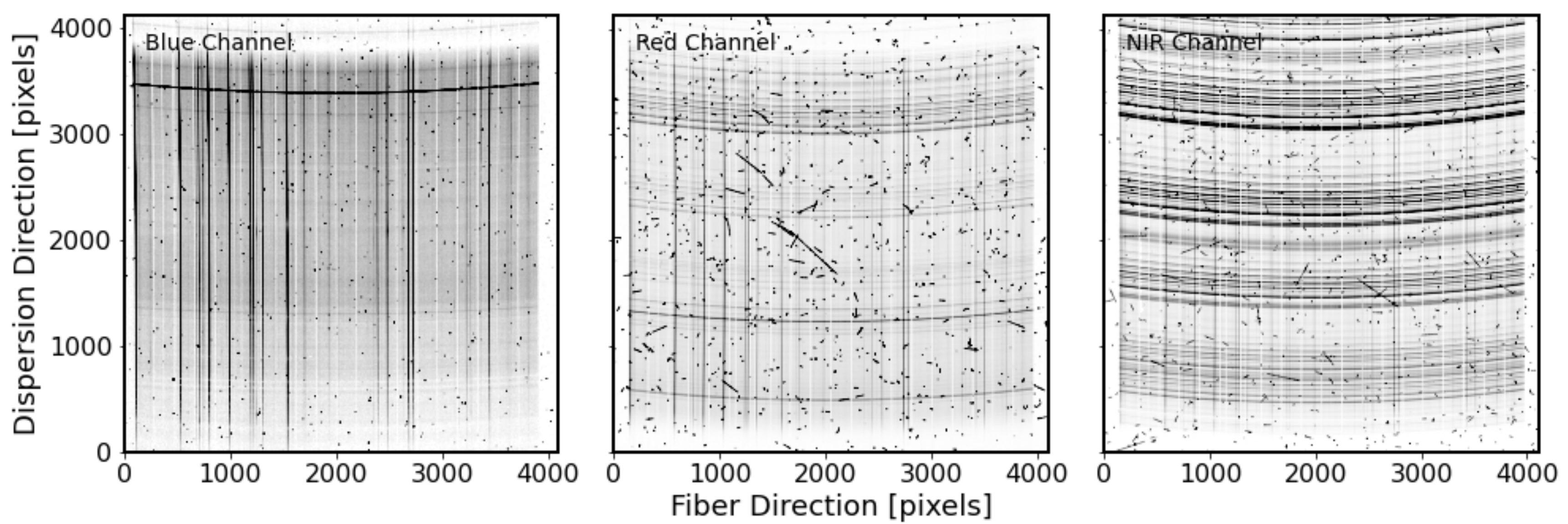}
 \caption{Example science images from spectrograph SP4. The three panels show inverted, grayscale images from the blue, red, and NIR channels after removal of the overscan and bias level, and before additional processing. Wavelength increases with pixel number along the vertical axis. Spectra of the same 500 fibers are parallel to the columns on each detector. The vertical, dark stripes correspond to the spectra of the brightest sources in this exposure, while the horizontal arcs correspond to night sky emission lines. There is a small gap between each of the 20 fiber blocks in the slithead, and these gaps appear as vertical bands with no flux between each set of 25 spectra. This is exposure ID 68879, which is a 900s exposure in dark conditions obtained on 19 December 2020. Note the greater sensitivity of the thicker CCDs in the red and NIR channels to cosmic rays. \label{fig:specdisp}}
 \end{figure*}

Spectrograph and cryostat FV proceeded as each spectrograph and its cryostats were installed in the Shack. The tests for each spectrograph included both a repeat of all pre-ship functional tests and a subset of the pre-ship qualification tests. The functional tests included the operation of the mechanisms, operation of the fiber illumination system, and readout of the detections. The qualification tests were measurements of the read noise, well capacity, gain, non-linearity, dark curent, and CCD cosmetics. Once the focal plane and calibration systems were installed, we measured light leaks, image quality, spectral resolution, cross talk, and stability. We also used the Hartmann doors to check the focus and make minor adjustments to each cryostat. As the last spectrograph was delivered after the start of commissioning, spectrograph FV was not complete until several months into the commissioning period (December 2019). By that time we had gained sufficient experience with the system that we established a daily calibration protocol. This includes zeros, darks, flat fields, and arc calibration exposures that are typically obtained in the late afternoon. The flats and arc calibrations use the calibration lamp system mounted to the top ring. Figure~\ref{fig:specdisp} shows an example of the data from one (SP4) of the ten spectrographs after the removal of the overscan and bias subtraction steps.  

FV of the corrector system included the two ADC motors and the hexapod. Both of these units were extensively tested with ICS in advance of installation and this integration was consequently straightforward. Upon installation of the Commissioning Instrument, both it and the FVC went through FV. The main step was illumination of the fidicuals on the Commissioning Instrument, and imaging those spots with the FVC. This produced the first recorded images through the new corrector. On-sky observations, described in \S\ref{sec:ci} and shown in Figure~\ref{fig:firstlight}, demonstrated the superb performance of the corrector optics. 

Functional verification of the focal plane and fiber system occurred in three phases. The first phase was a basic checkout of the cooling services. This included leak and pressure checks, the operation of the liquid chiller, the cooling fans, temperature sensors, and humidity sensors, and their integration with ICS. The second phase was a basic functional test of the positioners and GFAs. The positioner operation was first checked with visual inspection and then the illuminated positioners were imaged with the FVC. At this time we verified that $\geq 99$\% of the fibers were intact after installation. The GFAs were also turned on and operated under ICS control. Finally, the third phase combined the calibration of the positioner parameters with the mapping of the coordinate transformations from the pre-ship metrology of each petal to the FVC. This process included accounting for the distortion caused by both the corrector optics and the FVC optics. The positioner calibration included precise measurements of the six parameters per unit which are required to accurately transform between motor rotation angles and fiber tip coordinates on the focal surface. This mapping is critical for accurate alignment of the fiber tips with astrophysical sources and to avoid collisions between positioners. Larger uncertainties in these parameters would require larger tolerance regions around the moving components, which would impact fiber assignment. The instrument was ready for on-sky observations in October 2019 upon completion of FV for the focal plane and fiber system. 

\section{On-Sky Performance Results} \label{sec:results}
 
Commissioning began in October 2019 as soon as a sufficient fraction of the instrument was ready for on-sky observations. The purpose of this phase was to test and optimize the instrument and demonstrate that it met all performance requirements from $0^\circ$ to $60^\circ$ from zenith. The key instrument properties that could only be tested with on-sky observations were the image quality, total throughput and noise, fiber positioning accuracy, and guiding performance. During the commissioning period, we also worked to decrease the length of the inter-exposure sequence, measured the stability of the spectrograph PSF, and improved the telemetry data, access to these data, and alarm and error handling. In the following subsections we describe the measurement of the total throughput and noise, characterization of the guiding performance, validation of the fiber positioning accuracy, report on the stability of the spectrograph PSF and resolution, and demonstrate the superb data quality with observations of some faint  emission-line galaxies. We also describe efforts to decrease the inter-exposure sequence in order to maximize our observational efficiency. The validation of the image quality was described in \S\ref{sec:ci} as part of the Commissioning Instrument campaign (see also Figure~\ref{fig:corrperf}). The commissioning period led to improvements in nearly every aspect of instrument operations, including calibration procedures, data analysis, and hardware upgrades to improve performance. During commissioning we also uncovered several hardware issues that impacted performance. In the last subsection, we describe those issues and how we addressed them. Commissioning was interrupted from March through November 2020 due to the global COVID-19 pandemic and formally ended in December 2020 with the start of the Survey Validation phase.

\subsection{Throughput} \label{sec:throughput}

\begin{figure*}[ht!]
 \includegraphics[width=7in]{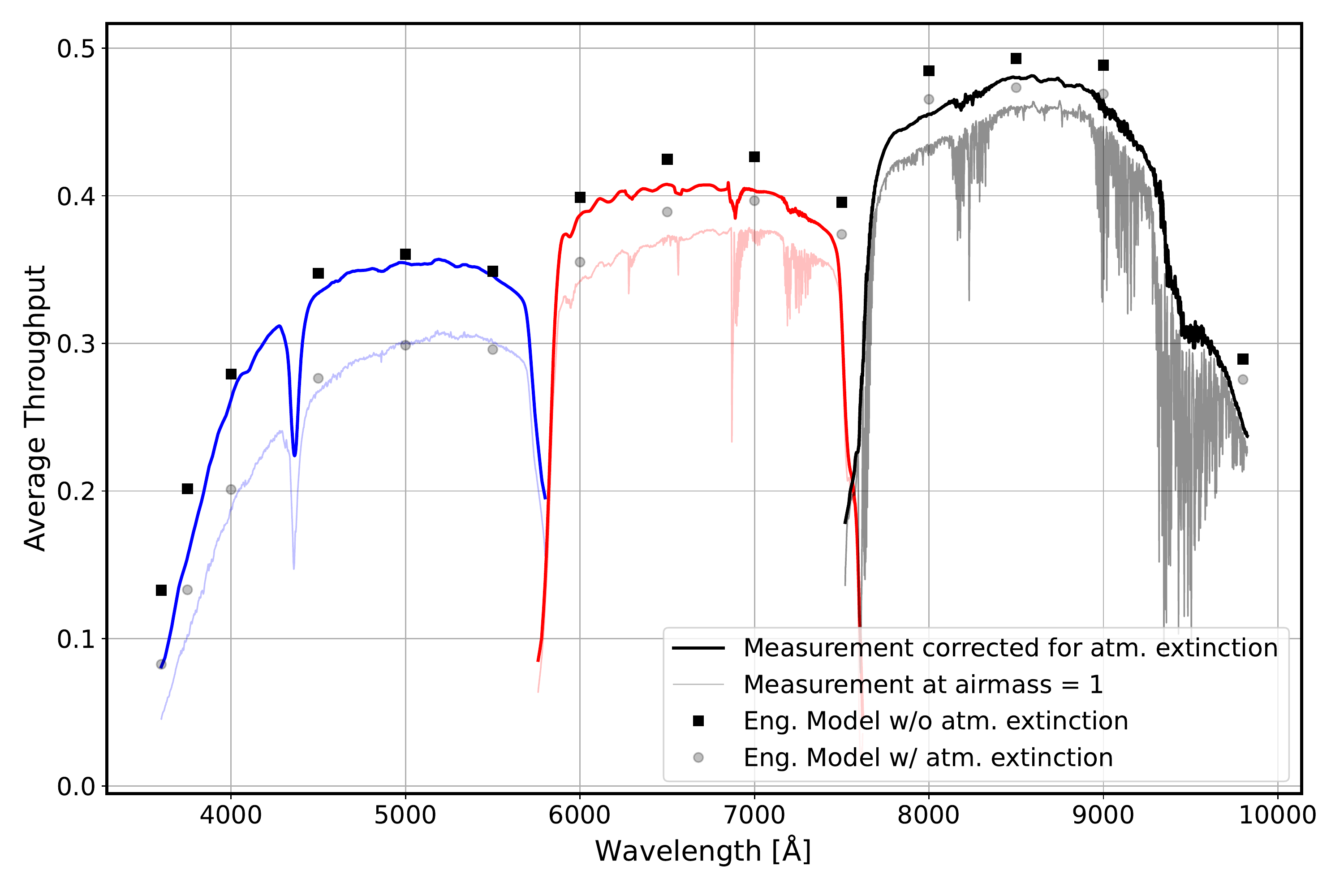}
 \caption{Average measured total throughput compared to the engineering model. The average throughput is measured with standard stars on photometric nights. All measurements have been corrected for losses due to the fiber acceptance fraction and therefore do not include the contribution of seeing. The darker lines ({\it blue, red, black}) show the throughput of the three spectrograph channels with a correction that removed atmospheric extinction, while the lighter lines show the throughput at the zenith. The points are the predicted throughput of the engineering model for both cases. The agreement between the measurement and the model is excellent for the blue channel and very good for the red and NIR channels. The minor differences between the model and the measurement are likely due to many small differences between estimates and the final performance. The dip at 440\,nm is due to an issue with the spectrograph collimator coating. The average throughput without atmospheric extinction has been smoothed to reduce the impact of individual telluric absorption features on the measured throughput. \label{fig:totthru}}
 \end{figure*}

The total system throughput was one of the most important parameters throughout the project. As described in \S\ref{sec:start}, we developed a detailed total throughput budget early in the development of DESI to set requirements on the instrument subsystems. We updated this budget throughout the project as we obtained empirical data for many contributions (e.g., optical coatings, FRD, detector response). We obtained our first opportunity to measure the total system throughput with observations of the standard stars that we include in all petals on every field we observe, and we gradually refined these throughput measurements as we acquired more data and improved the fiber positioning accuracy. Figure~\ref{fig:totthru} shows the total measured throughput from standard stars observed over many photometric nights, both with and without corrections for atmospheric absorption based on the extinction curve for Kitt Peak. The measurement with atmospheric extinction has been corrected to airmass $X = 1$. 

The measurements shown in the figure are based on observations over nights that exhibited excellent image quality (seeing between $0.7''$  and $0.9''$ FWHM). We determined the fractional flux in a fiber for each observation based on stacked images of the guide frames that were obtained in conjunction with the spectroscopic observations. This fiber aperture loss is of about 40\% for our typical seeing of $1.1''$ and increases as the seeing degrades. The total throughput shown in Figure~\ref{fig:totthru} is corrected for fiber aperture loss. We assessed that we obtain a consistent throughput using data observed with worse seeing, which imply a higher correction. The figure also shows the predicted throughput from the system engineering model, which is the total throughput budget based on the best empirical data or analytic estimate for individual contributions. The agreement between the data and the engineering model is extraordinary for the blue channel and very good, within 10\%, for both the red and NIR channels. The mild disagreement between the prediction and model for the two redder channels is likely due to many small discrepancies in the engineering model. 

\subsection{Guiding and Fiber Flux} \label{sec:guide}

\begin{figure*}[ht!]
\includegraphics[width=7in]{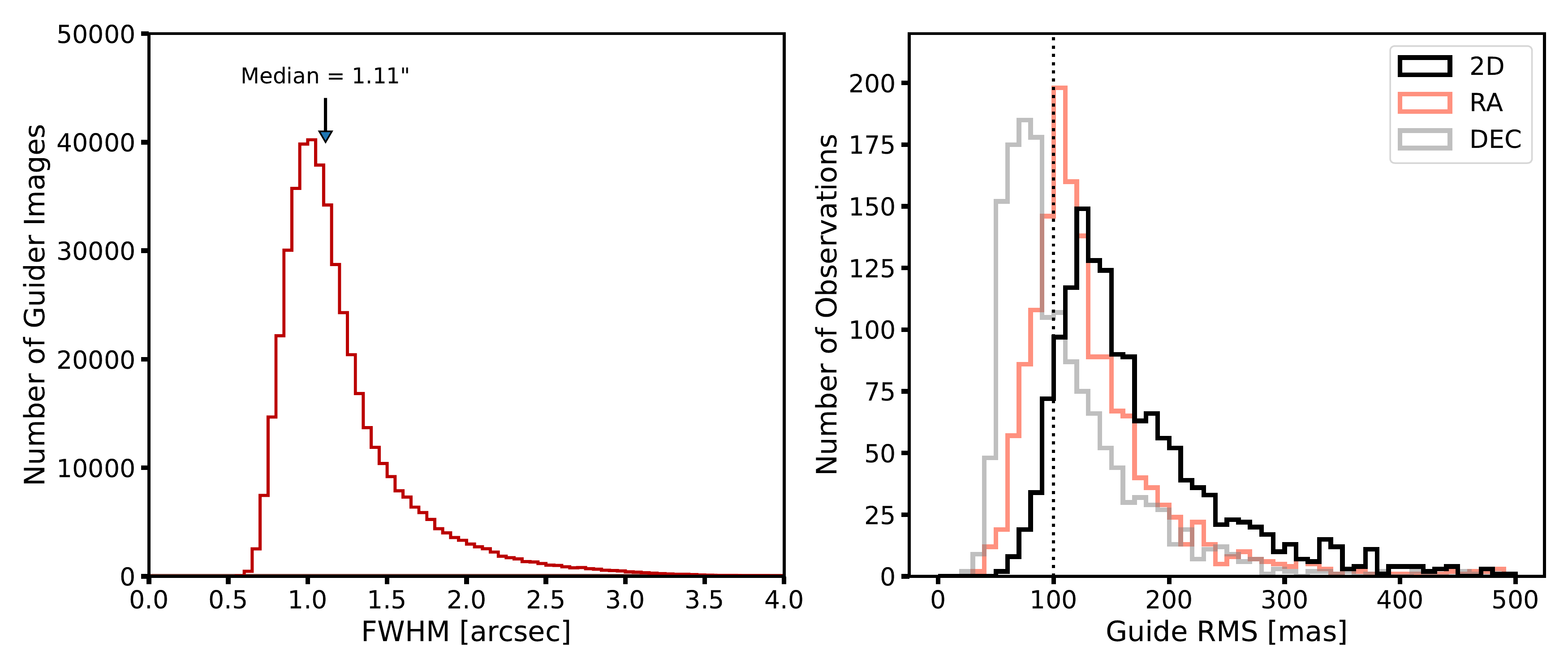}
\caption{({\it Left}) Image quality measured in guider images during commissioning in early 2020. These are measurements on individual 5\,s guider images over many nights and at a range of airmass. ({\it Right}) Guiding performance from observations between January and April 2021. The requirement is 100\,milliarcseconds rms on each axis ({\it vertical, dotted line}). The contribution in right ascension is larger than the contribution in declination, which is expected for an equatorial mount telescope.  \label{fig:guiding}}
\end{figure*}

 The previous subsection demonstrated how DESI provides exceptional throughput. The next subsections show how well the telescope, corrector, and focal plane delivers the light from our targets to the fibers. The top-level requirement on the telescope guiding performance (L3.3.3) is the assumption that the telescope guiding accuracy is 100\,mas rms 1-D. Multiple factors contribute to the guiding accuracy, including the telescope pointing model and tracking, the performance of the algorithm that identifies and measures sources on the guider chips, the algorithm that combines the measurements to create correction signals in right ascension and declination, and the correction for field rotation with the hexapod. 
 
\begin{figure*}[ht!]
 \includegraphics[width=7.0in]{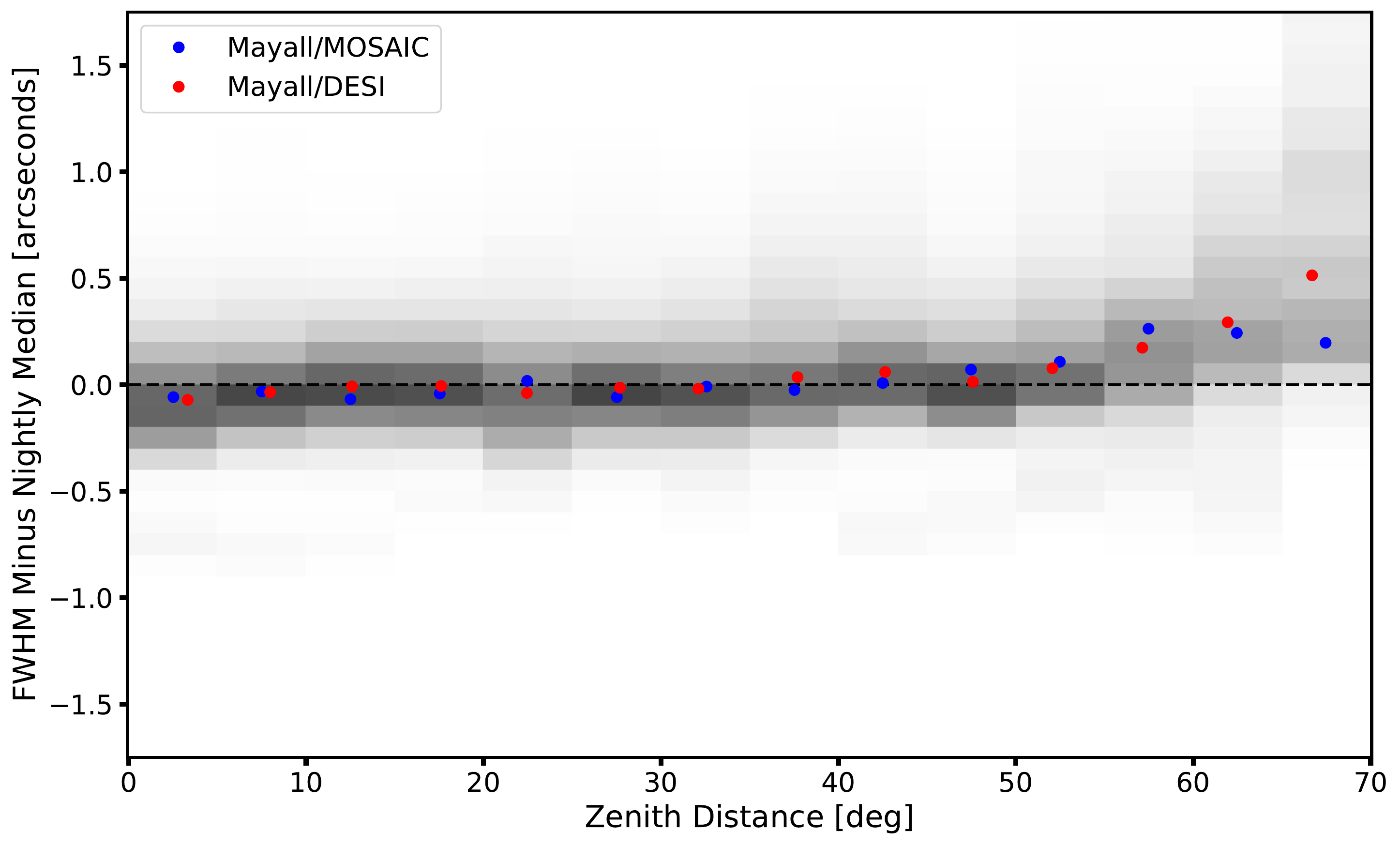}
 \caption{FWHM as a function of zenith distance. This figure is based on measurements of the FWHM from $\sim480,000$ guider images obtained during DESI commissioning, where the median nightly seeing has been subtracted from each measurement. The grayscale shows the number of points in each bin in zenith distance, and each bin in zenith distance is normalized to unity as there are many more observations near the zenith than at higher airmass. The FWHM has no dependence on zenith distance until $\sim 50^\circ$ deg, and the behavior with the DESI corrector and the previous MOSAIC corrector are very similar. This is a modified version of a figure shown in \citet{meisner20}. 
 \label{fig:fwhmvszd}}
 \end{figure*}

 We used the guider cameras to validate the performance of the corrector and telescope after the installation of the focal plane system. Figure~\ref{fig:guiding} shows the distribution of measured seeing values from guider images obtained between October 23, 2019 and March 15, 2020. These FWHM measurements were obtained from the measured radial profile of stars on a guider chip. The best FWHM measurements are between $0.6 - 0.65''$ and demonstrate the excellent performance of the corrector system. The median is $1.11''$, which is superior to the median of $1.2 - 1.3''$ measured by \citet{dey14}. For typical seeing of $1.1''$, about 60\% of the flux from a point source enters the fibers. Figure~\ref{fig:fwhmvszd} shows the FWHM as a function of zenith distance, which demonstrates the excellent performance of the ADC and the overall stability of the top end and hexapod control of the corrector system.

 We obtained several new data sets to update the pointing model before optimizing the guiding performance (see \S\ref{sec:tcs}). The latest model from April 1, 2021 yielded excellent pointing and tracking performance. The median performance was about 100\,mas from January through April of 2021 on each axis and this is shown in the right panel of Figure~\ref{fig:guiding}. The performance is slightly better along the declination axis, as expected for an equatorial mount telescope. The measurement is after a correction for field rotation during the spectroscopic exposure. We still plan to implement a range of filters on the raw guide signal to improve this performance.

 The overall performance of the GFAs has been excellent. We regularly monitor the dark current and readnoise, and these studies have shown very good overall stability. The dark current tracks the CCD temperature in a consistent manner over the typical operational temperature of $10 - 11^\circ$\,C. There is a modest amount of variation in the readnoise that we attribute to low-level pattern noise. The readnoise has consistently remained below the requirement of $25\,e^-$ rms per pixel. 

\subsection{Fiber Positioning} \label{sec:dither}

 \begin{figure*}[ht!]
 \includegraphics[width=7in]{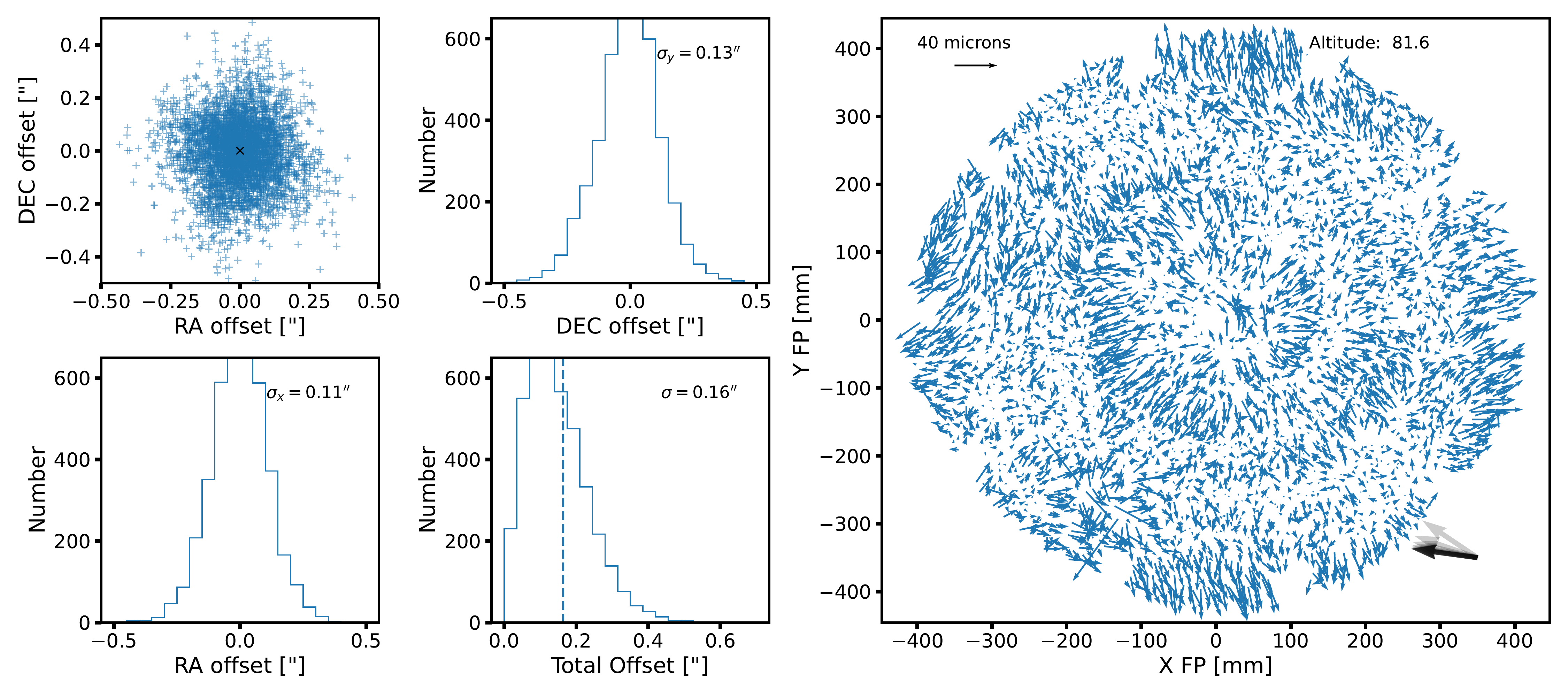}
 \caption{Results of a fiber dither test on December 19, 2021. ({\it Left}) The four, smaller panels show the offset of each fiber positioner from the best fit position determined by the dither test. The four panels show the offsets in right ascension and declination for all positioners ({\it top left}), the distribution of offsets in declination ({\it top right}) and right ascension ({\it bottom left}) and the total offset ({\it bottom right}). The total offset corresponds to the amplitude of the decenter of each positioner, and in this case the width of the distribution is $0.16''$ and the procedure is sensitive to offsets smaller than $0.05''$. ({\it Right}) Amplitude and direction of the offsets in focal plane coordinates. The primarily radial pattern may indicate a residual error in the distortion model. An offset of $40\,\mu$m is indicated by the arrow in the upper-left corner; for context, we note that the fiber diameter is $107\,\mu$m and the plate scale is approximately 70\,$\mu$m per arcsecond. This field was observed near zenith at an altitude of $81.4^\circ$, and the direction of zenith is indicated in the lower-right corner. The fluxes in the fibers at each dither position are from the red channel of the spectrographs. \label{fig:dither}}
 \end{figure*}
 
Our method to ensure the $107\,\mu$m diameter fibers are centered on astronomical targets combines many complex elements. These include the alignment of the fiber tips with the aspheric focal surface of the corrector, the tilts of positioners relative to the focal surface, precision metrology of illuminated fiducials and GFA pixels, optical distortion models of both the corrector and the FVC camera lens, and a model for atmospheric refraction. To verify all of these individual steps, we developed an end-to-end test based on a series of small, random offsets of each fiber that we refer to as fiber dithering. 

Fiber dithering treats the calculation of fiber positioning errors as a nonlinear optimization problem. The model parameters are the flux of each star assigned to a fiber, the two-dimensional misalignment of each fiber relative to the nominal location, the seeing, the transparency, and any telescope offsets between each exposure of a dither sequence. The basic scheme is that we take some number of exposures $N_{\rm exp}$, with some number of fibers $N_{\rm fibers}$ assigned to stars. Between each exposure, we dither each fiber in a separate, random direction by some known amount. We then measure the integrated flux of the source in each fiber in each exposure with the spectrographs, as well as measure the precise, relative motion of the fiber positioners with the FVC. The dither pattern needs to match the PSF size to get good centroid measurements for the positioners. If the positioning errors are much larger than the PSF size, then positioners may have no dither positions that collect the light of the target star and we would not measure their centroids. In this case, we still dither the fibers by the PSF size between exposures, but we also add larger ($\sim 5''$) dither steps onto each target to attempt to align an occasional positioner with the target star on different parts of the focal plane.

We used this dithering analysis on multiple occasions during commissioning as an integrated check of all of the components of fiber positioning. The analysis is especially sensitive to large-scale features, such as shifts between the GFA pixels and other focal plane metrology or errors in the distortion models. Based on the results of several fiber dither sequences at a range of zenith distances, we rechecked the raw focal plane metrology data and replaced the original FVC lens with the simpler design described in \S\ref{sec:fvc}. Within about two months, these improvements and checks led to improvements in fiber positioning from initially  $\sim10''$ rms to the current performance of $\sim0.1''$. For reference, offsets of $0.1''$ ($0.2''$) correspond to throughput loses of 0.07\% (3\%) for a galaxy with a half-light radius of $0.35''$ observed in $1.1''$ seeing. Figure~\ref{fig:dither} shows one illustrative result of the dither analysis from December 19, 2021 with data from the red channel of the spectrographs.  The dithering analysis also provides a direct way to measure the total throughput independent of fiber positioning errors. 

We attribute the remaining radial pattern to residual errors in the distortion model. These residual errors vary between fields and therefore have not been removed. Both these variations between fields, and variations between the blue, red, and NIR channels, are of the same order of magnitude as the variations shown in the figure. In addition to this distortion pattern, dome seeing and turbulence effects also contribute to fiber positioning inaccuracies. Non-moving devices with known stationary positions in the focal plane -- both fiducials and non-functional positioners -- provide a measure of the turbulence field at each iteration of the positioning loop. We fit this field and correct all measured positions for the turbulence field. Based on a dither analysis with and without these corrections, we find that the turbulence correction produces a $\sim10$\% improvement in positioning. 

\subsection{Spectroscopic PSF Stability and Resolution} \label{sec:psf}

 \begin{figure*}[ht!]
 \includegraphics[width=7in]{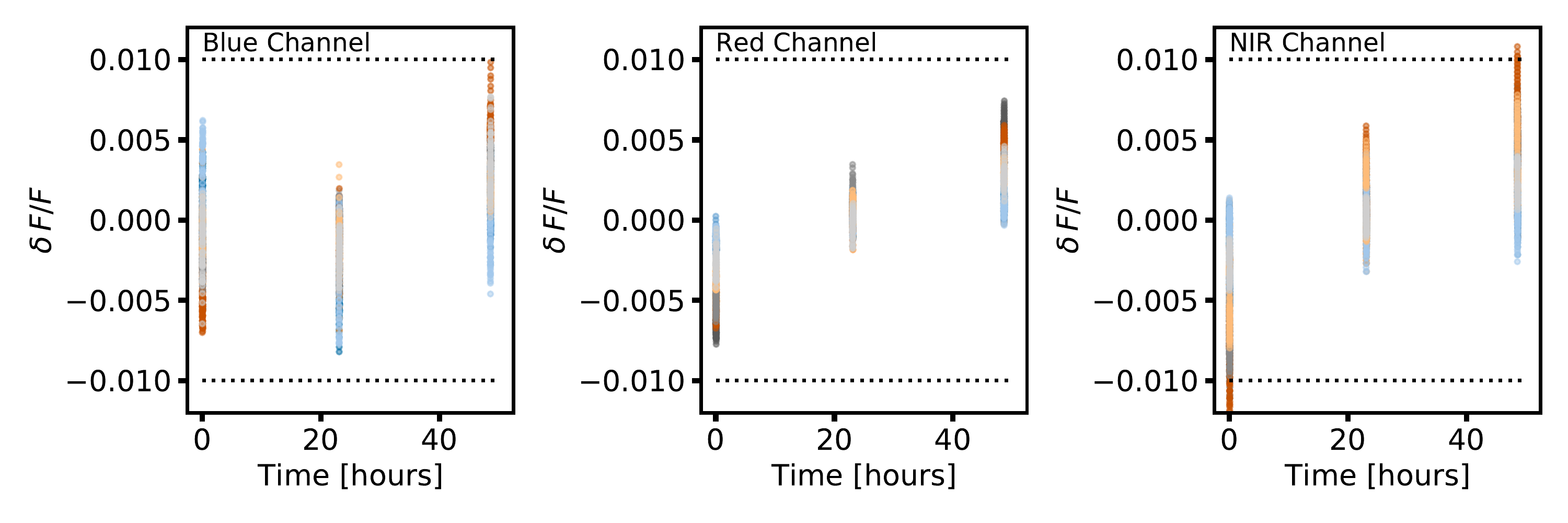}
 \caption{Stability of the PSF across three days of operations for the three spectrograph channels. The fractional change in the PSF is measured from three sets of arc calibrations obtained on three consecutive days, and the fractional change $\delta F/F$ is the change in the PSF relative to the average computed for a selection of fibers at a selection of wavelengths. Each point color corresponds to a different spectrograph, and the distribution of points for each spectrograph corresponds to the variation in the PSF for a range of fibers. There was a night of observations between each calibration set, and the fiber positioners are in different locations for each calibration. The variation is consequently indicative of the level of variation per night. \label{fig:psf}}
 \end{figure*}

The spectral extraction code employs a model of the spectrograph PSF to accurately measure the flux, noise, and spectral resolution of every spectrum and subtract night sky lines. Requirement L3.1.6 states that the PSF bias shall not exceed 1\%, where the PSF bias is $F = \Sigma p_i q_i / \Sigma q_i q_i - 1$ and the two quantities are the intensity distribution of the true PSF $p_i$ and the linearly-interpolated PSF $q_i$ based the arc calibration data. We measure the PSF from arc calibration frames obtained as part of the daily calibration protocol. The main factors that are expected to produce changes in the PSF are changes in forces on the fiber cables and environment-induced changes in the spectrographs. The forces on the fiber cables change when the telescope moves to different positions on the sky and when the positioners move the fiber tips within their patrol regions. Changes in the spectrographs could be produced by changes in the temperature and humidity of the Shack environment. 

\begin{figure*}[ht!]
 \includegraphics[width=7in]{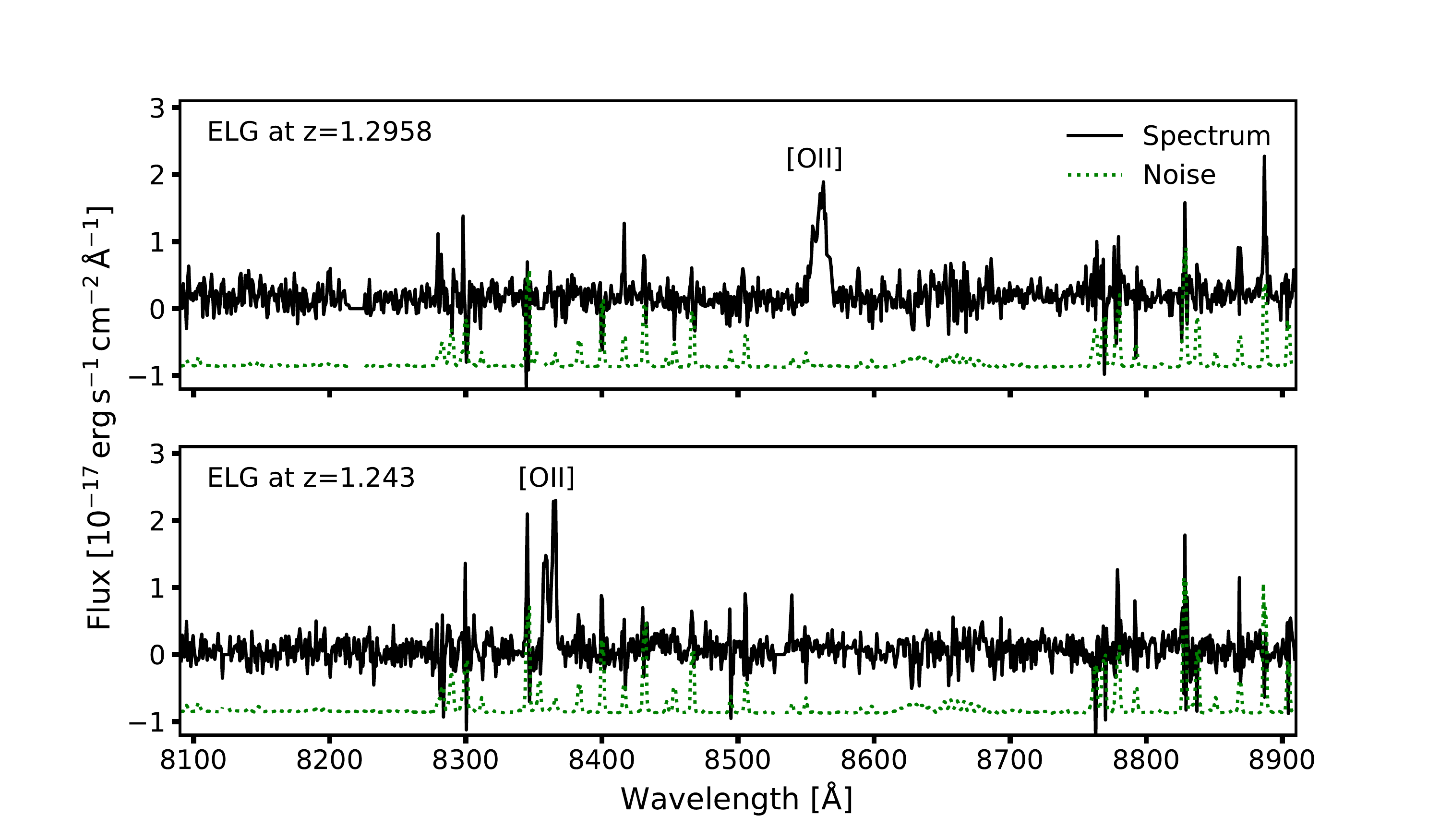}
 \caption{Example spectra of two emission-line galaxies that illustrate both the sensitivity of the instrument and the quality of the sky subtraction. These two galaxies are at sufficiently high redshift that the \oii\ doublet falls in a wavelength region with a substantial number of atmospheric emission lines. The noise spectrum in both panels illustrates the increase at discrete wavelengths due to the Poisson noise from these lines. The noise is shifted by $-10^{-17}$ in these units to make it more visible. These galaxies were both observed on 30 November 2021.  \label{fig:spectrum}} 
 \end{figure*}

We have measured the size of the PSF bias with numerous, dedicated tests to determine if daily (afternoon) measurements of the PSF from arc calibration data are sufficient. For one test we fixed the telescope at the altitude of the dome flat screen, moved the telescope to different azimuth locations in the dome, rotated the dome to place the flat field screen in front of the telescope, and took a sequence of arc calibrations. For a second test, we kept the telescope in the same position and took a sequence of arc calibrations with the fiber positioners in different configurations. Figure~\ref{fig:psf} shows the variation in the PSF bias measured with calibration data obtain on three consecutive afternoons after nighttime observations. After each night of observing, the fiber positioners are in different locations, as they are left in the configuration for the last observation of the night. The PSF bias is $<1$\% for nearly all fibers, and it is sufficiently stable to obtain a sky subtraction precision better than $1$\%. Figure~\ref{fig:spectrum} shows two examples of the superb sky subtraction with spectra of emission line galaxies at $z>1$. These are particularly challenging objects because redshift measurements require detection of the faint, marginally-resolved \oii\ doublet, which is in the midst of significant atmospheric emission at these redshifts. The sky subtraction algorithm and PSF stability will be described further in Guy et al. (2022), {\it in preparation}. 

\begin{figure*}[ht!]
 \includegraphics[width=7in]{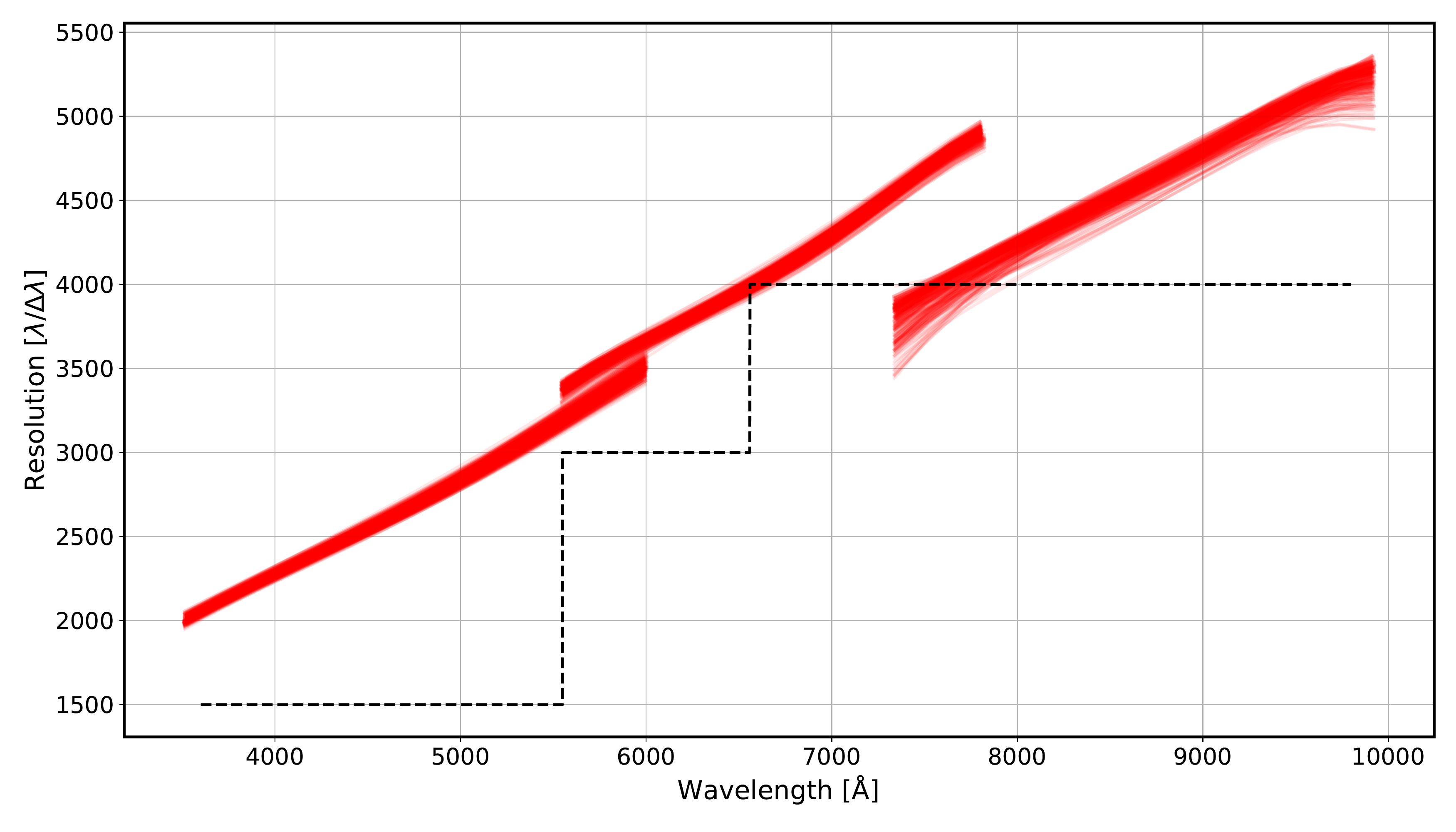}
 \caption{Spectral resolution as a function of wavelength for all ten spectrographs. The line width $\Delta\lambda$ is the FWHM of the 2D PSF model fit to nightly average calibration data after the PSF is projected along the dispersion axis. The range of lines correspond to all measurements from January 2022 for all ten spectrographs and a subset of fibers. The resolution exceeds the requirement over the entire wavelength range. The crossover region for the red pass dichroic is $566 - 593$\,nm and the crossover region for the NIR pass dichroic is $747 - 772$\,nm. \label{fig:specres}} 
 \end{figure*}

In addition to this superb stability, we assessed that the spectrographs delivered the expected spectral resolution. We measure the spectroscopic resolution with the afternoon arc calibration described in \S\ref{sec:fv}. Specifically, we measured the 2D PSF of each fiber in each spectrograph as a function of wavelength, convolved this model with the CCD pixels, projected the model onto the dispersion axis, and then measured the FWHM of the profile. Figure~\ref{fig:specres} shows a collection of resolution measurements for a subset of fibers in all ten spectrographs from January 2022. This figure demonstrates that the spectral resolution exceeds the requirement. In the red and NIR channels, this resolution is about twice that of SDSS and BOSS and this enables DESI to observe the faint continuum between the OH sky lines and resolve the \oii\ doublet from high-redshift star-forming galaxies. 

\subsection{Inter-Exposure Sequence} \label{sec:inter}
 
During commissioning and early in the early operations phase we worked to minimize the length of the inter-exposure sequence, which is described in \S\ref{sec:exposure} and summarized in Figure~\ref{fig:inter_exp}. Our requirement is that the inter-exposure sequence be less than two minutes, which is a Level 4 requirement that flows down from the requirement to maximize the survey efficiency. 

The times for the spectrograph CCD readout and the setup of the telescope are as short as practical, and both begin as soon as the shutter closes. The spectrograph readout time is 60\,s. This readout occurs in parallel with other tasks during the inter-exposure sequence, and those take sufficiently long that the inter-exposure sequence is not limited by the spectrograph readout time. The time to move the telescope, dome, and ADC depend on the relative locations of the current and next field. Early in the project we used the move parameters of the telescope's Hour Angle and Declination drives, the dome motors, and the ADC motors to simulate the distribution of times for each of these motions to complete for simulated survey observations. These actions happen in parallel, and the telescope dominates 88\% of the time, while the dome dominates the remaining 12\%. The mean (median) time for these moves to complete is 41\,s (27\,s). Based on our operations experience to date, moves complete somewhat more quickly, possibly because we have continued to optimize the survey software to minimize slew distances. While the telescope is in motion, we adjust the location of the hexapod for the new position based on a static look-up table and perform the blind of the fiber positioners to set up on the new field. The hexapod adjustment takes approximately 5\,s and the execution of the blind move takes 35\,s.

Once the telescope is in position, we obtain and analyze a 15\,s acquisition image with the guider GFAs. We have pre-established guide stars for every field, and we have pre-established where those guide stars need to be located on the GFAs in order to ensure that the fibers are aligned with their targets. It takes 4\,s to analyze the GFA images and about 7\,s to adjust the telescope pointing, start the hexapod rotator, and initialize the guider. In total, we typically complete the field acquisition and are ready to begin guiding about 30\,s after the end of the telescope slew. For very small slews, we are ready to begin guiding as soon as 40\,s after the spectrograph shutters closed to end the previous exposure. Once we begin guiding, we are ready to back-illuminate the fibers, turn on the fiducuals, obtain an FVC image, and calculate and execute the correction move. The correction move is typically complete 51\,s after guiding begins. Once this step is complete, we take a second FVC image to record the final locations of the fiber positioners, turn off the back-illumination and fiducials, and are ready to start the exposure 15\,s after the completion of the correction move. At present we have achieved inter-exposure times as short as 118s between fields that are close together on the sky. 

\subsection{Performance Problems and Solutions} \label{sec:problems}

 Early in the operations phase we experienced problems with the CAN communication to positioners. This would manifest as a loss of communication with an entire CAN bus and cause the temporary loss of communication to the devices on that bus. This would result in the loss of 28 to 78 devices. While we could recover from this error condition, the recovery procedure was sufficiently long that it proved more efficient to observe the rest of the night without these positioners and to perform the recovery procedure the next day. We traced this problem to communications problems with individual positioners that would disrupt an entire bus. To minimize disruption, we redesigned and changed all of the transverse boards to add a relay for each positioner. This provided the flexibility to switch off individual problematic positioners, rather than risk temporary loss of an entire bus. We also increased the number of CAN channels from 10 to 20, which reduced the maximum number of devices per CAN bus to 28. As the transverse boards are directly behind the petals, this required a major operation: the disassembly of the FPE and the removal of the petals to obtain sufficient access. We successfully conducted this upgrade during the summer 2021 monsoon season and we are now able to maintain higher observing efficiency. 
 
 The other main performance issue that we identified during early testing at the Mayall was that some positioners would occasionally stop moving the commanded amount, which we traced to performance problems with the motor for either the $\phi$ or $\theta$ arm. The most common failure mode was that the motor would start to move a fairly constant fraction of the requested move, and we referred to this as the ``linear $\phi$'' (or $\theta$) problem. When the petals were removed in the summer of 2021, we extracted several of these problematic positioners that were near the edge of a petal and consequently possible to extract with minimal risk to neighbors. High-resolution CT scans of these positioners showed the pinion gear inside of the motors had cracked radially through the wall thickness, which allowed the gear to slip relative to the motor shaft. The pinion gear had been attached to the motor shaft via a press fit. We hypothesize that some assemblies may have exceeded the (very challenging) tolerance limits which press fits at this small scale can require, and that the stress induced in these cases initiated a slow-growing crack in the affected gears. A particular challenge in the detection of the failures and then the identification of the root cause was that affected gearboxes may spin properly in the horizontal orientation (typical of our lab tests) and only become symptomatic when pointed downward (as they are at the zenith position of the telescope). As of January 2022, about 14\% of positioners are not used on targets due to mechanical or electrical problems, although most of these naturally fall on blank sky positions and are therefore still useful. While the cracked motors cannot be replaced without rebuilding the petals, we are exploring operational changes to regain the use of the motors that show linear behavior.   

 \section{Successes and Lessons Learned} \label{sec:lessons}
 
 In this section we describe the key management decisions and actions that were especially helpful, and in some cases critical, for the completion of DESI ahead of schedule and under budget. We also take advantage of hindsight to identify actions that we could have taken, or could have taken earlier, that would have helped with the schedule and/or budget. These lessons learned are most relevant within the broader context of a project that follows the DOE project management structure. This at minimum includes that the project has clearly defined requirements, seeks to follow best practices for project management, has a well-defined systems engineering approach, develops a risk management plan, and maintains a risk registry. 
 
 The most important early action was to assemble a capable leadership team with substantial experience with previous large projects. In this respect, and in many others, DESI drew in particular on the community of expertise that was developed by the earlier Stage III and II dark energy projects, most notably DES and the SDSS BOSS projects. This leadership group  made several important early decisions that played an outsized role in the success of the project. Potentially the most important was to define interfaces and requirements early, along with the verification methods and acceptance criteria for the requirements. This was especially valuable given the dispersed nature of the instrument development, as we used these requirements and their verification methods to set clear expectations for the scope of work of various partner institutions, as well as to negotiate contracts with vendors for critical early procurements. The requirements that we wrote at the start of the project in 2014 were sufficiently robust that we did not change them through the end of CD-4, which is an impressive testament to value of our early R\&D and a major success. A second important decision was to seek early funding to begin the longest-lead procurements, which were the optics, detectors, and spectrographs. We would not have completed the project on time and on budget without the early definition of the requirements for these key components, our success with early fundraising, and early R\&D. Another key feature of project planning was the allocation of adequate cost and schedule contingency, especially to accommodate the tendency of less experienced control account managers to underestimate costs. Two lessons learned regarding management were the value of experienced control account managers and that having fewer control account managers simplifies earned value management. 
 
 Our experience with large, complex procurements underscored the importance of best-value assessments of bids to avoid commitments to substandard, lowest bids. The most important considerations were minimum standards for vendors and any subcontractors and the award of contracts to vendors that had an established track record with similar work (if they existed). For example, we anticipated the most likely problems that could arise in a procurement and required that vendors possess the equipment they would need to solve such problems. The selection criteria need to be clearly specified in the Request for Proposals along with a clear description of the tests that should be conducted to demonstrate that requirements have been met. We recommend trying to award contracts for difficult, specialized work to experienced vendors, to consult with experts and previous customers to assess a vendors capabilities and responsiveness, and to try to back-load the payment schedule as much as practical to maintain a vendor's engagement. 
  
 Some specific successes with the instrumentation beyond procurement included the detailed finite element analysis of the corrector system, the decision to modularize the focal plane system into ten petals, the involvement of a production engineer in the fiber system work starting with the transition from the research and development stage to production, testing the complete spectrographs at the optics vendor, and the modular and scalable architecture for the ICS. Lastly, we benefited from the dedication of experienced teams on all of the subsystems, and especially note the extremely transferable experience of the teams at FNAL and UCL with the DECam corrector, CEA-Saclay for Megacam, UCB with optics and spectrographs, NAOC with LAMOST, the Ohio State team with the instrument control system for DECam and numerous spectrograph projects (MODS, OSMOS), and the experience of the LBNL team with optics, CCDs, fiber positioners, and the instrument upgrades of the BOSS spectrographs. 
 
 There were also several notable challenges. One of the larger challenges was the fabrication and testing of the fiber positioners, largely because of the sheer scale of the number of components. Each positioner contains dozens of parts that required careful quality control, inventory tracking, and precision tooling for assembly. We would have benefited from more resources for the quality assurance/quality control of the positioner production as early as the development and planning stage. We would also have benefited from a large-scale, pre-production run of at least 1000 units, as smaller batches are insufficient to uncover all issues and may produce insufficient disposable units for integrated tests. This would have helped to fine tune the manufacturing process and potentially have alleviated some delays and added costs. We should have conducted more positioner testing in the same gravity orientation as they experience in nighttime operations, as we may then have identified the issue with the press fit of the gears earlier. This testing should have included more resources to diagnose any failed positioners. We also underestimated the yield on the production of science grade CCDs for the spectrographs and ultimately needed to purchase additional devices. Lastly, we largely developed the positioner software independent of the ICS. We would have benefited from more software developer resources earlier in the project to methodically coordinate the interfaces between multiple subsystems and institutions. 
 
 The main successes of the Assembly, Integration, and Test phase included prototypes, thorough planning and test runs of critical/high value installation steps, and proper staffing levels. The pre-commissioning activities with ProtoDESI and the Commissioning Instrument were invaluable for technical demonstrations. The experience we gained with operations was also important, as this helped us more appropriately staff commissioning. We similarly gained valuable knowledge and risk reductions with test stands for numerous other components, most notably the testing of the completed petals and the completed spectrographs prior to shipment. As noted in \S\ref{sec:fpsinstall}, we built a complete petal installation setup on the ground floor of the Mayall and practiced petal installation with engineering dummy petals. This experience helped us to improve the installation procedures and led to some minor hardware modifications. Lastly, appropriate staffing levels were absolutely critical to the success of this very stressful and labor-intensive phase of the project. One potential area for improvement would have been to more directly address subsystem requirements in design reviews. In some instances the review documentation did not explicitly address the relevant subsystem requirements, which unnecessarily complicated the subsystem acceptance process. Another area for improvement would have been earlier integration of subsystems. More complete and integrated lab testing and higher-fidelity hardware simulators may have helped to identify integration challenges that we otherwise uncovered during nighttime commissioning. 
 
\section{Summary} \label{sec:sum}

The Dark Energy Spectroscopic Instrument (DESI) is the largest multi-object spectrograph constructed to date and was designed to efficiently conduct a comprehensive spectroscopic exploration into the nature of dark energy. DESI will survey 40 million galaxies and quasars in just five years, and we forecast it will measure dark energy parameters sufficiently well to become the first completed Stage IV survey according to the definition developed by the Dark Energy Task Force. 

DESI is able to conduct this unprecedentedly large survey in such a relatively short time due to the substantial instrumentation development described in this paper. The three most fundamental aspects that enable these goals are that DESI can record 5000 spectra in a single exposure, has very high throughput, and maintains excellent operational efficiency. In \S\ref{sec:reqs} we provided a detailed description of the science and survey requirements, how they were used to produce the technical requirements, and how the instrument development was managed. We then described the main components of the instrumentation. The most noteworthy components, requirements and technical challenges are: 

Corrector (Section \ref{sec:corr}): The prime focus corrector has six lenses that produce a $3.2^\circ$ diameter field of view. Four of the lenses are constructed from fused silica, two of which have an aspheric surface, and the remaining two form an atmospheric dispersion corrector constructed from borosilicate glass. The lenses are approximately a meter in diameter and are coated with extremely high-performance, broadband anti-reflection coatings. Two key achievements with the corrector are superb image quality and extraordinarily high throughput. Based on the as-built optics and alignment precision, the performance meets requirements, including a flux-weighted mean FWHM $<0.4''$ over the full $360 - 980$\,nm bandpass at zenith. The coatings meet their average requirement of $>98.5\%$ transmission over the same bandpass by approximately 0.5\% per coating, which corresponds to a net gain of about 5\%  for the product of all twelve surfaces. 

Corrector support system (Section \ref{sec:corrsupp}): The corrector barrel and lens cells meet very high static and dynamic tolerances, some of which are as challenging as $20\,\mu$m for lateral alignment and $20\,\mu$rad for tilt. The corrector support system includes a new top ring that was mounted on the existing Serrurier truss. The replacement of the old, massive split ring is the main reason that the total mass and moment of the Mayall top end is relatively unchanged in spite of the new corrector and other DESI instrumentation. The new ring has a set of vanes that support the prime focus cage, and the corrector is mounted in this cage with an extremely precise hexapod with six degrees of freedom. We use the hexapod in conjunction with a static lookup table and the analysis of intra- and extrafocal images to maintain the optical alignment of the corrector with the Mayall primary mirror. 

Focal plane system (Section \ref{sec:fps}): The focal plane system includes the FPA with the robotic fiber positioners, illuminated fiducials, and GFA cameras, as well as the enclosure and FVC. The FPA is mounted at the prime focus and is subdivided into ten identical, $36^\circ$ wedge-shaped petals that place the fiber tips on the aspheric focal surface of the corrector. Each of the ten petals has 502 robotic fiber positioners with fibers that connect to one of the spectrographs, 12--14 illuminated fiducials, and a GFA camera. The key technology that enables DESI to fit all of the fiber positioners in an 0.812\,m diameter focal surface are the two 4\,mm diameter brushless DC motors in each fiber positioner. Placing the fiber tips on astronomical targets with an accuracy of $\sim 10\,\mu$m ($0.2''$) requires many interconnected components to work well. These include the FVC, precision metrology of the petals, illuminated fiducials, the GFA cameras, back-illumination of the fiber cables, optical models of the FVC and corrector optics, careful calibration of the individual fiber positioners, and precise and accurate transformations from celestial coordinates to focal surface.

Fiber system (Section \ref{sec:fibers}): The fiber system achieved its key requirement of transmitting light from the focal surface to the spectrographs with minimal throughput loss. Several factors were critical to this success, including careful control of stresses on the fibers, precision cleaves of the fibers, AR-coating of the fiber tips, careful alignment of the fiber tips in the slitheads, and fusion splicing. Fusion splicing connected the short lengths of fibers in the positioners to the remainder of the fiber cables and slitheads once the positioners were assembled into the petals. This resulted in the creation of continuous, high-throughput fiber cables from the focal surface to the spectrographs.

Spectrograph system (Section \ref{sec:spec}): The spectrograph system includes ten spectrographs, their environmental enclosure, the calibration system, and the sky monitor system. Each spectrograph has three channels with a distinct grating, spectral resolution, and camera design that optimizes the performance of that channel. Together the channels record the flux of each object from $360 - 980$\,nm with extraordinarily high throughput. Specific contributors to the high throughput include very high efficiency volume phase holographic gratings and superb coatings on the collimator mirror, dichroics, optical elements in each camera, and very sensitive CCDs. The spectroscopic calibration is extraordinarily stable due to the tight temperature control of $<0.1^\circ$C provided by the environmental enclosure, and it is well determined due to the careful design of the calibration lamp system and dome screen. 

Instrument control system (Section \ref{sec:ics}): The ICS is responsible for the operation and coordination of all instrument components and the telescope. These include the selection of each field, telescope pointing and guiding, adjustment of the hexapod's six degrees of freedom, readout of the ten GFA cameras, movement of all fiber positioners, operation of the 123 illuminated fiducials, activation of the 20 shutters and their pneumatic seals, illumination of ten LEDs on the exposure shutters to back-illuminate the focal plane, readout of 30 spectrograph CCDs, operation of the two cameras in the sky monitor system, control of the FVC, and recording of myriad temperature sensors and other telemetry data. The ICS coordinates the operation of all of components to maximize survey efficiency. The ICS also provides a warning and alert system for potential problems, as well as a database to track performance over time. 

Data systems (Section \ref{sec:datasystems}): Data systems includes software for target selection, survey planning, fiber assignment, nighttime data quality assurance (Nightwatch), data transfer, and the spectroscopic pipeline. Some of the key achievements of data systems are data quality feedback to observers within minutes of each exposure, the near-realtime transfer of raw data from KPNO to NERSC, and processing of the raw data into sky-subtracted, flux-calibrated spectra with classifications and redshifts by the morning after observations are complete. Survey planning software includes tools to monitor progress relative to expectations and to evaluate alternative strategies. The target selection and fiber assignment modules include the tools to convert astronomical targets into focal plane configurations at each telescope pointing.

Mayall Upgrades (Section \ref{sec:mayall}): We upgraded numerous components of the 4\,m Mayall telescope to prepare for the installation of DESI. An especially important upgrade was the modernization of the TCS, including new encoders for both the hour angle and declination axes, new servo controllers for the telescope drives, and a new control computer. These changes led to improvements in the pointing from $10 - 20''$ to $2 - 3''$ rms and an improvement in the open loop tracking errors from $30''$/hour to $1.6''$/hour in RA and $0.1''$/hour in declination. 

The acceptance, integration, and functional verification of DESI largely occurred in 2018 and 2019, with the peak of installation during a dedicated shutdown of the Mayall telescope for most of February 2018 through September 2019 (see \S\ref{sec:ait}). We then proceeded with commissioning to both test and optimize the complete instrument. During commissioning we verified a number of requirements that could only be tested with on-sky observations, such as the total throughput. The full list of key commissioning tests are described in \S\ref{sec:results}. Throughout the instrument development process, we tracked many key decisions and actions that kept the project on schedule and on budget. We also conducted a retrospective exercise to identify other actions and decisions that could have improved the schedule and/or budget. These successes and lessons learned are described in \S\ref{sec:lessons} and are intended to be a resource for other, similar projects. After commissioning, DESI started Survey Validation on 14 December 2020. During this phase, we validated the scientific requirements and specifically the target selection algorithms. The target selection for Survey Validation is by design more expansive than will be employed for the main survey, and includes longer spectroscopic exposures to better characterize redshift measurements with nominal survey exposures. The target selection algorithms were then frozen before the start of the main survey.  

The main survey began on 14 May 2021 and we are routinely observing 80,000 galaxies and quasars on every clear night with minimal technical downtime and very high on-sky efficiency. DESI measured over 15\,million extragalactic redshifts in the first year of main survey operations and is already the largest spectroscopic survey ever conducted. Data from DESI will lead to improved measurements of cosmic acceleration, more sensitive searches for time evolution of the dark energy component, stronger constraints on modified gravity models, and significantly improved upper limits on the sum of the neutrino masses. Upon the completion of the survey, DESI will have measured redshifts for approximately 14 million galaxies in the BGS, 8 million LRGs, 17 million ERGs, and 2.8 million QSOs. With these data we expect to measure the cosmic distance scale with at least $0.22$\% aggregate precision from $0 < z < 1.1$, $0.31$\% from $1.1 < z < 1.9$, and the Hubble parameter to $0.84$\% precision from $1.9 < z < 3.7$. DESI may distinguish between the inverted and normal neutrino mass hierarchy, should provide extraordinary new information about the nature of gravity through measurements of anisotropic clustering, and probe the nature of inflation through the amplitude of primordial non-Gaussianty in the matter power spectrum.

\acknowledgments

The DESI collaboration is indebted to the late John Donaldson, Robin Lafever, Tammie Lavoie, and Glenn Roberts for their many contributions to the success of this project and mourns their passing.

This research is supported by the Director, Office of Science, Office of High Energy Physics of the U.S. Department of Energy under Contract No. DE–AC02–05CH11231, and by the National Energy Research Scientific Computing Center, a DOE Office of Science User Facility under the same contract; additional support for DESI is provided by the U.S. National Science Foundation, Division of Astronomical Sciences under Contract No. AST-0950945 to the NSF’s National Optical-Infrared Astronomy Research Laboratory; the Science and Technologies Facilities Council of the United Kingdom; the Gordon and Betty Moore Foundation; the Heising-Simons Foundation; the French Alternative Energies and Atomic Energy Commission (CEA); the National Council of Science and Technology of Mexico (CONACYT); the Ministry of Science and Innovation of Spain (MICINN), and by the DESI Member Institutions: \url{https://www.desi.lbl.gov/collaborating-institutions}.

The authors are honored to be permitted to conduct scientific research on Iolkam Du’ag (Kitt Peak), a mountain with particular significance to the Tohono O’odham Nation.

For more information, visit \url{https://desi.lbl.gov}.

This work has made use of data from the European Space Agency (ESA) mission {\it Gaia} (\url{https://www.cosmos.esa.int/gaia}), processed by the {\it Gaia} Data Processing and Analysis Consortium (DPAC, \url{https://www.cosmos.esa.int/web/gaia/dpac/consortium}). Funding for the DPAC has been provided by national institutions, in particular the institutions participating in the {\it Gaia} Multilateral Agreement.

\section*{Data Availability}

All data points shown in the published graphs are available in a machine-readable form on the following website: https://zenodo.org/record/6397426. 

\vspace{5mm}
\facility{Mayall (DESI)}



\end{document}